# Programa de Erlangen para o Espaço-Tempo por meio Álgebra Geométrica do Espaço-Tempo Induzida pela Característica *R* Vetorial do Anel dos Números Híbridos Z[1]

## *Erlangen's Program for Space-Time through Space-Time Geometric Algebra Induced by the R Vector Characteristic of the Ring of Hybrid Numbers Z*

**Ricardo Capiberibe Nunes**[2]

**Universidade Federal de Mato Grosso do Sul**

**Resumo:**
Este ensaio sintetiza os esforços necessários para se construir um programa de uma topologia de baixa dimensão unificada que permita caracterizar todos estes espaços-tempos planos. Como as variedades espaço-temporais são espaços topológicos munidos de métrica, suas propriedades são caracterizadas pelas álgebras de Clifford em anéis hipercomplexos associativos com unidade, de forma que as transformações de Galileu são induzidas por um número dual; as transformações de Lorentz, por um número perplexo e as transformações de Euclides, por um número complexo. Esse fato nos levou a estabelecer um automorfismo interno no anel dos números híbridos que atua como um mapa da variedade e induz a métrica do espaço-tempo a partir da qualidade (característica) da unidade hipercomplexa associada. A partir desse automorfismo construímos funções trigonométricas híbridas, que chamamos de funções de Poincaré e que permitiram deduzir propriedades gerais do espaço-tempo, das geometrias hiperbólicas, parabólicas e elípticas e dos grupos SO(3), SO(4) e SO(1,3). Essa abordagem permite destacar as propriedades globais do espaço-tempo, sugere métodos para modelos geodinâmicos e permite interpretar a anti-matéria como a matéria em um espaço-tempo euclidiano onde o a natureza do tempo é imaginária.

**Abstract:**
This essay summarizes the efforts required to build a program of a unified, low-dimension topology that allows characterizing all these flat space-times. Since spatiotemporal manifolds are topological spaces equipped with metrics, their properties are characterized by Clifford algebras in hypercomplex rings associative with unity, so that Galileo's transformations are induced by a dual number; the Lorentz transformations, by a perplexed number and the Euclid transformations, by a complex number. This fact led us to establish an internal automorphism in the ring of hybrid numbers that acts as a map of the manifolds and induces the space-time metric based on the quality (characteristic) of the associated hypercomplex unit. From this automorphism, we built hybrid trigonometric functions, which we call Poincaré functions, which allowed us to deduce general properties of space-time, hyperbolic, parabolic and elliptical geometries and the groups SO (3), SO (4) and SO (1, 3). This approach allows us to highlight the global properties of space-time, suggests methods for geodynamic models and allows us to interpret anti-matter as matter in a Euclidean space-time where the nature of time is imaginary.

---

[1] "O presente trabalho foi realizado com apoio da Universidade Federal de Mato Grosso do Sul - UFMS/MEC - Brasil"
[2] Contato: ricardo.capiberibe@ufms.br

## Introdução

O Programa de Erlangen (Erlanger Programm) foi uma proposta de unificação das geometrias por meio de grupos simetria, realizada em 1872 pelo matemático alemão, Félix Klein. Motivados por esse trabalho, nós propomos um programa de unificação das álgebras geométricas de variedades espaço-temporais planas. Motivados pelo programa de Erlangen, este trabalho é a síntese de uma ampla pesquisa teórica que objetivava responder a seguinte questão: como é possível unificar todos os espaço-tempos planos em uma única estrutura? Por espaço-tempo plano entendemos qualquer variedade ou espaço topológico que satisfaça o princípio da relatividade e as conexões de Riemann-Christofell se anulem sobre todos os pontos da variedade. A organização desse trabalho pode ser categorizado em duas partes.

Na primeira parte apresentamos os nove parâmetros de Cayley e as geometrias associadas. O exame das propriedades do espaço-tempo apresentadas por Poincaré (1905, 1906), Einstein (1905, 1920) e Minkowski (1909), reduzimos os espaço-tempos possíveis a três possibilidades: euclidiano, galileano e minkowskiano. Empregando os métodos da álgebra geométrica, provamos que cada um destes espaço-tempos pode ser associado a uma unidade hipercomplexa $R$, a saber: imaginária (euclidiano), dual (galileano), perplexo (minkowskiano). Por meio de uma variante de um método apresentado por Brown (2017) e Josipovic (2019) deduzimos as transformações gerais do espaço e do tempo entre diferentes observadores inerciais, das quais as transformações de Lorentz e Galileu são apenas casos particulares. Ao final da primeira parte analisamos cada um destes espaço-tempos e suas propriedades particulares.

Na segunda parte, apresentamos uma proposta de uma estrutura unificada a partir do emprego do anel dos números (quartenions) híbridos. A estrutura dos números híbridos foi amplamente discutido em duas publicações recentes de Özdemir (2018, 2019), por isso nos limitamos a apresentar as suas principais propriedades. A execução desse programa exige a construção de uma função híbrida trigonométrica que denominamos de função de Poincaré. As principais propriedades dessa função são discutidas em detalhes. Deduzimos a métrica geral do espaço-tempo e mostramos que as funções de Poincaré produzem rotações no plano híbrido e coincidem com as transformações do espaço e tempo que deduzimos nas seções anteriores. Por fim, analisamos as (des)igualdades triangulares no espaço-tempo e seu significado geométrico.

A terceira parte, trata-se de uma discussão sucinta sobre porque a variedade minkowskiana é a melhor configuração para a descrição dos fenômenos físicos. Minkowski (1909), comparou as variedades galileana e minkowskiana, e justificou a escolha com base em argumentos de inteligibilidade matemática, mas não apresentou nenhum contra-argumento as variedades euclidianas. Mais recentemente, Brown (2017) rejeitou a estrutura euclidiana ao afirmar que não há evidências empíricas sobre curvas fechadas no tempo. Nosso argumento baseia-se que o método de sincronização de relógios propostos por Poincaré (1898, 1900, 1904) e Einstein (1905) exige uma estrutura minkowskiana.

Acreditamos que qualquer leitor familiarizado com os aspectos físicos-matemáticos da teoria da Relatividade poderá acompanhar esse texto sem dificuldades. Detalhes conceituais podem ser vistos em Langevin (1913, 1922) e Martins (2012). Por outro lado, indicamos uma ampla literatura sobre o formalismo das álgebras geométricas, números hipercomplexos e números híbridos, que o leitor poderá consultar para sanar as eventuais dúvidas que apareçam durante o texto.

## Parâmetros de Cayley

De acordo com os parâmetros de Cayley, todas as geometrias do espaço podem ser caracterizadas por um par ordenado (*m, n*), onde *m* é a medida da distância e *n,* a medida do ângulo e cada uma destas medidas pode assumir um dos seguintes valores, a saber: -1 (elíptica), 0 (parabólica), +1 (hiperbólica) (YAGLOM, 1968, 1979). Uma geometria será definida pelo conjunto de todos de todos pares ordenados G: $\{(m, n), \forall\ m, n \in (-1, 0, +1)\}$. Desta forma teremos nove geometrias, com mostra a figura abaixo:

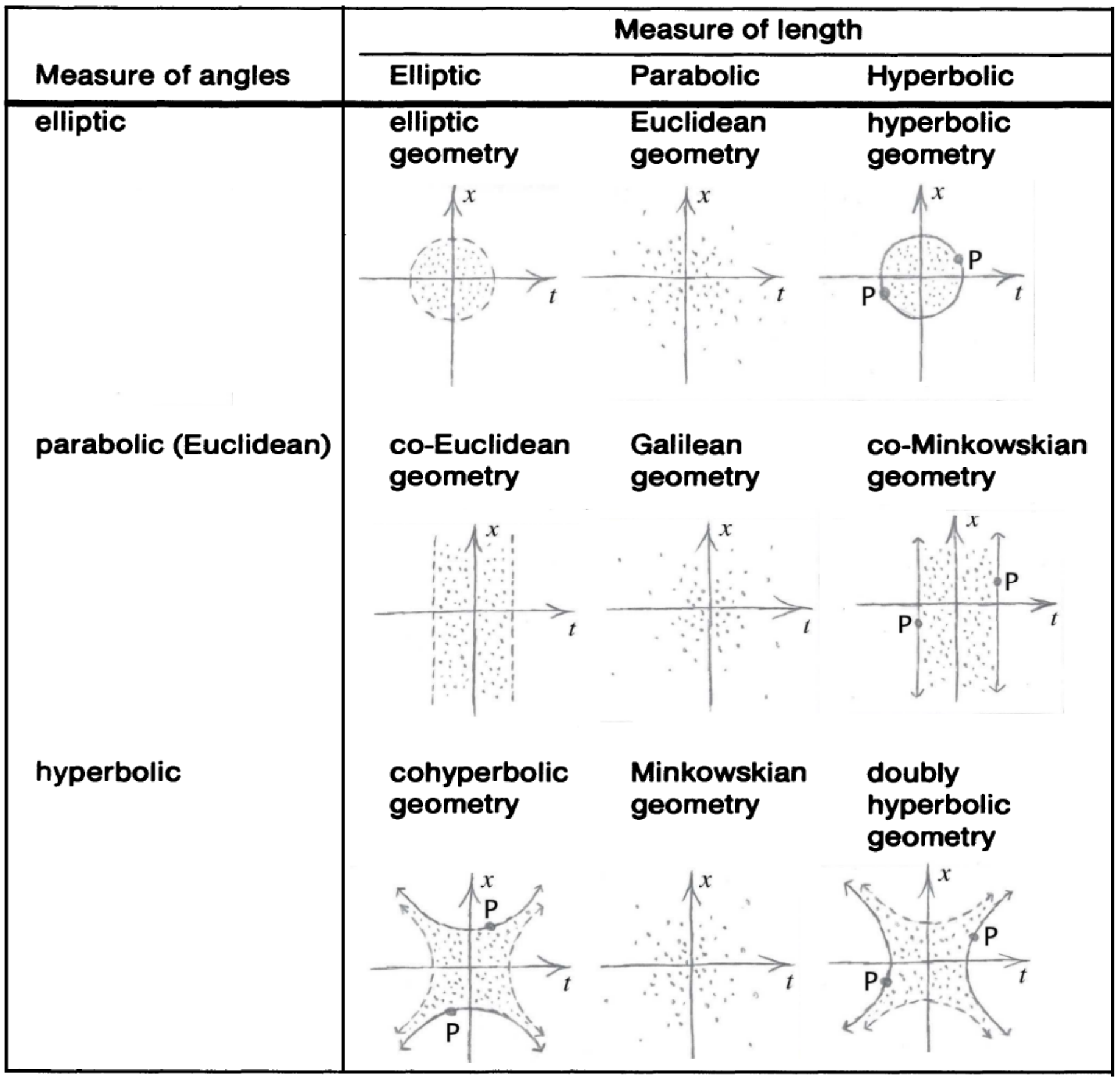

As 9 geometrias planas que podem ser construídas pela combinação dos parâmetros de Cayley.
**Fonte**: Adaptado de Yaglom (1968, 1979) e MCRAE (2007).

Observe que podemos definir o conjunto G como: G: $\{(R_x^2, R_y^2) \mid R_i \in (i, \varepsilon, h), \text{com } (i = x, y)\}$, onde $i$ é a unidade imaginária ou elíptica; $\varepsilon$, a unidade dual ou parabólica; $h$ é a unidade perplexa ou hiperbólica. Esta escolha permite caracterizar a geometria do espaço por meio da característica do ideal do anel dos números hipercomplexos Z (Özdemir, 2018):

$$\mathbb{HC}: \left\{ Z = a + b\mathbf{R}, \quad \mathbf{R} \in \{i, \varepsilon, h\}, \mathbf{R}^2 \in \{-1, 0, +1\}, \; a, b \in \mathbb{R} \right\}$$
$$\mathbb{R}[z] / \left\langle z^2 - R_i^2 \right\rangle \quad (i = x, y)$$

A partir dessa formulação, propomos uma álgebra geométrica do espaço-tempo plano que é induzida pela característica e derivamos suas propriedades gerais. Esse formalismo se mostra particularmente útil, porque a álgebra do espaço-tempo se torna isomorfa a álgebra dos números híbridos, discutidas em detalhes por Özdemir (2018, 2019).

# Espaço-Tempo Plano

O espaço-tempo construído por Poincaré (1905, 1906) e Minkowski (1909) é definido como um espaço vetorial afim $M$ sobre um corpo de números reais, de dimensão 4 (três espaciais, que formam um contínuo euclidiano; e uma temporal) munido de uma forma bilinear $g$ que define a sua (pseudo-)métrica (Vaz Jr., 1999). Os pontos nesse espaço são chamados de *eventos.* Uma linha que conecta dois eventos é chamado de *linha de mundo.* Uma geodésica nesse espaço-tempo é uma linha reta que conecta dois eventos, portanto a *medida da distância* é *parabólica*. O espaço-tempo $M$ satisfaz cinco princípios (ou leis) (Poincaré, 1902, Einstein, 1920):

1) Princípio da Homogeneidade
2) Princípio da Isotropia
3) Princípio da Inércia
4) Princípio da Relatividade
5) Princípio da Perda de Memória[3]

Estas propriedades implicam que as leis da física devem ser covariantes para todos os referenciais inerciais e as transformações das medidas de espaço e tempo entre estes referenciais devem ser descritas por transformações lineares ortogonais. Em termos mais técnicos, tais transformam devem formar um grupo SO com uma álgebra de Lie so associada. Embora o espaço-tempo seja uma estrutura quadridimensional (1-3), restrigiremos restringir nossa análise ao espaço-tempo plano bidimensional (1-1), visto que generalizar os resultados para espaço-tempo (3-1) não é um exercício difícil (Vaz Jr., 2007).

[3] O princípio da perda de memória alega que qualquer informação do passado de um sistema físico não pode ser recuperado a partir das condições que o sistema apresenta no presente.

Formalmente, definimos o **espaço-tempo** $\mathcal{E}$ como (GOURGOULHON, 2013, p. 03)

> [...] Um espaço afim de dimensão 4 em *R*. Observaremos E o espaço vetorial subjacente, que é isomórfico para $R^4$. Os elementos de $\mathcal{E}$ são chamados de eventos e os de *E* são chamados de **vetores**, ou **quadrivetores**, abreviados como **4-vetores**.

Sobre o termo 4-vetor, Gourgoulhon (2013, p. 03) faz uma importante observação:

> O termo *quadrivetor* ou *4-vetor* introduzido pelo físico não representa nada além de um vetor para o matemático, ou seja, o elemento de um espaço vetorial (*E* no presente caso). O prefixo "4-" simplesmente lembra que esse vetor pertence a um espaço vetorial de dimensão 4 em *R*. Esses vetores são, portanto, diferenciados dos vetores de espaços vetoriais tridimensionais usualmente manipulados pelo físico não-relativista.

Essa distinção é necessária pois nas álgebras de extensão e, por conseguinte, nas álgebras geométricas é possível construir por meio do produto exterior de vetores (de *n* dimensões) objetos novos como os *p*-vetores e os multivetores. Por essa nomenclatura, percebemos que existe um objeto também cominado de *4-vetor na álgebra de extensão (e geométrica)*, porém esse objeto *não corresponde aos 4-vetores da Teoria da Relatividade Especial*. Com efeito, o *4-vetor relativístico corresponde ao 1-vetor*, com dimensão 4. Como neste trabalho usaremos álgebras geométricas, é preciso registrar que a nomenclatura 4-vetor refere-se ao 1-vetor de Minkowski.

Tsamparlis (2010, p. 94-96) elenca oito propriedades geométricas do espaço-tempo:

> Além de seus pontos, o espaço-tempo é caracterizado por sua geometria. Na Relatividade Especial, supõe-se que a geometria do espaço-tempo seja absoluta, no sentido de que as relações que o descrevem permanecem as mesmas - são independentes dos vários fenômenos físicos que ocorrem nos sistemas físicos. A geometria do espaço-tempo é determinada em termos de várias suposições que estão resumidas abaixo:
>
> (1) O espaço-tempo é um espaço linear real quadridimensional.
>
> (2) O espaço-tempo é homogêneo. Do ponto de vista geométrico, isso significa que todos os pontos no espaço-tempo podem ser usados equivalentemente como a origem das coordenadas. No que diz respeito à física, isso significa que onde e quando um experimento (isto é, evento) ocorre não afeta a qualidade e os valores das variáveis dinâmicas que descrevem o evento.
>
> (3) No espaço-tempo, existem retas, curvas sem limites, que são descritas geometricamente com equações da forma
>
> $$\boldsymbol{r} = \boldsymbol{a}t + \boldsymbol{b},$$
>
> onde $t \in R$, $\mathbf{a}$, $\mathbf{b} \in R^4$. Do ponto de vista da física, essas curvas são as trajetórias de movimentos especiais dos sistemas físicos em $R^4$, que chamamos de movimentos inerciais relativísticos. Todos os movimentos que não são movimentos inerciais relativísticos são chamados de movimentos aceleradores. A cada movimento acelerado, associamos uma quatro força de uma maneira a ser definida posteriormente.
>
> (4) O espaço-tempo é isotrópico, ou seja, todas as direções em qualquer ponto são equivalentes. Os pressupostos de homogeneidade e isotropia implicam que o espaço-tempo da Relatividade Especial é um espaço plano ou, equivalente, tem curvatura zero. Na prática, isso significa que é possível definir um sistema de coordenadas que cubra todo o espaço-tempo ou, equivalentemente, o espaço-tempo é difeomórfico (isto é, parece) ao espaço linear $R^4$.

(5) O espaço-tempo é um espaço afim, isto é, se for dada uma linha reta ou um hiperplano (= subespaço linear tridimensional com curvatura zero) no espaço-tempo, então (axioma!) Existe pelo menos um hiperplano paralelo a eles, no sentido de encontrar a linha reta ou o outro hiperplano no infinito.

(6) Se considerarmos uma linha reta no espaço-tempo, há uma sequência contínua de hiperplanos paralelos que cortam a linha reta uma vez e preenchem todo o espaço-tempo. Dizemos que esses hiperplanos foliam o espaço-tempo. Devido ao fato de não haver linhas retas absolutas no espaço-tempo, existem infinitas foliações. No espaço newtoniano, existe a linha reta absoluta do tempo. Portanto, há uma foliação preferida (a do tempo cósmico; veja a Fig. 4.3).

(7) O espaço-tempo é um espaço vetorial métrico. A necessidade da introdução da métrica é dupla: (a) Seleciona um tipo especial de sistemas de coordenadas (os sistemas cartesianos da métrica) que são definidos pelo requisito de que nesses sistemas de coordenadas a métrica tenha sua forma canônica (isto é, diagonal com componentes $\pm 1$). Associamos esses sistemas de coordenadas aos sistemas inerciais relativísticos. (b) Cada linha reta semelhante ao tempo define a foliação na qual os planos paralelos são normais para essa linha.

(8) A métrica do espaço-tempo é a métrica de Lorentz, ou seja, a métrica do espaço real quadridimensional cuja forma canônica é $(-1, 1, 1, 1)$. A seleção da métrica de Lorentz é uma consequência do Princípio da Relatividade de Einstein, como será mostrado na Seção. 4.7 O espaço-tempo dotado da métrica de Lorentz é chamado de espaço Minkowski (veja também a Seção 1.6). A seguir, preferimos nos referir ao espaço de Minkowski do que ao espaço-tempo, porque a Relatividade Especial não é uma teoria do espaço e do tempo, mas uma teoria de muito mais quantidades físicas, descritas com os tensores definidos sobre o espaço de Minkowski. Além disso, é melhor reservar a palavra espaço-tempo para a Relatividade Geral.

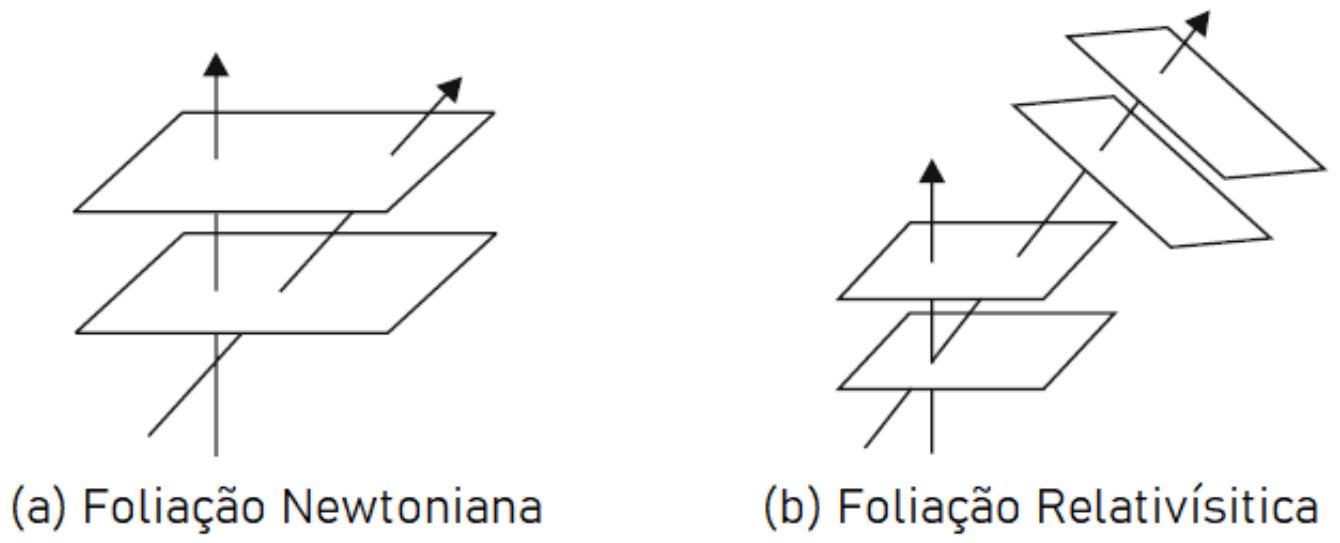

Foliações newtonianas e relativísticas do espaço-tempo

Sobre as colocações de Tsamparlis gostaríamos apenas de fazer uma pequena objeção ao item (8). Como veremos na próxima seção, a métrica de Lorentz não apenas é uma consequência do Princípio da Relatividade, de fato, o espaço de Galileu e o Espaço de Euclides também são consequências deste Princípio. A métrica de Lorentz é uma consequência da invariância das equações de Maxwell-Lorentz. De fato, historicamente, Lorentz (1904) deduziu suas transformações sem fazer referência ao Princípio da Relatividade. Essa junção foi realizada por Poincaré (1905, 1906) por meio da teoria de grupos.

# A Medida do Ângulo do Espaço-Tempo (1-1)[4]

No espaço-tempo (1-1), a medida do ângulo é calculado pelo quadrado do bivetor $e_x e_t$:

$$(e_x e_t)^2 = (e_x e_t)(e_x e_t) = -(e_t e_x)(e_x e_t) = -e_t (e_x)^2 e_t = -(e_t)^2 (e_x)^2$$

Como a componente espacial é euclidiana, então $(e_x)^2 = 1$, então:

$$(e_x e_t)^2 = -(e_t)^2$$

Desta forma, será o quadrado do vertor temporal que irá definir a medida do ângulo do espaço-tempo. Com base nesse resultado, vamos introduzir o vetor $e_r$,

$$\begin{cases} (e_r)^2 = -1 \\ e_t = (cR)e_r \end{cases}$$

com $R$ sendo a característica do ideal do anel dos números híbridos Z; $c$, a velocidade da luz no vácuo. Por simplicidade, escolheremos um sistema de unidades onde $c = 1$, de forma que nosso vetor temporal seja escrita como:

$$e_t = Re_r$$

Nestas condições, a medida do ângulo será definida por:

$$(e_x e_t)^2 = -(e_t)^2 = -(Re_r)^2 = -R^2 (e_r)^2 = -R^2(-1) = R^2$$

Consequentemente, a norma do bivetor será igual ao quadrado da característica $R$.

$$(e_x e_t)^2 = R^2$$

E a medida do ângulo e a natureza geométrica do espaço-tempo serão induzidas por $R$. A Tabela abaixo sintetiza algumas destas propriedades induzidas:

| R | Espaço-Tempo | Medida do Ângulo | Geometria | Trigonometria | $(m, n)$ |
|---|---|---|---|---|---|
| $i$ | **Euclidiano** | **Elíptica** | **Euclidiana** | **Polar** | **(0, -1)** |
| $\varepsilon$ | **Galileano** | **Parabólica** | **Não-Euclidiana** | **Parabólica** | **(0, 0)** |
| $h$ | **Minkowskiano** | **Hiperbólica** | **Pseudo-Euclidiana** | **Hiperbólica** | **(0, +1)** |

**Tabela 1:** Espaço-tempos e propriedades induzidas pela característica *R*. **Fonte**: Autoral.

[4] Como abordaremos alguns elementos de álgebras geométricas para detalhes recomendamos: Jancewicz (1989). Vaz Jr. (1997, 2000), Hestenes (1998, 2002, 2003, 2015) Doran, Lasenby (2003), Arthur (2011), Vaz Jr, Rocha Jr. (2012), Kanatani (2015), Vaz Jr., Mann, (2018), Josipović (2019).

# Transformações Lineares Ortogonais no Espaço-Tempo (1-1)

Diferentes observadores em diferentes referenciais inerciais realizam medidas com réguas (hastes rígidas graduadas) e relógios idênticos, cujas medidas não são afetadas por informações do passado, e que, segundo o princípio da relatividade, devem ser equivalentes. Como mencionamos anteriormente, essa equivalência entre os referenciais inercias, em um espaço homogêneo e isotrópico, exigem um conjunto de transformações lineares ortogonais, que para o espaço-tempo são chamadas de transformações de Galileu, para o espaço-tempo Galileano, e transformações de Lorentz, para o espaço-tempo Minkowskiano.

Para obtermos a regra geral, das quais as transformações de Galileu e Lorentz são apenas casos particulares, induzidos pela característica *R,* vamos descrever a situação contemplada por dois observadores inerciais O e O' que se encontram na origem do sistema de coordenadas ($x = x' = 0$) e se deslocam com velocidade relativa $v$ na direção longitudinal (que coincide com os eixos $x$, $x'$, ambos orientados no mesmo sentido). Da perspectiva do observador O', o observador O se desloca com velocidade $v$ no sentido positivo de $x$. Reciprocamente, na perspectiva O, é o observador O' que se desloca com velocidade $v$, mas no sentido negativo de $x$.

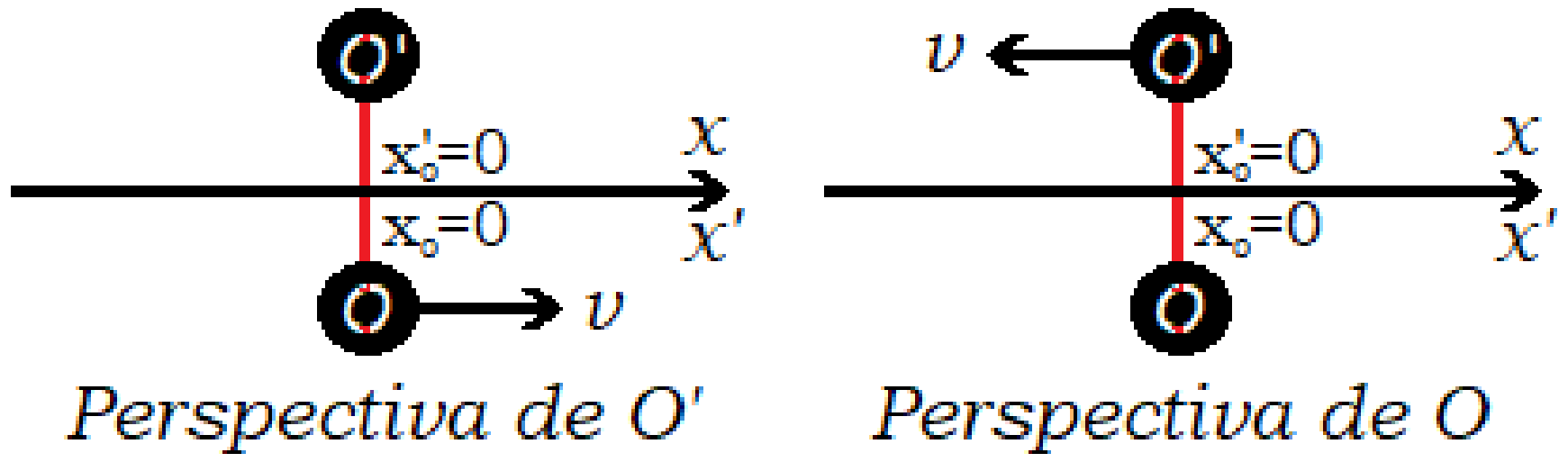

As transformações lineares ortogonais entre as medidas de espaço e tempo destes dois observadores deve ser da forma:

$$\begin{cases} X' = AX + BT \\ T' = CX + DT \end{cases}$$

que devido a ortogonalidade apresentam a seguinte restrição:

$$\det\begin{bmatrix} A & B \\ C & D \end{bmatrix} = 1$$

$$AD - BC = 1$$

Como temos três equações e quatro parâmetros, o sistema apresenta um grau de liberdade, isso implica que deveremos determinar o valor das incógnitas em função de uma delas, no nosso caso, optaremos por A.

Inicialmente, vamos considerar a situação descrita pelo observador O'. Da sua perspectiva, ele se encontra na origem do sistema de coordenadas e o observador O se desloca com velocidade *v* no sentido positivo de *x*.

$$0 = AX + BT$$
$$-BT = AX$$
$$B = -A\frac{X}{T}$$

Como a razão do espaço pelo tempo é sempre constante e igual a velocidade, concluímos que B deve ser igual a:

$$A = -Bv$$

Agora vamos determinar o valor da incógnita *D*. Para isso, invocaremos o ponto de vista cinemático do observador O. Do ponto de vista de O, ele se encontra na origem do sistema de coordenadas enquanto O' se desloca com velocidade constante no sentido negativo de *x*.

$$\begin{cases} X' = A0 + BT \\ T' = C0 + DT \end{cases} \sim \begin{cases} X' = BT \\ T' = DT \end{cases}$$

Dividindo as duas equações, encontramos:

$$\frac{X'}{T'} = \frac{B}{D}$$

Levando em conta o valor de *B* que calculamos e que a velocidade é constante, no sentido negativo de *x*:

$$-v = -\frac{Av}{D}$$

De onde concluímos que $A = D$.

Para determinarmos *C* em função de A, usamos a condição de ortogonalidade:

$$AD - BC = 1$$
$$AA - (-vA)C = 1$$
$$A^2 + (vA)C = 1$$
$$C = \frac{1 - A^2}{vA}$$

Por uma questão de conveniência, vamos escrever *C* da seguinte forma:

$$C = -KAv, \qquad K = \frac{A^2 - 1}{(vA)^2}$$

Convém, por razões que ficarão mais claras adiante, escrever $A^2$ em função de $K$:

$$Kv^2A^2 = A^2 - 1$$
$$A^2 - Kv^2A^2 = 1$$
$$A^2\left(1 - Kv^2\right) = 1$$
$$A^2 = \left(1 - Kv^2\right)^{-1}$$

Portanto, as transformações entre sistemas inerciais são:

$$\begin{cases} X' = A\left(X - vT\right) \\ T' = A\left(T - KvX\right) \end{cases}$$

Resta apenas determinarmos o valor de $A$, para isso recorreremos a outro aspecto matemático do espaço-tempo: sua álgebra de Lie.

## Álgebra de Lie do Espaço-Tempo (1-1)

Como as transformações entre referenciais inerciais são lineares e ortogonais, elas formam um grupo SO. Todo grupo SO apresenta uma álgebra de Lie so que permite a construção de invariantes que preservam a estrutura do espaço. Este foi o método empregado por Poincaré (1905, 1906) e Minkowski (1909). Para construirmos um invariante da álgebra de Lie do espaço-tempo empregaremos o método proposto por Hestenes (1998, 2002, 2003, 2015) conhecido como STA (Space-Time Algebra). Inicialmente vamos construir dois 1-vetores **S** e **S'**:

$$\begin{cases} \boldsymbol{S} = Xe_x + Te_t \\ \boldsymbol{S}' = X'e_x' + T'e_t' \end{cases}$$

O quadrado da norma de 1-vetor é um invariante da Álgebra de Lie:

$$\begin{cases} \boldsymbol{S}^2 = \left(Xe_x + Te_t\right)\left(Xe_x + Te_t\right) \\ \boldsymbol{S}'^2 = \left(X'e_x + T'e_t\right)\left(X'e_x + T'e_t\right) \end{cases}$$

$$\begin{cases} \boldsymbol{S}^2 = X^2\left(e_x'\right)^2 + XT\left(e_x'e_t' + e_t'e_x'\right) + T^2\left(e_t'\right)^2 \\ \boldsymbol{S}'^2 = X'^2\left(e_x'\right)^2 + X'T'\left(e_x'e_t' + e_t'e_x'\right) + T'^2\left(e_t'\right)^2 \end{cases}$$

Destas expressões, definimos a Álgebra de Clifford do espaço tempo:

$$\begin{cases} \left(e_x\right)^2 = \left(e_x'\right)^2 = 1, \quad \left(e_t\right)^2 = \left(e_t'\right)^2 = -R^2, \\ e_xe_t + e_te_x = 0, \qquad e_x'e_t' + e_t'e_x' = 0. \end{cases}$$

Nestas condições, o invariante do espaço-tempo pode ser escrito como:

$$\begin{cases} \mathbf{S}^2 = X^2 - R^2T^2 \\ \mathbf{S}'^2 = X'^2 - R^2T'^2 \\ \quad \mathbf{S}'^2 = \mathbf{S}^2 \end{cases}$$

Essa expressão é chamada de *Forma Quadrática Fundamental do Espaço-Tempo*.

Para determinarmos *A,* determinaremos o valor *K.* Para isso utilizaremos uma variante de um método proposto por Lugonov (2004). Primeiro, façamos uma mudança de variáveis:

$$\begin{cases} t = T - KvX \\ T = t + KvX \end{cases}$$

Substituindo na forma quadrática fundamental, obtemos:

$$\mathbf{S}'^2 = X^2 - R^2\left(t + KvX\right)^2$$
$$\mathbf{S}'^2 = X^2 - R^2\left(t^2 + 2KvXT + K^2v^2X^2\right)$$
$$\mathbf{S}'^2 = X^2\left(1 - R^2K^2v^2\right) - 2R^2KvXT - R^2t^2$$

Denotando o termo em parêntesis de $\Gamma^2$:

$$\mathbf{S}'^2 = \Gamma^2X^2 - R^2KvXT - R^2t^2$$

Vamos completar o quadrado, em relação a variável *X*:

$$\mathbf{S}'^2 = \Gamma^2X^2 - 2R^2KvXT + R^4K^2\Gamma^{-2}v^2t^2 - R^4K^2\Gamma^{-2}v^2t^2 - R^2t^2$$
$$\mathbf{S}'^2 = \left(\Gamma X - R^2K\Gamma^{-1}vt\right)^2 - R^2\left(1 - R^2K^2v^2\Gamma^{-2}\right)t^2$$
$$\mathbf{S}'^2 = \Gamma^{-2}\left(\Gamma^2X - R^2Kvt\right)^2 - R^2\Gamma^{-2}\left(\Gamma^2 - R^2K^2v^2\right)t^2$$

Calculemos o termo no segundo parêntesis:

$$\Gamma^2 - R^2K = 1 - R^2K^2v^2 - R^2K^2v^2$$
$$\Gamma^2 - R^2K = 1$$

Substituindo esse valor,

$$\mathbf{S}'^2 = \Gamma^{-2}\left(\Gamma^2X - R^2Kvt\right)^2 - R^2\Gamma^{-2}t^2$$

Substituindo o valor de *t* e Γ*,* obtemos:

$$\mathbf{S}'^2 = \Gamma^{-2}\left(\Gamma^2X - R^2Kv\left[T - KvX\right]\right)^2 - R^2\Gamma^{-2}\left(T - KvX\right)^2$$
$$\mathbf{S}'^2 = \Gamma^{-2}\left(\left[1 - R^2K^2v^2\right]X - R^2Kv\left[T - KvX\right]\right)^2 - R^2\Gamma^{-2}\left(T - KvX\right)^2$$

$$\boldsymbol{S}'^2 = \Gamma^{-2}\left(\left[1 - R^2K^2v^2\right]X + R^2K^2v^2X - R^2KvT\right)^2 - R^2\Gamma^{-2}\left(T - KvX\right)^2$$
$$\boldsymbol{S}'^2 = \Gamma^{-2}\left(\left[1 - R^2K^2v^2 + R^2K^2v^2\right]X - R^2KvT\right)^2 - R^2\Gamma^{-2}\left(T - KvX\right)^2$$
$$\boldsymbol{S}'^2 = \Gamma^{-2}\left(X - R^2KvT\right)^2 - R^2\Gamma^{-2}\left(T - KvX\right)^2$$

Escrevendo o lado direito da equação, obtemos:

$$X'^2 - R^2T'^2 = \Gamma^{-2}\left(X - R^2KvT\right)^2 - R^2\Gamma^{-2}\left(T - KvX\right)^2$$

Desta igualdade, obtemos as transformações de *X* e de *T*:

$$X' = \Gamma^{-1}\left(X - R^2KvT\right)$$
$$T' = \Gamma^{-1}\left(T - KvX\right)$$

Por outro lado, as expressões das transformadas gerais que obtivemos são:

$$X' = A\left(X - vT\right)$$
$$T' = A\left(T - KvX\right)$$

Comparando as duas equações para a transformação do tempo, obtemos:

$$A = \Gamma^{-1} = \left(1 - R^2K^2v^2\right)^{-1/2}$$

Elevando ao quadrado, obtemos:

$$A^2 = \Gamma^{-2}$$

Pela equação de *X,* obtemos os valores de *K.* Porém, devido a possibilidade de *R* poder ser nilpotente de ordem 2, vamos analisar esse problema em duas partes.

## Caso 1. Quando $R^2 \neq 0$

Nesta circunstância, temos que:

$$R^2 \in \{-1, +1\}, \quad R^4 = 1$$

Igualando as transformações de *X,* obtemos:

$$A\left(X - vT\right) = A\left(X - R^2KvT\right)$$
$$X - vT = X - R^2KvT$$
$$\left(R^2K - 1\right)vT = 0$$

Como o fator $vT$ é arbitrário, resulta que o termo em parêntesis deve ser nulo:

$$R^2K = 1$$

Multiplicando por $R^2$, e levando em conta que $R^4 = 1$, obtemos o valor de $K$:

$$K = R^2$$

Portanto o valor de $A^2$ é único e igual à:

$$A^2 = \frac{1}{1 - R^2 v^2}$$

Se extrairmos a raiz, teremos dois valores possíveis, correspondente da remoção do módulo de $A$. Porém, devemos escolher o valor positivo para garantir que as transformações formem um grupo[5].

$$A = \frac{1}{\sqrt{1 - R^2 v^2}}$$

Se escolhermos, $R^2$ igual à 1, obtemos as transformações de Lorentz. Se tomarmos $R^2$=-1, obtemos um novo tipo de transformação que chamaremos de transformações de Euclides. As transformadas do espaço se torna:

$$X' = A\left(X - R^2 R^2 vT\right)$$
$$X' = A\left(X - R^4 vT\right)$$
$$X' = A\left(X - vT\right)$$

que é a transformação do espaço, que obtivemos anteriormente.

Desta forma, as transformações do espaço e do tempo para essas duas variedades serão:

$$X' = A\left(X - vT\right) \qquad A = \frac{1}{\sqrt{1 - R^2 v^2}}$$
$$T' = A\left(T - R^2 vX\right)$$

## Caso 2. Quando $R^2 = 0$

Da forma quadrática fundamental, obtemos a seguinte igualdade:

$$\begin{cases} \boldsymbol{S}^2 = X^2 \\ \boldsymbol{S}'^2 = X'^2 \end{cases} \qquad \begin{matrix} \boldsymbol{S}'^2 = \boldsymbol{S}^2 \\ X'^2 = X^2 \end{matrix}$$

Por meio da relação entre $A$ e o fator $\Gamma$, podemos determinar o valor de $K$, para isso vamos usar as expressões elevadas ao quadrado::

$$A^2 = \Gamma^{-2}$$

---

[5] Para detalhes ver Poincaré (1906), Einstein (1905, 1907), Minkowski (1909), Miller (1986, 1997), Neto (2008), Bassalo e Cattani (2013), Hall (2016).

$$\left(1-Kv^2\right)^{-1}=\left(1-R^2K^2v^2\right)^{-1}$$
$$1-Kv^2=1-R^2K^2v^2$$
$$\left(1-R^2K\right)Kv^2=0$$

Para *v* arbitrário, obtemos:

$$\left(1-R^2K\right)K=0$$

Como $R^2$ é nulo, resulta que:

$$K=0=R^2$$

Levando em consideração que para $R^2$ não-nulo, *K* também é igual a $R^2$, então para todos os valores possíveis de *R,* temos que $K = R^2$. Também é fácil ver que nesse caso, $A^2$ é igual a unidade:

$$A^2=1 \quad \rightarrow \quad A=+1$$

onde escolhemos o valor positivo, novamente para que as transformações formem um grupo. Desse fato, decorre que:

$$X'=X, \qquad T'=T$$

A transformação de *X* decorre da invariância da forma quadrática fundamental do espaço-tempo de Galileu. Porém, ainda não provamos que a expressão

$$X'=A\left(X-vT\right)$$

mantém invariante a forma quadrática fundamental. Façamos a demonstração:

$$X^2=A^2\left(X-vT\right)^2$$
$$X^2=A^2X^2-2A^2vTX+A^2v^2T^2$$
$$A^2X^2-X^2+A^2v^2T^2-2A^2vTX=0$$

Vamos completar os quadrados em *X*:

$$A^2X^2-2X^2+X^2-2A^2vTX+A^2v^2T^2=0$$
$$X^2-2A^2vTX+A^2v^2T^2=2X^2-A^2X^2$$
$$\left(X-AvT\right)^2=\left(2-A^2\right)X^2$$

Como *A* é igual a unidade, as equações se tornam:

$$\left(X-1vT\right)^2=\left(2-1\right)X^2$$
$$\left(X-vT\right)^2=X^2$$

mas, o lado direito é justamente *X'*:

$$X'^2 = X^2$$

que demonstra que a transformação de *X'* mantém invariante a forma quadrática.

## Invariância de *R*

Vamos provar que o fator *R* é um invariante relativístico. Para isso, definamos a velocidade da partícula no sistema O':

$$\frac{x'}{t'} = \frac{A_v\left(x - vt\right)}{A_v\left(t - R_v^2 vx\right)}$$

$$\frac{x'}{t'} = \frac{x - vt}{t - R_v^2 vx}$$

Essa é a forma fundamental do grupo. Vamos agora construir um terceiro sistema de coordenadas O'' e estabelecer as transformações para o sistema O' e O.

$$\frac{x''}{t''} = \frac{x' - ut'}{t' - R_u^2 ux'}$$

$$\frac{x''}{t''} = \frac{A_v\left(x - vt\right) - uA_v\left(t - R_v^2 vx\right)}{A_v\left(t - R_v^2 vx\right) - R_u^2 uA_v\left(x - vt\right)}$$

$$\frac{x''}{t''} = \frac{\left[x - vt - ut + R_v^2 uvx\right]}{\left[t - R_v^2 vx - R_u^2 ux + R_u^2 uvt\right]}$$

Agora devemos reorganizar os fatores para que eles assumam a forma fundamental do grupo:

$$\frac{x''}{t''} = \frac{\left(1 + R_v^2 uv\right)x - \left(u + v\right)t}{\left(1 + R_u^2 uv\right)t - \left(R_v^2 v + R_u^2 u\right)x}$$

Evidenciando o primeiro fator do denominador, obtemos a forma fundamental:

$$\frac{x''}{t''} = \frac{x - \dfrac{\left(u + v\right)}{\left(1 + R_v^2 uv\right)}t}{\dfrac{\left(1 + R_u^2 uv\right)}{\left(1 + R_v^2 uv\right)}t - \dfrac{\left(R_v^2 v + R_u^2 u\right)}{\left(1 + R_v^2 uv\right)}x}$$

$$\frac{x''}{t''} = \frac{x - wt}{t - R_w^2 wx}$$

Comparando as equações, a primeira parcela no denominador deve ser a unidade:

$$\frac{\left(1+R_u^2uv\right)}{\left(1+R_v^2uv\right)}=1$$

$$1+R_u^2uv=1+R_v^2uv$$

$$\therefore \quad R_u^2=R_v^2\equiv R^2$$

A velocidade *w* é a lei de composição de velocidades relativísticas:

$$w=\frac{u+v}{1+R^2uv}$$

Por fim, vamos obter $R^2_w$:

$$\frac{\left(R^2v+R^2u\right)}{\left(1+R^2uv\right)}=R_w^2w$$

$$R^2\frac{\left(u+v\right)}{\left(1+R^2uv\right)}=R_w^2w$$

$$R^2w=R_w^2w$$

$$\therefore \quad R_w^2=R^2$$

Isso demonstra que o fator *R* não depende da escolha do sistema de coordenadas.

# Números Hipercomplexos e a Caracterização do Espaço- Tempo

Nas seções anteriores mostramos que as unidades hipercomplexas (imaginária, dual, perplexa) *R,* induzem o comportamento do espaço-tempo. Mais precisamente, essas unidades operam diretamente sobre o tempo. Nessa seção discutiremos cada um dos três espaço-tempos induzidos por *R*. Mais precisamente, mostraremos que o efeito de cada unidade está em modelar a forma (topologia do tempo), preservando a ortogonalidade entre o tempo e o espaço, conforme as seguintes regras:

1) A unidade imaginária induz o tempo a ter a forma de uma esfera $S^1$.
2) A unidade dual induz o tempo a se contrair em um ponto sem dimensão.
3) A unidade perplexa induz o tempo a ter forma de uma linha reta *R*.

Por essa razão, iniciamos o texto com uma breve revisão sobre a topologia do tempo.

## A Topologia do Tempo

Apesar dos esforços empreendidos, a questão da natureza do tempo ainda se encontra sem solução, principalmente porque ainda não fomos capaz de construir uma topologia para o tempo.

> Os filósofos debateram questões como as seguintes: O tempo pode ser "fechado" ou "circular"? Os debates geralmente assumem que é claro o que significa dizer, por exemplo, que o tempo é "circular", e o debate se concentra em saber se a possibilidade de tal estrutura de tempo é uma possibilidade real. Acontece, no entanto, que a noção de topologia do tempo assume uma qualidade bastante ilusória no contexto dos espaços-tempos relativísticos. (EARMAN, 1977, p. 211)

Portanto, o problema ontológico do tempo se torna um problema topológico. Como construir uma topologia para o tempo?

> Um procedimento que sempre pode ser aplicado é atribuir a topologia de projeção do espaço quociente *M/T(V)*, ou seja, se *p: M → M/T(V)* denota o mapa de projeção natural, então um subconjunto $X \subset M/T(V)$ é considerado aberto se $p^{-1}(X)$ estiver aberto em *M*. É um lema padrão que a topologia de projeção seja a maior topologia em que *p* é $C^o$. A topologia de projeção também possui o bom recurso de se comportar bem em relação aos subespaços. Seja *Y* um subconjunto de *M/T(V)*. Há duas maneiras de atribuir uma topologia a *Y*: a topologia do subespaço que ela herda de *M/T(V)* ou a topologia de projeção que ela herda de $p: p^{-1}(Y) \rightarrow Y$. Devido ao fato de que p: *M → M/T(V)* é um mapa aberto, essas duas topologias para *Y* coincidem. (EARMAN, 1977, p. 212)

Essa abordagem, embora seja um caminho viável para construir uma topologia do tempo apresenta um inconveniente ela exclui todas soluções em que o tempo não pode ser separado em classes laterais, como observa Earman:

> Essa abordagem assume que o espaço-tempo *<M, g>* em consideração admite uma família *s* de fatias de tempo que particionam *M*. Essa é uma forte suposição. Isso exclui muitos modelos cosmológicos interessantes. Por exemplo, o espaço-tempo de Godel não possui uma única fatia de tempo; e há outros tempos espaciais que possuem alguns intervalos de tempo, mas não podem ser particionados por eles. Mas se a abordagem de projeção estiver no caminho certo, pode-se argumentar que a pergunta "Qual é a topologia do tempo?" não está bem posicionado quando a suposição falha. Quando a suposição é válida, podemos considerar a topologia do tempo, dada por *s*, como a topologia da projeção de *M/s*. (EARMAN, 1977, p. 214)

O método da projeção pode ser enunciado em duas definições equivalentes, propostas por Earman (1977, p. 214-215):

> ***Definição 1*****:** O tempo em *<M, g>* pode ser considerado *linear* se houver uma família s tal que *M*/s ≅ *R*. Da mesma forma, o tempo em *<M, g>* pode ser considerado circular se houver uma família s' tal que M/s' ≅ $S^1$.
>
> ***Definição* 2:** Função *t* é uma *função de tempo linear* para *<M, g>* se *t: M → R* é $C^o$ e aumenta ao longo de cada curva temporal futura direcionada. A função *c* é uma função de tempo circular para *<M, g>* se *c*: $M \rightarrow S^1$ é $C^o$ e para, e para qualquer $w, x, y, z \in M$ distinto, se houver uma curva *timelike* direcionada futura que vá de *w* a *x* a *y* a *z* que não re-intercepte a superfície nivelada de *c* de onde ela começa, então *c(w)*, *c(y)* par separa *c(x), c(z)* em $S^1$.

Em nossa topologia qualquer dimensão associada ao eixo dos números reais e perplexos será positiva e isomórfica a uma linha reta. As dimensões associadas a um número nilpotente de ordem dois (dual) terá dimensionalidade zero. Por fim, as dimensões associadas à números imaginários terão dimensões negativas e, portanto, serão curvas fechadas isomórficas à $S^1$. Por simplicidade, partiremos da premissa que as dimensões do espaço são positivas e associadas à $R^3$.

$$\mathbb{R} \mapsto R \qquad \mathbb{D} \mapsto 0$$
$$\mathbb{P} \mapsto R \qquad \mathbb{C} \mapsto S^1$$

O nosso modelo topológico será construído para variedades do tipo plana, isto é, variedades onde as conexões afins são nulas em todos os pontos.

## Espaço-Tempo de Galileu e os Números Duais[6]

Em nossa análise anterior, vimos que a dimensionalidade do tempo no espaço de Galileu era nula, pois o fator de escala do tempo deveria ser zero. Nestas condições, o tempo seria um eixo nulo e pela álgebra linear sabemos que a dimensão de um vetor nulo é sempre zero. Porém, há uma maneira diferente de caracterizar esse espaço, usando os números duais, também chamados de nilpotentes ou parabólicos.

Definimos o anel dos números duais da seguinte forma:

$$D:\left\{d = r + \varepsilon\tau \mid r,\tau \in \mathbb{R},\ \varepsilon^2 = 0,\ \bar{\varepsilon} = -\varepsilon\right\}$$

Observe que o número e elevado a segunda potência é zero. Todos os elementos que para uma potência *n* são nulos são chamados de elementos nilpotentes de ordem *n* e definem uma álgebra de Grasmann de ordem *n* sobre um espaço *E* e seu dual:

$$\varepsilon^n = 0 \rightarrow \begin{cases} \Lambda^n E \\ \Lambda^n E^* \end{cases}$$

Definimos o número dual de *d* pela seguinte relação:

$$d^* = r + \bar{\varepsilon}\tau = r - \varepsilon\tau$$

A norma de um número dual é calculada a partir do produto de *d* por seu dual:

$$\begin{aligned} d^2 &= dd^* = (r+\varepsilon\tau)(r-\varepsilon\tau) \\ d^2 &= r^2 - \varepsilon^2\tau^2 \\ d^2 &= r^2 \end{aligned}$$

Em uma dimensão, essa equação corresponde a parametrização de uma parábola. Por isso esses números também podem ser chamados de parabólicos. Procuremos agora o automorfismo que define as transformações nesse espaço, isto é, as matrizes que satisfazem a equação do automorfismo interno:

$$G^{-1}dG = d$$

Como os números duais são um anel, eles apresentam classes laterais, o que nos permite efetuar o produto tanto pela esquerda quanto pela direita. Se multiplicarmos a equação do automorfismo por G a esquerda, obteremos:

---

[6] Para detalhes sobre números duais, o leitor poderá consultar as seguintes autores: Todd (1936), Brand (1947), Semple, Roth, (1949), Eastham (1961) Deakin (1966), Miller, Boehning (1968), Yaglom, (1968, 1979), Veldkam (1976), Mccarthy (1986), Fischer (1998), Herranz, Ortega, Santander (1999), Cerejeiras, Kähler, Sommen (2005), Kisil (2007, 2008, 2012), Babusci, Dattoli, Palma, Sabia (2011), Vasantha, Smarandache (2012), Dattoli (2018) Ozdemir (2018, 2019). Uma abordagem mais avançada pode ser encontrada em McCarthy (2001) e Behr, Dattoli, Lattanzi, Licciardi (2019).

$$\left(GG^{-1}\right)dG = Gd$$
$$IdG = Gd$$
$$dG = Gd$$

Agora vamos montar a equação matricial:

$$\begin{pmatrix} r & \tau \end{pmatrix}\begin{pmatrix} a & b \\ c & d \end{pmatrix} = \begin{pmatrix} a & b \\ c & d \end{pmatrix}\begin{pmatrix} r & \tau \end{pmatrix}$$
$$\begin{pmatrix} ar + c\tau & br + d\tau \end{pmatrix} = \begin{pmatrix} ar + b\tau & cr + d\tau \end{pmatrix}$$

Essa igualdade nos leva ao sistema de equações:

$$ar + c\tau = ar + b\tau$$
$$br + d\tau = cr + d\tau$$

De onde tiramos a igualdade:

$$c = b$$

A equação de automorfismo exige que a transformação seja ortogonal:

$$G^{-1} = G^{T} = \begin{pmatrix} a & b \\ b & d \end{pmatrix}$$

Impondo que o produto de uma matriz por sua inversa é a identidade:

$$GG^{-1} = I$$
$$\begin{pmatrix} a & b \\ b & d \end{pmatrix}\begin{pmatrix} a & b \\ b & d \end{pmatrix} = \begin{pmatrix} 1 & 0 \\ 0 & 1 \end{pmatrix}$$
$$\begin{pmatrix} a^2 + b^2 & ab + bd \\ ab + bd & b^2 + d^2 \end{pmatrix} = \begin{pmatrix} 1 & 0 \\ 0 & 1 \end{pmatrix}$$

Que nos leva a um sistema de equações:

$$\begin{cases} a^2 + b^2 = 1 \\ b\left(a + d\right) = 0 \end{cases} \qquad \begin{matrix} b^2 + d^2 = 1 \\ a^2 - d^2 = 0 \end{matrix}$$

Onde a última equação foi obtida pela subtração da primeira equação pela terceira equação. Da segunda linha podemos obter duas soluções:

$$\begin{cases} b = 0 \\ a = -d \end{cases}$$

A primeira solução satisfaz a condição de um grupo e como as transformações automórficas são um grupo, então a solução que procuramos é *b = 0*.

$$\begin{cases} a^2 = 1 \\ b = 0 \\ d^2 = 1 \end{cases}$$

Outra condição para que a matriz G forme um grupo é que seu determinante seja próprio, isto é, positivo. Esta condição admite duas soluções:

$$I : \begin{cases} a = +1 \\ b = 0 \\ d = +1 \end{cases} \qquad II : \begin{cases} a = -1 \\ b = 0 \\ d = -1 \end{cases}$$

A escolha dos sinais apenas altera o sentido da transformação, por isso adotaremos, sem perda de generalidade, o sinal positivo. Portanto a matriz de transformação dual é a matriz identidade.

$$G = \begin{pmatrix} 1 & 0 \\ 0 & 1 \end{pmatrix}$$

Este é justamente o gerador da álgebra de Lie da variedade de Galileu. Portanto, podemos concluir que o anel dos números duais é o anel característico da álgebra de Galileu. Outro fato que corrobora nossa hipótese é a matriz de rotação parabólica:

$$P = \begin{pmatrix} 1 & 1 \\ 1 & 0 \end{pmatrix}$$

Se aplicarmos o operador *P* sobre um número dual, ele sofrerá uma rotação parabólica:

$$Pd = \begin{pmatrix} 1 & 1 \\ 0 & 1 \end{pmatrix} \begin{pmatrix} r & \tau \end{pmatrix}$$
$$Pd = \begin{pmatrix} r + \tau & \tau \end{pmatrix}$$

Essa aplicação também é um automorfismo interno de D.

$$P^{-1}dP = \begin{pmatrix} 1 & -1 \\ 0 & 1 \end{pmatrix} \begin{pmatrix} r & \tau \end{pmatrix} \begin{pmatrix} 1 & 1 \\ 0 & 1 \end{pmatrix}$$
$$P^{-1}dP = \begin{pmatrix} r - \tau \\ \tau \end{pmatrix} \begin{pmatrix} 1 & 1 \\ 0 & 1 \end{pmatrix}$$
$$P^{-1}dP = \begin{pmatrix} r & \tau \end{pmatrix}$$
$$P^{-1}dP = d$$

Se tomarmos o τ como o produto da velocidade pelo tempo, *vt,* a transformação *P* corresponde a uma transformação de Galileu. Em outras palavras, uma transformação de Galileu equivale a uma rotação parabólica na variedade dual, onde o ângulo parabólico é a velocidade entre os referenciais.

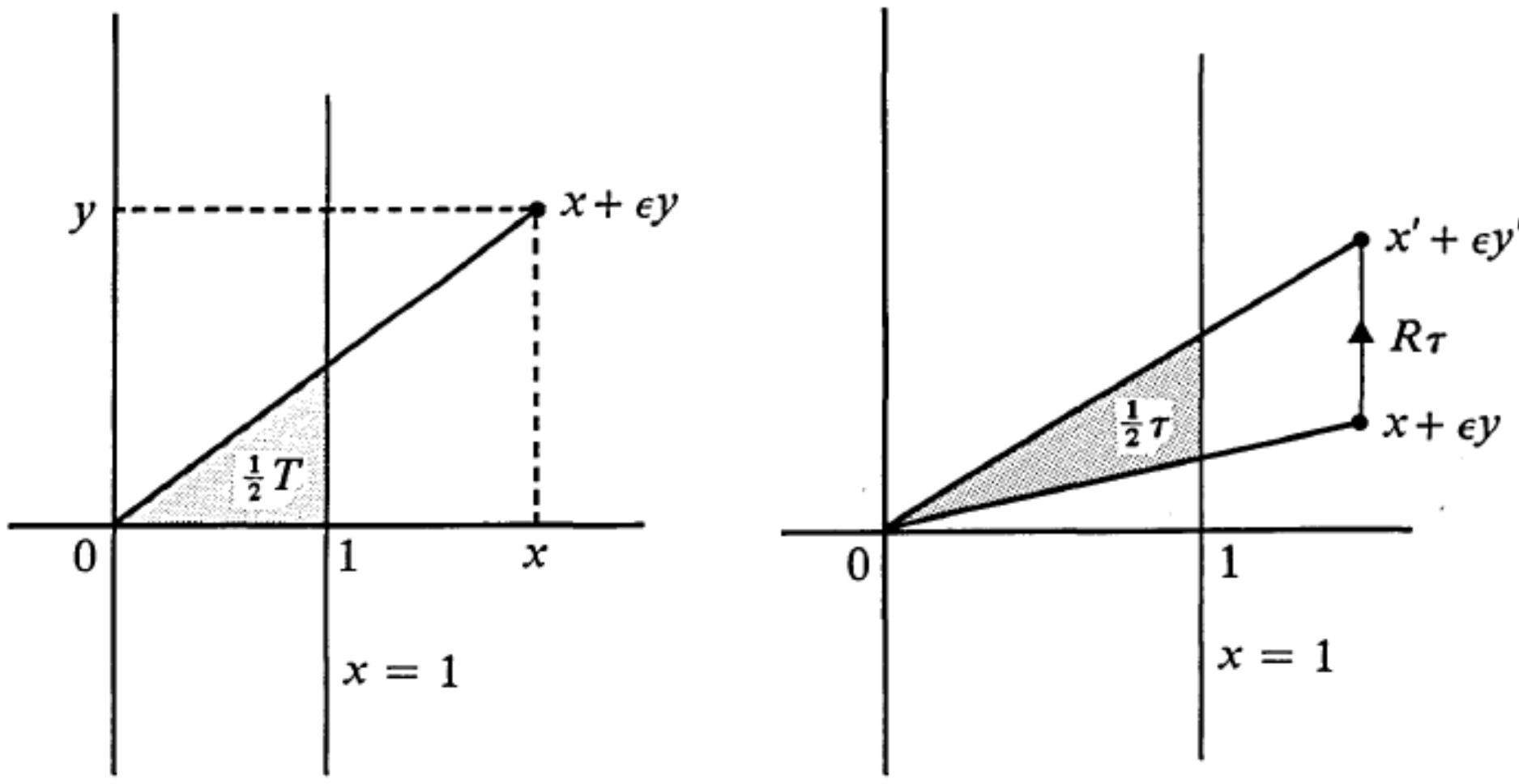

Transformação Parabólica no plano dual. **FONTE**: Rooney (1978, p. 95)

Nas transformações parabólicas, os eventos simultâneos são aqueles cujo ângulo de rotação é nulo. Além disso, sinais emitidos em sentidos contrários são observados da mesma maneira por todos referenciais inerciais, sem que o espaço necessite se contrair e o tempo se dilatar.

Se adotarmos que a variedade de Galileu tem como anel característico o anel dual, fica fácil entender porque o tempo não apresenta dimensionalidade. O tempo está associado ao eixo do número $\varepsilon$ que é nilpotente de segunda ordem. Quando tomamos a sua forma quadrática fundamental por meio da norma, o tempo não participa da equação devido ao caráter nilpotente de ε. Por exemplo, se definirmos a distância entre dois pontos na variedade de Galileu,

$$d(x,t) = \sqrt{\Delta x^2 + \varepsilon^2 \Delta t^2}$$
$$d(x,t) = \sqrt{\Delta x^2}$$

Portanto, o fato do tempo (e da velocidade) entre dois referenciais inerciais não dependerem do tempo é uma consequência do anel característico da variedade que é o dos números duais que tem uma unidade nilpotente de segunda ordem. Do ponto de vista qualitativo, essa é uma interpretação diferente da usual, pois agora a variedade de Galileu é um espaço-tempo, há um eixo temporal não-nulo, mas nilpotente. Como demonstrarei, o que define a dimensionalidade real, isto é, aquele que podemos associar a um número real inteiro, do tempo é o quadrado do seu número caraterístico, como o número característico do tempo na variedade de Galileu é um número nilpotente de segunda ordem, a dimensão real é zero. Há uma outra consequência interessante dessa nova forma de interpretar o espaço de Galileu. No século XIX, o matemático francês P. Laplace forneceu uma importante teoria que ficou conhecida como teoria do potencial. Na ausência de fontes, a equação do potencial é:

$$\nabla^2 \varphi = 0$$

Em nosso nova interpretação, deveríamos escrever a equação do potencial de forma ligeiramente diferente:

$$\nabla^2\varphi - \varepsilon^2 \frac{\partial^2\varphi}{\partial\tau^2} = 0$$

Essa é a equação se assemelha a equação da onda derivada por D'Alambert, mas como o número dual ao ser elevado quadrado se torna zero, por sua natureza nilpotente, retornamos a equação de Laplace. Alguém poderá argumentar que essa modificação é apenas uma forma "elegante" de se escrever a equação de Laplace. De fato, quantitativamente as duas equações são idênticas, porém, como veremos ao estudar os espaços euclidianos e lorentzianos, a modificação tem severas consequências qualitativas e é a partir delas que derivaremos a natureza dimensional do tempo.

Registre que essa não é a única equação associada aos números duais. Com efeito, como veremos, a equação euclidiana é uma equação elíptica temporal, a equação minkowskiana, é uma equação hiperbólica temporal. Por outro lado, a equação de Laplace é uma equação elíptica no espaço. A equação parabólica temporal corresponde a equação da difusão de Fourier, essa é a segunda equação associada aos números duais. A demonstração que faremos é uma variante daquela sugerida por Cerejeiras, Kähler e Sommen (2005, p. 1717) e é realizada usando os operadores parabólicos de Dirac. Primeiro vamos introduzir dois números duais e a sua álgebra de Clifford:

$$\nabla \equiv \sum e_j\partial_{x_j} \qquad \{f, f^+\} \equiv ff^+ + f^+f = 1$$

$$\nabla^2 \equiv \left(\sum e_j\partial_{x_j}\right)^2 \qquad \{f, e_j\} \equiv fe_j + e_jf = 0$$

$$f^2 = \left(f^+\right)^2 = 0 \qquad \{f^+, e_j\} \equiv f^+e_j + f^+e_j = 0$$

Agora, consideremos o operador definido por:

$$D_{x,t} = \sum e_j\partial_{x_j} + f\partial_t - f^+$$

Elevando ao quadrado:

$$D_{x,t}^2 = \left(\sum e_j\partial_{x_j} + f\partial_t \pm f^+\right)^2 = \left(\sum e_j\partial_{x_j}\right)^2 + f^2\partial_t^2 + \left(f^+\right)^2 - \{f, f^+\}\partial_t + \sum\left(\{f, e_j\}\partial_t - \{f^+, e_j\}\right)\partial_{x_j}$$

Aplicando as relações da álgebra da Clifford, obtemos a forma do operador $D^2_{x,t}$:

$$D_{x,t}^2 = \nabla^2 - \partial_t$$

Aplicando esse operador a uma função $\varphi$ e igualando a zero, obtemos a equação da difusão:

$$D_{x,t}^2\varphi = 0 \quad \leftrightarrow \quad \nabla^2\varphi - \partial_t\varphi = 0$$

$$\therefore \qquad \nabla^2\varphi = \partial_t\varphi$$

## Espaço-Tempo de Minkowski e os Números Perplexos[7]

O espaço-tempo de Poincaré-Minkowski, ou apenas de Minkowski, é uma variedade diferenciável chamada de Minkowskiana ou Lorentziana, pois as transformações automórficas são as transformações de Lorentz descobertas por H. Lorentz em 1904 e aprimoradas por H. Poincaré em 1905. Nesta variedade o tempo tem dimensionalidade unitária e é acompanhado de um fator de escala, a velocidade da luz. Assim como ocorre com o espaço de Galileu, há uma maneira diferente de caracterizar esse espaço, usando os números perplexos, também chamados de hiperbólicos.

Definimos o anel dos números perplexos da seguinte forma:

$$\mathbb{P}:\left\{w=r+h\tau \mid r,\tau\in\mathbb{R},\ h^2=1,\ \bar{h}=-h\right\}$$

Definimos o conjugado perplexo de *w* pela seguinte relação:

$$\bar{w}=r+\bar{h}\tau$$
$$\bar{w}=r-h\tau$$

A norma de um perplexo é calculada a partir do produto de *w* por seu conjugado:

$$w^2=w\bar{w}=(r+h\tau)(r-h\tau)$$
$$w^2=r^2-h^2\tau^2$$

Em uma dimensão, essa equação corresponde a parametrização de uma hipérbole. Por isso esses números também podem ser chamados de hiperbólicos. Nas seções anteriores estudamos a exaustão as propriedades desse tipo de configuração, por isso iremos nos atentar em outros fatos. O primeiro que gostaríamos de registrar é que esse anel admite divisor em zero e norma negativa. Esse é o resultado que esperaríamos, já que o espaço-tempo admite vetores do tipo-tempo, tipo-luz e tipo-espaço. O segundo fato é que o automorfismo que define as transformações nesse espaço são as matrizes de Lorentz:

$$\Lambda^{-1}w\Lambda=w$$

Essas matrizes de Lorentz correspondem as rotações hiperbólicas no espaço-tempo, que são o equivalente as rotações parabólicas no espaço-tempo de Galileu. Se adotarmos que a variedade de Lorentz tem como anel característico o anel perplexo, fica fácil entender porque o tempo apresenta dimensionalidade. O tempo está associado ao eixo do número *p*. Quando tomamos a sua forma quadrática fundamental por meio da norma, o tempo participa da equação devido ao de *h*.

---

[7] Para informações mais detalhadas sobre o anel dos números perplexos e suas aplicações (em particular na relatividade especial), o leitor deve consultar: Miller, Boehning (1968), Hawkins (1972), Fjelstad (1986), Assis, (1991), Herranz, Ortega, Santander (1999), Borota, Osler (2002), Harkin, Harkin (2004), Khrennikov, Segre (2005), Catoni, Boccaletti, Cannata, Catoni, Nichelatti, Zampetti (2008), Sabadini, Shapiro, Sommen, (2009), Poodiack (2009), Dattolia, Franco (2010), Babusci, Dattoli, Palma, Sabia (2011), Sabadini, Sommen (2011), Catoni, Boccaletti, Cannata, Catoni, Zampetti (2011), Catoni, Zampetti (2012), Kisil, (2012, 2013), Gargoubi, Kossentini (2016), Amorim, Santos, Carvalho, Massa (2018), Boccaletti, Catoni, Catoni (2018), Dattoli (2018), Ozdemir (2018, 2019);

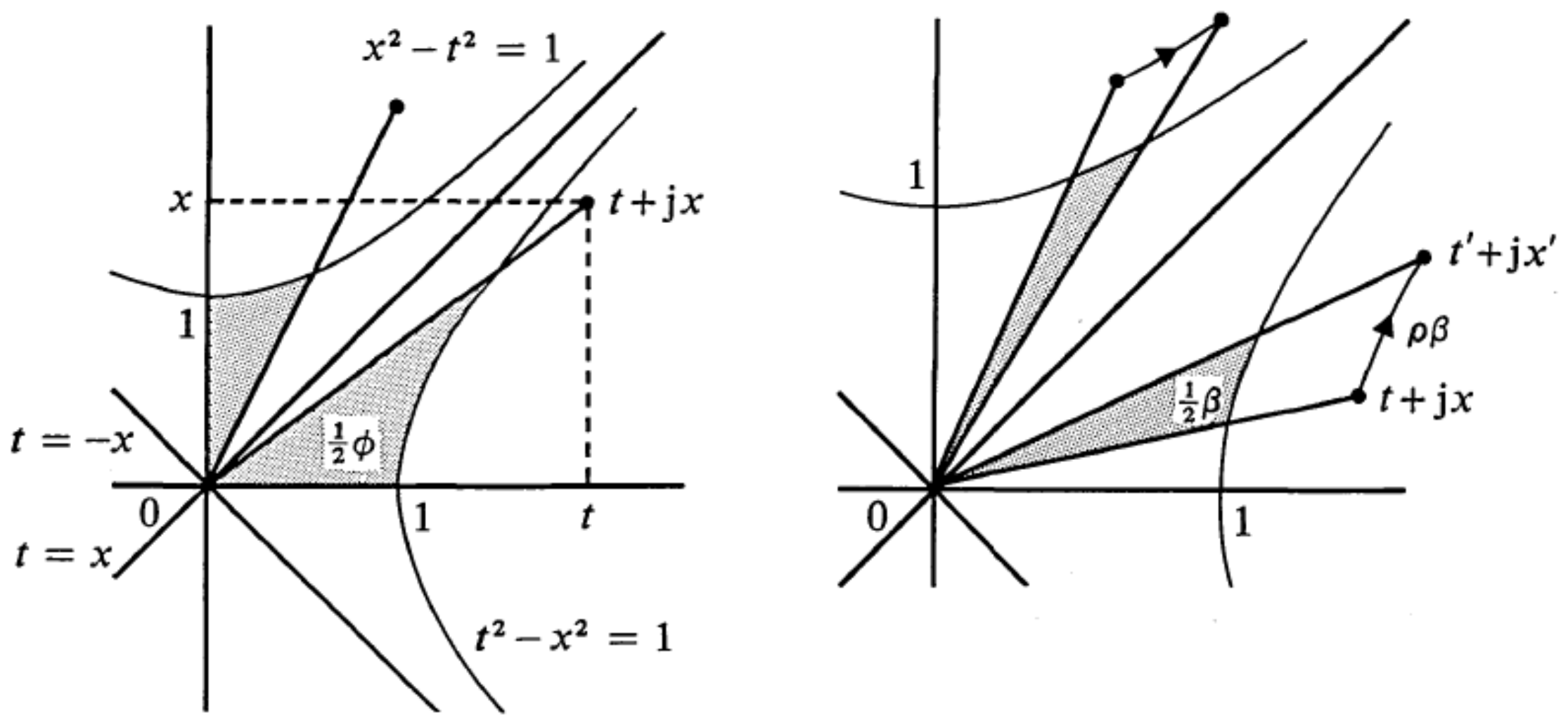

Transformação Hiperbólica no plano perplexo. **FONTE**: Rooney (1978, p. 95)

Por exemplo, se definirmos a distância entre dois pontos na variedade de Lorentz, teremos:

$$d(x,t)=\sqrt{\Delta x^2 - p^2 \Delta t^2}$$
$$d(x,t)=\sqrt{\Delta x^2 - \Delta t^2}$$

Portanto, o fato do tempo (e da velocidade) entre dois referenciais inerciais dependerem do tempo é uma consequência do anel característico da variedade que é o dos números perplexos. Há outras consequências interessantes envolvendo os números perplexos quando passamos para o campo da análise. Vamos definir uma função perplexa como uma aplicação de $R^2$ em P, expressa como:

$$p:\mathbb{R}\times\mathbb{R}\to\mathbb{P}$$
$$p(r,\tau)=\varphi(r,\tau)+h\phi(r,\tau)$$

A analiticidade dessa função é definida pelas condições de Cauchy-Riemann que nos implicam que o limite de *h* em um ponto da variedade deve ser o mesmo por qualquer caminho que pertença ao domínio conexo. Nós definimos a derivada na variedade perplexa, por meio do seguinte limite:

$$\frac{dp}{dw}=\lim_{\substack{\Delta r\to 0\\ \Delta\tau\to 0}}\frac{\varphi(r+\Delta r,\tau+\Delta\tau)+h\phi(\Delta r+r,\tau+\Delta\tau)}{\Delta r+h\Delta\tau}$$

A derivada da função existe e é continua (classe $C^1$) se satisfizer a condição de Cauchy-Riemann:

$$\lim_{\substack{\Delta r\to 0\\ \Delta\tau=0}}\frac{\varphi(r+\Delta r,\tau)+h\phi(\Delta r+r,\tau)}{\Delta r}=\lim_{\substack{\Delta r=0\\ \Delta\tau\to 0}}\frac{\varphi(r,\tau+\Delta\tau)+h\phi(r,\tau+\Delta\tau)}{h\Delta\tau}$$

Abrindo as frações:

$$\lim_{\Delta r\to 0}\left[\frac{\varphi(r+\Delta r,\tau)}{\Delta r}+h\frac{\phi(\Delta r+r,\tau)}{\Delta r}\right]=\lim_{\Delta\tau\to 0}\left[\frac{\varphi(x,\tau+\Delta\tau)}{h\Delta\tau}+\frac{h\phi(r,\tau+\Delta\tau)}{h\Delta\tau}\right]$$

Multiplicando o primeiro termo da direita por *p/p* e simplificando a segunda parcela:

$$\lim_{\Delta r\to 0}\left[\frac{\varphi(r+\Delta r,\tau)}{\Delta r}+h\frac{\phi(\Delta r+r,\tau)}{\Delta r}\right]=\lim_{\Delta\tau\to 0}\left[h\frac{\varphi(x,\tau+\Delta\tau)}{h^2\Delta\tau}+\frac{\phi(r,\tau+\Delta\tau)}{\Delta\tau}\right]$$

Levando em consideração que $p^2$ é a unidade e a definição de derivada parcial:

$$\frac{\partial\varphi}{\partial r}+h\frac{\partial\phi}{\partial r}=h\frac{\partial\varphi}{\partial\tau}+\frac{\partial\phi}{\partial\tau}$$

Igualando as partes perplexas e reais obtemos as condições de Cauchy-Riemann:

$$\begin{cases}\dfrac{\partial\varphi}{\partial r}=\dfrac{\partial\phi}{\partial\tau}\\[2ex]\dfrac{\partial\phi}{\partial r}=\dfrac{\partial\varphi}{\partial\tau}\end{cases}$$

Se diferenciarmos a primeira equação em relação a *r* e a segunda equação em relação à τ, obtemos:

$$\begin{cases}\dfrac{\partial^2\varphi}{\partial r^2}=\dfrac{\partial^2\phi}{\partial r\partial\tau}\\[2ex]\dfrac{\partial^2\phi}{\partial\tau\partial r}=\dfrac{\partial^2\varphi}{\partial\tau^2}\end{cases}$$

Como a função é analítica, as derivadas parciais comutam:

$$\begin{cases}\dfrac{\partial^2\varphi}{\partial r^2}=\dfrac{\partial^2\phi}{\partial r\partial\tau}\\[2ex]\dfrac{\partial^2\phi}{\partial r\partial\tau}=\dfrac{\partial^2\varphi}{\partial\tau^2}\end{cases}$$

Substituindo a primeira equação na segunda, obtemos a equação diferencial associado a variedade:

$$\frac{\partial^2\varphi}{\partial r^2}=\frac{\partial^2\varphi}{\partial\tau^2}$$

$$\frac{\partial^2\varphi}{\partial r^2}-\frac{\partial^2\varphi}{\partial\tau^2}=0$$

$$\nabla^2\varphi-\frac{\partial^2\varphi}{\partial\tau^2}=0$$

Como t está associado a velocidade da luz, em situações não relativísticas, a parcela temporal tende a zero e a equação tende ao laplaciano. Agora, iremos mostrar que essa é a equação do potencial de Laplace-Beltrami na variedade de Lorentz. Para isso usaremos o conceito de 4-vetor covariante em Lorentz. O vetor nabla ou vetor *del* em sistemas gerais de coordenadas é um vetor covariante. Em análise em variedades pode-se provar que nabla são os vetores da base do espaço cotangente e, portanto, o dual do vetor diferencial *dr* e, em coordenadas ortogonais, a base recíproca. Nós definimos que o 4-vetor gradiente covariante a partir das regras:

$$\nabla_i = \left(\frac{1}{c}\partial_t, \nabla\right), \quad \nabla'_i = \left(\frac{1}{c}\partial'_t, \nabla'\right), \quad \partial_i \equiv \frac{\partial}{\partial x_i}$$

Nabla é um vetor covariante, então sua transformação de Lorentz direta e inversa são, respectivamente:

$$\begin{aligned}
\frac{\partial'_t}{c} &= \frac{1}{c}\cosh a\partial_t - \sinh a\ \partial_x & \qquad \frac{\partial_t}{c} &= \frac{1}{c}\cosh a\partial'_t + \sinh a\ \partial'_x \\
\partial'_x &= \cosh a\ \partial_x - \frac{1}{c}\sinh a\ \partial_t & \qquad \partial_x &= \cosh a\partial'_x + \frac{1}{c}\sinh a\ \partial'_t \\
\partial'_y &= \partial_y & \qquad \partial_y &= \partial'_y \\
\partial'_z &= \partial_z & \qquad \partial_z &= \partial'_z
\end{aligned}$$

Multiplicando a primeira equação por *c* e abrindo as funções hiperbólicas:

$$\begin{aligned}
\partial'_t &= \gamma(\partial_t - v\partial_x) & \qquad \partial_t &= \gamma(\partial'_t + v\partial'_x) \\
\partial'_x &= \gamma\left(\partial_x - \frac{v}{c^2}\partial_t\right) & \qquad \partial_x &= \gamma\left(\partial'_x + \frac{v}{c^2}\partial'_t\right) \\
\partial'_y &= \partial_y & \qquad \partial_y &= \partial'_y \\
\partial'_z &= \partial_z & \qquad \partial_z &= \partial'_z
\end{aligned}$$

Portanto nosso 4-vetor nabla tem a seguinte forma:

$$\nabla'_i = \partial'_i = \left(\gamma(\partial_t - v\partial_x), \gamma\left(\partial_x - \frac{v}{c^2}\partial_t\right), \partial_y, \partial_z\right)$$

No referencial próprio, podemos definir o 4-vetor nabla próprio:

$$\partial_i^o = \left(\frac{\partial_t}{c}, \nabla^o\right)$$

E com as coordenadas no sistema estacionário, temos que:

$$\partial_i = \left(\frac{\partial^t}{c}, \nabla\right), \quad \partial^i = \left(\frac{\partial_t}{c}, -\nabla\right)$$

Aplicando a regra de construção de invariantes, teremos que:

$$\frac{1}{c^2}\frac{\partial^2}{\partial t^2}-\nabla^2=\frac{1}{c^2}\frac{\partial^2}{\partial t'^2}-\nabla'^2$$

Que é a expressão da equação da onda e do D'Alambertiano. Portanto, o operador D'Alambertiano é um invariante relativístico:

$$\square=\frac{1}{c^2}\frac{\partial^2}{\partial t^2}-\nabla^2, \qquad \square'=\square$$

Definindo $\tau = ct$, então o potencial relativístico será:

$$\square \equiv \nabla^2\varphi-\frac{\partial^2\varphi}{\partial\tau^2}=0$$

Essa equação pode ser escrita em função da sua característica perplexa:

$$\nabla^2\varphi-h^2\frac{\partial^2\varphi}{\partial\tau^2}=0$$

## Espaço-Tempo de Euclides e os Números Complexos

O espaço-tempo de Euclides é uma variedade diferenciável supersimétrica, pois as transformações automórficas são elementos do grupo de rotações SO(4) curvas fechadas no espaço e no tempo. Nesta variedade o tempo tem dimensionalidade unitária negativa (ou imaginária) e é acompanhado de um fator de escala, a velocidade da luz. Assim como ocorre com o espaço de Galileu, há uma maneira diferente de caracterizar esse espaço, usando os números complexos. Visto que os números complexos são bastante difundidos nas teorias físicas, iremos apenas recapitular algumas de suas propriedades elementares. Definimos o anel dos números perplexos da seguinte forma:

$$\mathbb{C}:\left\{z=r+i\tau \mid r,\tau\in\mathbb{R},\ i^2=-1,\ \bar{i}=-i\right\}$$

Definimos o conjugado perplexo de *w* pela seguinte relação:

$$\bar{z}=r+\bar{i}\tau$$
$$\bar{z}=r-i\tau$$

A norma de um complexo é calculada a partir do produto de *z* por seu conjugado:

$$z^2=z\bar{z}=\left(r+i\tau\right)\left(r-i\tau\right)$$
$$z^2=r^2-i^2\tau^2$$
$$z^2=r^2+\tau^2$$

Em uma dimensão, essa equação corresponde a parametrização de uma circunferência ou de uma elipse. Por isso esses números também podem ser chamados de polares. O automorfismo que define as transformações nesse espaço são as matrizes de Lorentz, como já provamos anteriormente:

$$R^{-1}zR = z$$

onde as matrizes de rotação são:

$$R = \begin{pmatrix} \cos\theta & \sin\theta \\ -\sin\theta & \cos\theta \end{pmatrix}, \qquad R^{-1} = \begin{pmatrix} \cos\theta & -\sin\theta \\ \sin\theta & \cos\theta \end{pmatrix}$$

Essas matrizes de Lorentz correspondem as rotações hiperbólicas no espaço-tempo, que são o equivalente as rotações parabólicas no espaço-tempo de Galileu. Se adotarmos que a variedade de Euclides tem como anel característico o anel complexo, fica fácil entender porque o tempo apresenta dimensionalidade imaginária ou negativa. O tempo está associado ao eixo do número *i*. Quando tomamos a sua forma quadrática fundamental por meio da norma, o tempo participa da equação devido ao de *i*.

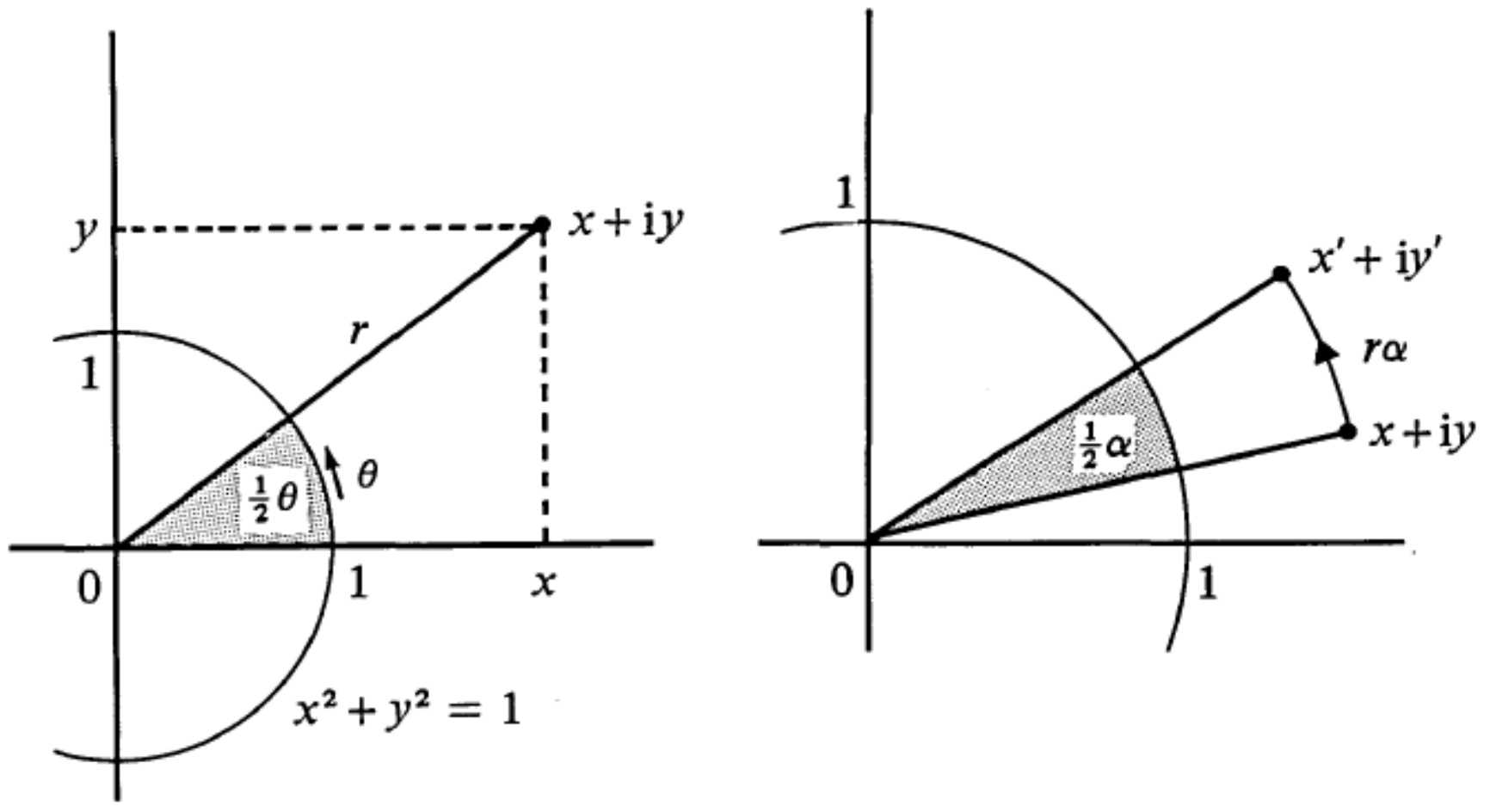

Transformação Elíptica no plano complexo. **FONTE**: Rooney (1978, p. 95)

Por exemplo, se definirmos a distância entre dois pontos na variedade de Euclides, teremos:

$$d(x,t) = \sqrt{\Delta x^2 - i^2 \Delta t^2}$$

Mas como o número $i^2$ é a unidade negativa:

$$d(x,t) = \sqrt{\Delta x^2 + \Delta t^2}$$

Que é a expressão da diagonal de um hipercubo de quatro dimensões e, portanto, a generalização 4-dimensional do Teorema de Pitágoras. Observe que se o tempo perplexo contribui com uma dimensão positiva, pois o número perplexo ao quadrado é a unidade. O tempo imaginário deve contribuir com uma dimensão negativa, pois o número imaginário ao quadrado é a unidade negativa. Portanto, o fato do tempo (e da velocidade) entre dois referenciais inerciais dependerem do tempo é uma consequência do anel característico da variedade que é o dos números complexos. Assim como fizemos com os perplexos, vamos estudar a analiticidade definindo uma função perplexa como uma aplicação de $R^2$ em C, expressa como:

$$
\begin{aligned}
&q:\mathbb{R}\times\mathbb{R}\to\mathbb{C} \\
&q(r,\tau)=\varphi(r,\tau)+i\phi(r,\tau)
\end{aligned}
$$

A analiticidade dessa função é definida pelas condições de Cauchy-Riemann que nos implicam que o limite de $q$ em um ponto da variedade deve ser o mesmo por qualquer caminho que pertença ao domínio conexo (plano complexo). Definimos a derivada na variedade complexa, por meio do limite:

$$\frac{dq}{dz}=\lim_{\substack{\Delta r\to 0\\ \Delta\tau\to 0}}\frac{\varphi(r+\Delta r,\tau+\Delta\tau)+i\phi(\Delta r+r,\tau+\Delta\tau)}{\Delta r+i\Delta\tau}$$

A derivada da função existe e é continua (classe $C^1$) se satisfizer a condição de Cauchy-Riemann:

$$\lim_{\substack{\Delta r\to 0\\ \Delta\tau=0}}\frac{\varphi(r+\Delta r,\tau)+i\phi(\Delta r+r,\tau)}{\Delta r}=\lim_{\substack{\Delta r=0\\ \Delta\tau\to 0}}\frac{\varphi(r,\tau+\Delta\tau)+i\phi(r,\tau+\Delta\tau)}{i\Delta\tau}$$

$$\lim_{\Delta r\to 0}\left[\frac{\varphi(r+\Delta r,\tau)}{\Delta r}+i\frac{\phi(\Delta r+r,\tau)}{\Delta r}\right]=\lim_{\Delta\tau\to 0}\left[\frac{\varphi(x,\tau+\Delta\tau)}{i\Delta\tau}+\frac{i\phi(r,\tau+\Delta\tau)}{i\Delta\tau}\right]$$

Multiplicando o primeiro termo da direita por *i/i* e simplificando o segundo.

$$\lim_{\Delta r\to 0}\left[\frac{\varphi(r+\Delta r,\tau)}{\Delta r}+i\frac{\phi(\Delta r+r,\tau)}{\Delta r}\right]=\lim_{\Delta\tau\to 0}\left[i\frac{\varphi(x,\tau+\Delta\tau)}{i^2\Delta\tau}+\frac{\phi(r,\tau+\Delta\tau)}{\Delta\tau}\right]$$

Como $i^2$ é a unidade negativa e a definição de derivada parcial:

$$\frac{\partial\varphi}{\partial r}+i\frac{\partial\phi}{\partial r}=-i\frac{\partial\varphi}{\partial\tau}+\frac{\partial\phi}{\partial\tau}$$

Igualando as partes perplexas e reais obtemos as condições de Cauchy-Riemann:

$$\begin{cases}\dfrac{\partial\varphi}{\partial r}=\dfrac{\partial\phi}{\partial\tau}\\[2ex] \dfrac{\partial\phi}{\partial r}=-\dfrac{\partial\varphi}{\partial\tau}\end{cases}$$

Toda função que satisfaz estas condições são chamadas de holomórficas. As funções holomórficas permitem construir mapas por meio das transformações conformes, que também são chamados de holomorfismos. Uma transformação conforme é uma aplicação 1-1 que a cada segmento da variedade, faz corresponder um novo segmento, que sofre uma rotação elíptica constante.

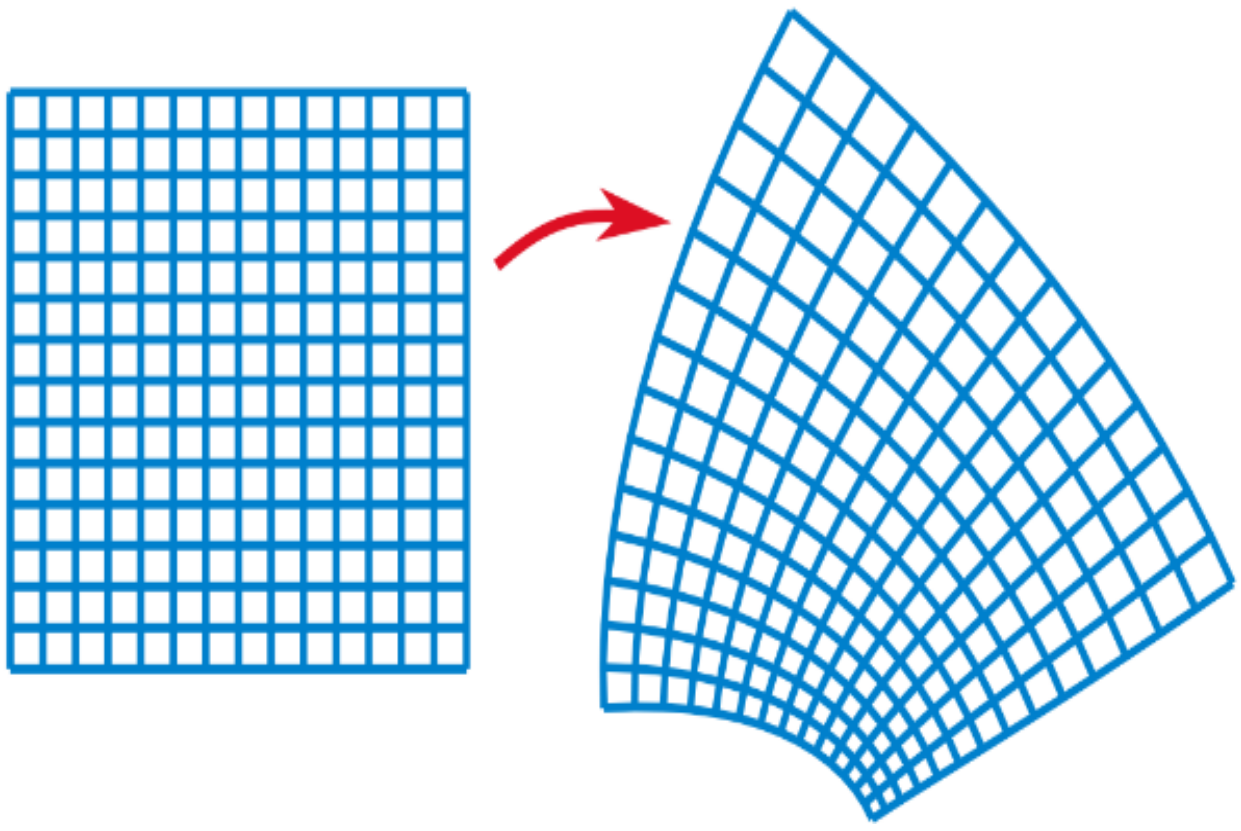

Transformação holográfica de uma superfície plano retangular.

Se diferenciarmos a primeira equação em relação a $r$ e a segunda equação em relação à τ, obtemos:

$$\begin{cases} \frac{\partial^2 \varphi}{\partial r^2} = \frac{\partial^2 \phi}{\partial r \partial \tau} \\ \frac{\partial^2 \phi}{\partial \tau \partial r} = -\frac{\partial^2 \varphi}{\partial \tau^2} \end{cases}$$

Como a função é analítica, as derivadas parciais comutam:

$$\begin{cases} \frac{\partial^2 \varphi}{\partial r^2} = \frac{\partial^2 \phi}{\partial r \partial \tau} \\ \frac{\partial^2 \phi}{\partial r \partial \tau} = -\frac{\partial^2 \varphi}{\partial \tau^2} \end{cases}$$

Substituindo a primeira equação na segunda, obtemos a equação diferencial associado a variedade:

$$\frac{\partial^2 \varphi}{\partial r^2} = -\frac{\partial^2 \varphi}{\partial \tau^2}$$

$$\frac{\partial^2 \varphi}{\partial r^2} + \frac{\partial^2 \varphi}{\partial \tau^2} = 0$$

$$\nabla^2 \varphi + \frac{\partial^2 \varphi}{\partial \tau^2} = 0$$

Essa é a expressão do potencial na variedade euclidiana que apresenta uma estrutura semelhante ao laplaciano de Laplace e Beltrami. Para fenômenos locais a grandeza $t$ tende a zero devido a magnitude da velocidade da luz e, novamente, obtemos a expressão galileana do potencial. Agora, permita-me escrever esse novo potencial usando a unidade imaginária:

$$\nabla^2 \varphi - i^2 \frac{\partial^2 \varphi}{\partial \tau^2} = 0$$

## Espaço-Tempo (1-1) e o Anel dos Números Híbridos

Na introdução desse ensaio definimos o parâmetro *R* como a característica do ideal do anel do anel do números hipercomplexos. Isso permitiu associar cada espaço-tempo a um anel de números hipercomplexos: dual, perplexo, complexo. Porém, essa não é única forma de definir *R,* de fato, podemos definir *R* a partir de uma estrutura muito mais rica e que permitirá unificar todos estes espaços, o anel dos números híbridos Z.

Para atingirmos esse objetivo, precisaremos introduzir alguns conceitos associados aos números híbridos. A discussão que apresentaremos aqui de forma sumarizada foi baseada no trabalho *Introduction to Hybrid Numbers* (Özdemir, 2018)[8] a qual o leitor deverá consultar para mais detalhes.

Definimos um número híbrido como um número quarteniônico da seguinte forma:

$$\mathbb{K}\colon\left\{Z = a + b\mathbf{i} + c\boldsymbol{\varepsilon} + d\,\mathbf{h},\, \mathbf{i}^2 = -1,\, \boldsymbol{\varepsilon}^2 = 0,\, \mathbf{h}^2 = +1,\, \mathbf{ih} = -\mathbf{hi} = \boldsymbol{\varepsilon} + \mathbf{i},\ a,b,c,d \in \mathbb{R}\right\}$$

O número real *a* é chamado de parte escalar e é denotado por **S**(Z). A parte $b\mathbf{i} + c\boldsymbol{\varepsilon} + d\mathbf{h}$ é chamada de parte vetorial e é denotada por **V**(Z).

$$Z = \mathbf{S}(Z) + \mathbf{V}(Z)$$

O conjugado de um número híbrido é definido da seguinte forma:

$$\bar{Z} = a - b\mathbf{i} - c\boldsymbol{\varepsilon} - d\,\mathbf{h}$$

Multiplicando um número híbrido por seu conjugado, obtemos seu módulo:

$$Z\bar{Z} = a^2 + (b-c)^2 - c^2 + d^2$$

A partir desse produto definimos o **vetor representação de *Z*,** denotado por $\mathcal{V}_z$.

$$\mathcal{V}_z = \big(a,(b-c),c,d\big)$$
$$C(Z) \equiv \langle \mathcal{V}_z, \mathcal{V}_z \rangle_{\mathbb{E}_2^4} = -Z\bar{Z},$$
$$Sign\,\mathbb{E}_2^4 = (+,+,-,-)$$

O produto interno do vetor representação permite classificar o número híbrido em três categorias: tipo-tempo ($C_{\mathcal{V}}(\mathbf{Z}) > 0$); tipo-luz ($C_{\mathcal{V}}(\mathbf{Z}) = 0$); tipo-espaço ($C_{\mathcal{E}}(\mathbf{Z}) < 0$).

Analogamente, podemos definir um novo vetor, usando apenas a parte vetorial de *Z,* que denominamos de **vetor híbrido** e denotamos por $\mathcal{E}_Z$:

---

[8] Uma extensão desse trabalho aparece em Özdermir (2019). Para uma abordagem alternativa dos números híbridos, porém equivalente, ver: Dattoli (2018).

$$\mathcal{E}_z = \left(\left(b-c\right), c, d\right)$$
$$C_{\mathcal{E}}\left(Z\right) \equiv \left\langle \mathcal{E}_z, \mathcal{E}_z \right\rangle_{\mathbb{E}_1^3} = -\left(b-c\right)^2 + c^2 + d^2,$$
$$Sign\,\mathbb{E}_1^3 = (-,+,+,)$$

O produto interno do vetor híbrido permite classificar o número híbrido em mais três categorias: hiperbólico ou tipo-hiper ($C_{\mathcal{E}}(\mathbf{Z}) > 0$); parabólico ou tipo-dual ($C_{\mathcal{E}}(\mathbf{Z}) = 0$); elíptico ou tipo-complexo ($C_{\mathcal{E}}(\mathbf{Z}) < 0$). Por meio do produto interno, podemos definir a norma do vetor híbrido $\mathcal{N}(\mathbf{Z})$:

$$\mathcal{N}\left(Z\right) = \sqrt{C_{\mathcal{E}}\left(Z\right)}$$

Há uma importante relação entre as categorias definidas por $C_{\mathfrak{v}}(\mathbf{Z})$ e $C_{\mathcal{E}}(\mathbf{Z})$, que sintetizamos na tabela abaixo:

| Tipo-Espaço | Tipo-Luz | Tipo-Tempo |
|---|---|---|
| **Hiperbólico (Tipo-Hiper)** | **Hiperbólico** | **Hiperbólico** |
| ---------------------------------- | **Parabólico (Tipo-Dual)** | **Parabólico** |
| ---------------------------------- | ---------------------------- | **Elíptico (Tipo-Complexo)** |

Tabela 2: Relação entre as categorias dos números híbridos. **Fonte:** Özdemir (2018, p. 09)

Por meio da parte vetorial e da norma do vetor híbrido, podemos definir uma nova grandeza que chamaremos de **versor híbrido:**

$$\mathbf{V}_0 = \frac{\mathbf{V}\left(Z\right)}{\mathcal{N}\left(Z\right)}$$

Elevando ao quadrado esse versor, pode-se demonstrar que ele pode assumir apenas três valores:

$$\mathbf{V}_0^2 = \begin{cases} -1, & se\ \ Z\ \ for\ elíptico \\ 0, & se\ \ Z\ \ for\ \ parabólico \\ +1, & se\ \ Z\ \ for\ hiperólico \end{cases}$$

Portanto, o versor $V_0$ corresponde ao parâmetro $R$:

$$R \equiv \mathbf{V}_0 = \frac{\mathbf{V}\left(Z\right)}{\mathcal{N}\left(Z\right)}$$

Portanto, ao invés de descrevermos cada espaço-tempo como sendo gerado pela característica do ideal de um anel hipercomplexo, podemos dizer que o espaço-tempo é uma estrutura gerada pela álgebra dos números híbridos. Desta forma não necessitamos de três anéis para gerar todos os espaço-tempos planos, mas apenas um anel: o do números híbridos. Por meio do estudo dos números híbridos podemos deduzir de maneira puramente algébrica como cada espaço-tempo se estrutura e as transformações de coordenadas.

| **Forma Polar** | **Característica** | | |
|---|---|---|---|
| **Categoria** | Tipo-Espaço | Tipo-Luz | Tipo-Tempo |
| Elíptico | ∅ | ∅ | $\cos\theta + \mathbf{V}_0 \sin\theta$ |
| Hiperbólico | $\sinh\theta + \mathbf{V}_0 \cosh\theta$ | $a\,(1+\mathbf{V}_0)$ | $\cosh\theta + \mathbf{V}_0 \sinh\theta$ |
| Parabólico | ∅ | ∅ | $(\epsilon + \mathbf{V}_0),\ \ \epsilon = sgn\ \mathbf{S}(Z)$ |

Tabela 3: Representação polar dos números híbridos. **Fonte:** Özdemir (2018, p. 17)

Por esta tabela podemos ver que o espaço-tempo hiperbólico (Minkowskiano) apresenta três regiões: a região do tipo-tempo, que representa os eventos conectados for fenômenos que se propagam mais devagar que a luz. A região do tipo luz, que consiste em eventos conectados por interações que se propagam com a mesma velocidade que a luz no vácuo. A região do tipo-espaço, que corresponde uma região onde os eventos interagem por vínculos mais rápidos que a velocidade da luz. Por outro lado, o espaço-tempo elíptico (Euclidiano) e o espaço-tempo parabólico (Galileano) apresentam apenas regiões com vínculo casual (tipo-tempo).

## As Funções de Poincaré

Como o espaço-tempo é uma variedade induzida por versor híbrido *R*, para descrever o espaço e o tempo propomos a construção de funções híbridas de classe $C^{\infty}$, que nós chamaremos de funções geométricas ou funções de Poincaré, em homenagem ao físico-matemático Henri Poincaré. Tomemos as transformações do espaço e do tempo:

$$\begin{cases} x' = A(x - vt) \\ t' = A(t - R^2 vx) \end{cases}$$

Chamaremos de função par de Poincaré, a função definida por:

$$P_R^+(v) = A$$

Analogamente, Chamaremos de função ímpar de Poincaré, a função definida por:

$$P_R^-(v) = vA$$

Nestas condições, as transformações de coordenadas assume a seguinte forma:

$$\begin{cases} x' = P_R^+(v)x - P_R^-(v)t \\ t' = P_R^+(v)t - R^2 P_R^-(v)x \end{cases}$$

Pela restrição imposta pela condição de automorfismo, teremos:

$$AD - BC = 1$$
$$\left[P_R^+(v)\right]^2 - R^2\left[P_R^-(v)\right]^2 = 1$$

Essa é a identidade fundamental da trigonometria da geometria.

Vamos agora definir a função tangente de Poincaré, como a razão da função ímpar pela função par de Poincaré:

$$P_R^{\mp}(v) = \frac{P_R^-(v)}{P_R^+(v)}$$

Assim como na variedade de Galileu, a função tangente de Poincaré determina a velocidade do corpo conforme o ângulo de inclinação na variedade:

$$P_R^{\mp}(v) = v$$

Devido à similaridade da função de Poincaré com as funções trigonométricas convencionais, somos induzidos a sumir que a seguinte identidade é válida:

$$e^{\pm Rv} = P_R^+(v) - RP_R^-(v)$$

De onde podemos derivar as seguintes identidades:

$$P_R^+(v) = \frac{e^{Rv} + e^{-Rv}}{2}, \qquad P_R^-(v) = \frac{e^{Rv} - e^{-Rv}}{2R}$$

Se estas relações são verdadeiras, elas devem satisfazer a equação fundamental da trigonometria. Para isso, tomemos o seu quadrado:

$$\left[P_R^+(v)\right]^2 = \left[\frac{e^{Rv} + e^{-Rv}}{2}\right]^2 \qquad \left[P_R^-(v)\right]^2 = \left[\frac{e^{Rv} - e^{-Rv}}{2R}\right]^2$$
$$\left[P_R^+(v)\right]^2 = \frac{e^{2Rv} + 2e^{Rv}e^{-Rv} + e^{-2Rv}}{4} \qquad \left[P_R^-(v)\right]^2 = \frac{e^{2Rv} - 2e^{Rv}e^{-Rv} + e^{-2Rv}}{4R^2}$$
$$\left[P_R^+(v)\right]^2 = \frac{e^{2Rv} + 2 + e^{-2Rv}}{4} \qquad \left[P_R^-(v)\right]^2 = \frac{e^{2Rv} - 2 + e^{-2Rv}}{4R^2}$$

Substituindo na relação fundamental:

$$\left[P_R^+(v)\right]^2 - R^2\left[P_R^-(v)\right]^2 = 1$$

$$l=\frac{e^{2Rv}+2+e^{-2Rv}}{4}-R^2\frac{e^{2Rv}-2+e^{-2Rv}}{4R^2}$$

$$l=\frac{e^{2Rv}+2+e^{-2Rv}-e^{2Rv}+2-e^{-2Rv}}{4}$$

$$l=1, \qquad\qquad Q.E.D.$$

Agora que conhecemos a forma analítica das funções de Poincaré, podemos obter as suas expansões em série de Taylor:

$$P_R^+(v)=\frac{1}{2}\left[1+\sum_{n=1}^{\infty}\frac{(vR)^n}{n!}+1+\sum_{n=1}^{\infty}\frac{(-vR)^n}{n!}\right]$$

Vamos separar as duas funções em suas partes pares e ímpares:

$$P_R^+(v)=\frac{1}{2}\left[2+\sum_{n=1}^{\infty}\frac{(vR)^{2n}}{(2n)!}+\sum_{n=1}^{\infty}\frac{(vR)^{2n+1}}{(2n+1)!}+\sum_{n=1}^{\infty}\frac{(-vR)^{2n}}{(2n)!}+\sum_{n=1}^{\infty}\frac{(-vR)^{2n+1}}{(2n+1)!}\right]$$

Realizando as operações algébricas, obtemos a expansão da função par de Poincaré:

$$P_R^+(v)=1+\sum_{n=1}^{\infty}\left(R^2\right)^n\frac{v^{2n}}{(2n)!}$$

Agora, vamos obter a expansão da função ímpar de Poincaré:

$$P_R^-(v)=\frac{1}{2R}\left[\sum_{n=1}^{\infty}\frac{(vR)^n}{n!}-\sum_{n=1}^{\infty}\frac{(-vR)^n}{n!}\right]$$

$$P_R^-(v)=\frac{1}{2R}\left[2+\sum_{n=1}^{\infty}\frac{(vR)^{2n}}{(2n)!}+\sum_{n=1}^{\infty}\frac{(vR)^{2n}(vR)}{(2n+1)!}+\sum_{n=1}^{\infty}\frac{\left[(-vR)^2\right]^n}{(2n)!}+\sum_{n=1}^{\infty}\frac{(-vR)^{2n}(-vR)}{(2n+1)!}\right]$$

$$P_R^-(v)=\frac{1}{2R}\left[2+\sum_{n=1}^{\infty}\frac{(vR)^{2n}}{(2n)!}+vR\sum_{n=1}^{\infty}\frac{(vR)^{2n}}{(2n+1)!}+\sum_{n=1}^{\infty}\frac{(vR)^{2n}}{(2n)!}-vR\sum_{n=1}^{\infty}\frac{(vR)^{2n}}{(2n+1)!}\right]$$

Vamos separar as duas funções em suas partes pares e ímpares:

$$P_R^-(v)=\frac{1}{2R}\left[\sum_{n=0}^{\infty}\frac{(vR)^{2n}}{(2n)!}+\sum_{n=0}^{\infty}\frac{(vR)^{2n+1}}{(2n+1)!}-\sum_{n=0}^{\infty}\frac{(-vR)^{2n}}{(2n)!}-\sum_{n=0}^{\infty}\frac{(-vR)^{2n+1}}{(2n+1)!}\right]$$

$$P_R^-(v)=\frac{1}{2R}\left[\sum_{n=0}^{\infty}\frac{(vR)^{2n}}{(2n)!}+\sum_{n=0}^{\infty}\frac{(vR)^{2n}(vR)}{(2n+1)!}-\sum_{n=0}^{\infty}\frac{\left[(-vR)^2\right]^n}{(2n)!}-\sum_{n=0}^{\infty}\frac{(-vR)^{2n}(-vR)}{(2n+1)!}\right]$$

$$P_R^-(v)=\frac{1}{2R}\left[\sum_{n=0}^{\infty}\frac{(vR)^{2n}}{(2n)!}+vR\sum_{n=0}^{\infty}\frac{(vR)^{2n+1}}{(2n+1)!}-\sum_{n=0}^{\infty}\frac{(vR)^{2n}}{(2n)!}+vR\sum_{n=0}^{\infty}\frac{(vR)^{2n+1}}{(2n+1)!}\right]$$

Realizando as operações algébricas, obtemos a expansão da função ímpar de Poincaré:

$$P_R^-(v) = v + \sum_{n=1}^{\infty} \left(R^2\right)^n \frac{v^{2n+1}}{(2n+1)!}$$

É fácil ver que a elas se aplicam a fórmula de duplicação de arcos:

$$P_+^R(\theta_1 \pm \theta_2) = P_+^R(\theta_1) P_+^R(\theta_2) \pm R^2 P_-^R(\theta_1) P_-^R(\theta_2)$$
$$P_-^R(\theta_1 \pm \theta_2) = P_+^R(\theta_1) P_-^R(\theta_2) \pm P_-^R(\theta_1) P_+^R(\theta_2)$$

Além disso, essas funções satisfazem a identidade trigonométrica generalizada:

$$\left[P_+^R(v)\right]^2 - R^2 \left[P_-^R(v)\right]^2 = 1$$

E, estão garantidas os valores das funções em 0:

$$P_+^R(0) = 1 \qquad\qquad P_-^R(0) = 0$$

## A Função Tangente de Poincaré

Assim como na trigonometria elementar e hiperbólica, convém introduzir uma função tangente definida como a razão das funções de Poincaré par e ímpar. Também introduziremos as funções arco, das quais o arco tangente irá desempenhar um papel fundamental na construção do teorema da composição das velocidades. Formalmente, definimos as funções tangentes pelos seguintes homeomorfismos:

$$P_{\mp}^R(\theta) : \mathbb{R} \to \mathbb{K} \qquad\qquad P_{\pm}^R(\theta) : \mathbb{R} \to \mathbb{K}$$
$$P_{\mp}^R(\theta) = \frac{P_-^R(\theta)}{P_+^R(\theta)} \qquad\qquad P_{\pm}^R(\theta) = \frac{P_+^R(\theta)}{P_-^R(\theta)}$$

A primeira será denominada de tangente de Poincaré e a segunda função será denominada de cotangente de Poincaré. Os sinais duplos indicam a ordem da razão entre as funções de Poincaré. É fácil ver que, estas funções são iguais ao fator beta de Lorentz generalizado:

$$P_{\mp}^R(\theta) = \frac{v}{k} \qquad\qquad P_{\pm}^R(\theta) = \frac{k}{v}$$
$$P_{\mp}^R(\theta) = \mathrm{B} \qquad\qquad P_{\pm}^R(\theta) = \mathrm{B}^{-1}$$

Por questão de comodidade, vamos introduzir uma nova função tangente, que chamaremos de tangente e cotangente de Poincaré *R,* que será definida como:

$$P_{\odot}^R(\theta) : \mathbb{R} \to \mathbb{K} \qquad\qquad P_{\otimes}^R(\theta) : \mathbb{R} \to \mathbb{K}$$
$$P_{\odot}^R(\theta) = R^2 P_{\mp}^R(\theta) \qquad\qquad P_{\otimes}^R(\theta) = R^2 P_{\pm}^R(\theta)$$

A partir da identidade trigonométrica generalizada, podemos estabelecer relações fundamentais envolvendo as tangentes de Poincaré *R*.

$$1-\left[P_{\odot}^{R}(\theta)\right]^{2}=\left[P_{+}^{R}(\theta)\right]^{-2} \qquad A=\frac{1}{\sqrt{1-\left[P_{\odot}^{R}(\theta)\right]^{2}}}$$
$$1-\left[P_{\otimes}^{R}(\theta)\right]^{2}=\left[P_{-}^{R}(\theta)\right]^{-2}$$

Para desenvolvermos a análise nas variedades espaço-temporais, devemos deduzir a regra de soma de arcos para a tangente e a cotangente de Poincaré:

$$P_{\mp}^{R}(\theta_1\pm\theta_2)=\frac{P_{-}^{R}(\theta_1\pm\theta_2)}{P_{+}^{R}(\theta_1\pm\theta_2)}$$

$$P_{\mp}^{R}(\theta_1\pm\theta_2)=\frac{P_{+}^{R}(\theta_1)P_{-}^{R}(\theta_2)\pm P_{-}^{R}(\theta_1)P_{+}^{R}(\theta_2)}{P_{+}^{R}(\theta_1)P_{+}^{R}(\theta_2)\pm R^2P_{-}^{R}(\theta_1)P_{-}^{R}(\theta_2)}$$

Evidenciando as funções pares do numerador e do denominador:

$$P_{\mp}^{R}(\theta_1\pm\theta_2)=\frac{P_{+}^{R}(\theta_1)P_{+}^{R}(\theta_2)\left[\frac{P_{-}^{R}(\theta_2)}{P_{+}^{R}(\theta_2)}\pm\frac{P_{-}^{R}(\theta_1)}{P_{+}^{R}(\theta_1)}\right]}{P_{+}^{R}(\theta_1)P_{+}^{R}(\theta_2)\left[1\pm R^2\frac{P_{-}^{R}(\theta_1)P_{-}^{R}(\theta_2)}{P_{+}^{R}(\theta_1)P_{+}^{R}(\theta_2)}\right]}$$

$$P_{\mp}^{R}(\theta_1\pm\theta_2)=\frac{\left[\frac{P_{-}^{R}(\theta_2)}{P_{+}^{R}(\theta_2)}\pm\frac{P_{-}^{R}(\theta_1)}{P_{+}^{R}(\theta_1)}\right]}{\left[1\pm R^2\frac{P_{-}^{R}(\theta_1)P_{-}^{R}(\theta_2)}{P_{+}^{R}(\theta_1)P_{+}^{R}(\theta_2)}\right]}$$

Usando a função tangente de Poincaré, obtemos:

$$P_{\mp}^{R}(\theta_1\pm\theta_2)=\frac{P_{\mp}^{R}(\theta_2)\pm P_{\mp}^{R}(\theta_1)}{1\pm R^2P_{\mp}^{R}(\theta_1)P_{\mp}^{R}(\theta_2)}$$

Para a cotangente teremos:

$$P_{\pm}^{R}(\theta_1\pm\theta_2)=\frac{1\pm R^2P_{\pm}^{R}(\theta_1)P_{\pm}^{R}(\theta_2)}{P_{\pm}^{R}(\theta_2)\pm P_{\pm}^{R}(\theta_1)}$$

Para as tangentes *R,* teremos:

$$P_{\odot}^{R}(\theta_1\pm\theta_2)=\frac{R^4}{R^2}\,\frac{P_{\mp}^{R}(\theta_2)\pm P_{\mp}^{R}(\theta_1)}{1\pm R^2P_{\mp}^{R}(\theta_1)P_{\mp}^{R}(\theta_2)}$$

$$P_{\odot}^{R}(\theta_1\pm\theta_2)=R^2\,\frac{R^2P_{\mp}^{R}(\theta_2)\pm R^2P_{\mp}^{R}(\theta_1)}{R^2\pm R^2P_{\mp}^{R}(\theta_1)R^2P_{\mp}^{R}(\theta_2)}$$

$$P_{\odot}^{R}\left(\theta_1 \pm \theta_2\right) = R^2 \frac{P_{\odot}^{R}\left(\theta_2\right) \pm P_{\odot}^{R}\left(\theta_1\right)}{R^2 \pm P_{\odot}^{R}\left(\theta_1\right) P_{\odot}^{R}\left(\theta_2\right)}$$

E para a cotangente *R,*

$$P_{\otimes}^{R}\left(\theta_1 \pm \theta_2\right) = R^2 \left( \frac{R^2 \pm P_{\odot}^{R}\left(\theta_1\right) P_{\odot}^{R}\left(\theta_2\right)}{P_{\odot}^{R}\left(\theta_2\right) \pm P_{\odot}^{R}\left(\theta_1\right)} \right)$$

# O Teorema de Adição de Velocidades

Um dos mais importantes resultados da Teoria da Relatividade Especial é a composição das velocidades. Há duas formas de deduzir esse resultado, para as componentes paralelas. Nessa seção apresentaremos um destes métodos que consiste em utilizar uma função bijetora entre o espaço matemático dos ângulos de rotação e o espaço das grandezas físicas associadas: velocidade, fator beta e fator gama. Vamos construir três sistemas inerciais *K, K' e K".* Sem perda de generalidade, convencionaremos que o sistema *K* é o sistema estacionário e o sistema *K'* se desloca na direção *x* com velocidade $v_1$ em relação ao sistema *K* e velocidade $v_2$, também na direção *x,* em relação ao sistema *K",* enquanto o sistema *K"* se de desloca com velocidade $v_3$ na direção *x* em relação ao referencial *K.* Queremos determinar a velocidade $v_3$ em função das velocidades $v_1$ e $v_2$. Cada deslocamento produz uma rotação de Poincaré *a.* Portanto a rotação de Poincaré total $a_3$ entre o referencial *K* e o referencial *K",* é a soma das rotações hiperbólicas $a_1$, entre os sistemas *K* e *K',* e $a_2$, entre os sistemas *K'* e *K",* isto é, $a_3 = a_1 + a_2$.

## Método das Funções de Poincaré

Vamos agora procurar uma aplicação bijetora que a cada valor de *a* associa a um valor de *v.* Como vimos, a função tangente de Poincaré transforma um vetor do espaço *A* em um vetor do espaço das velocidades *V* dividido pela velocidade *k.* Assim a nossa aplicação pode ser definida como:

$$\begin{array}{l} L : A \rightarrow V \\ L(a) \mapsto v \end{array} \qquad v = kP_{\mp}^{R}\left(a\right)$$

Portanto a velocidade $v_3$ é definida pela seguinte regra:

$$v_3 = kP_{\mp}^{R}\left(a_3\right) = kP_{\mp}^{R}\left(a_1 + a_2\right)$$

Agora podemos aplicar a regra de soma de arcos da tangente de Poincaré:

$$v_3 = k\left[ \frac{P_{\mp}^{R}\left(a_2\right) + P_{\mp}^{R}\left(a_1\right)}{1 + R^2 P_{\mp}^{R}\left(a_1\right) P_{\mp}^{R}\left(a_2\right)} \right]$$

Substituindo os valores da tangente de Poincaré, obtermos o teorema de adição de velocidades em função dos fatores beta:

$$v_3 = \left[ \frac{B_1 + B_2}{1 + R^2 B_1 B_2} \right],$$

$$B_3 = \frac{B_1 + B_2}{1 + R^2 B_1 B_2}$$

Abrindo os fatores B, obtemos a lei de composição de velocidades:

$$v_3 = \frac{v_1 + v_2}{1 + R^2 v_1 v_2}$$

Pelo mesmo método podemos calcular a transformação do fator gama entre o sistema *K* e o sistema *K”*. Desta vez usaremos a função par de Poincaré:

$$R : K \to \mathfrak{F}$$
$$R(a) \mapsto A$$

onde,

$$A = P_{\mp}^{R}(a)$$

Portanto o fator $A_3$ é definida pela seguinte regra:

$$A_3 = P_{+}^{R}(a_3)$$
$$A_3 = P_{+}^{R}(a_1 + a_2)$$
$$A_3 = P_{+}^{R}(a_1) P_{+}^{R}(a_2) + R^2 P_{-}^{R}(a_1) P_{-}^{R}(a_2)$$
$$A_3 = A_1 A_2 + R^2 A_1 A_2 B_1 B_2$$

Evidenciando os fatores comuns, obtemos a transformação *A*:

$$A_3 = A_1 A_2 \left(1 + R^2 A_1 A_2\right)$$

Também poderíamos ter utilizado a função ímpar de Poincaré:

$$B_3 A_3 = P_{-}^{R}(a_3)$$
$$B_3 A_3 = P_{-}^{R}(a_1 + a_2)$$
$$B_3 A_3 = P_{+}^{R}(a_1) P_{-}^{R}(a_2) + P_{-}^{R}(a_1) P_{+}^{R}(a_2)$$
$$B_3 A_3 = A_1 A_2 B_1 + A_1 A_2 B_2$$
$$\frac{(A_1 + A_2)}{1 + R^2 B_1 B_2} A_3 = A_1 A_2 (B_1 + B_2)$$
$$\Gamma_3 = \Gamma_1 \Gamma_2 \left(1 + R^2 B_1 B_2\right)$$

## Método da Álgebra Geométrica Induzida

Na seção anterior deduzimos o teorema de adição de velocidades usando as funções híbridas de Poincaré. Esse mesmo resultado pode ser deduzido por meio da álgebras geométrica. Nessa seção mostraremos esse procedimento. Para detalhes o leitor deverá consultar Josipovic (2019, p. 102).

Inicialmente vamos introduzir o conceito de paravetor do espaço tempo. Um paravetor é um multivetor construído a partir do produto direto das álgebras de extensão $\wedge^0(\mathbb{R})$ e $\wedge^1(\mathbb{R})$, que pode ser escrito como a seguinte aplicação bilinear:

$$\psi : \mathbb{R}\times\mathbb{E} \to \wedge^0(\mathbb{R}) \oplus \wedge^1(\mathbb{R})$$
$$\psi(v_o, \vec{v}) = v_o + R\vec{v}$$

Chamaremos de involutor de um paravetor como o paravetor construído a partir da operação involução, definida por:

$$\bar{\psi} : \mathbb{R}\times\mathbb{E} \to \wedge^0(\mathbb{R}) \oplus \wedge^1(\mathbb{R})$$
$$\bar{\psi}(v_o, \vec{v}) = v_o - R\vec{v}$$

O produto de um paravetor por seu involutor define um invariante denominado norma ao quadrado:

$$\psi^2 : \wedge^0(\mathbb{R}) \oplus \wedge^1(\mathbb{R}) \times \wedge^0(\mathbb{R}) \oplus \wedge^1(\mathbb{R}) \to \mathbb{R}$$
$$\psi^2 : \psi\bar{\psi} = v_o^2 - R^2 v^2$$

Para nossa dedução introduziremos dois paravetores: tempo próprio e a velocidade própria, definidos, respectivamente, como:

$$\begin{cases} \tau = t + R\vec{x} \\ \bar{\tau} = t - R\vec{x} \\ \tau^2 = t^2 - R^2 x^2 \end{cases} \qquad \begin{cases} u = A(1 + R\vec{v}) \\ \bar{u} = A(1 - R\vec{v}) \\ u^2 = A^2(1 - R^2 v^2) \end{cases}$$

Evidenciando o valor $t^2$ do invariante do tempo próprio, obtemos:

$$\tau^2 = t^2\left(1 - R^2\, x^2/t^2\right)$$
$$\tau^2 = t^2\left(1 - R^2 v^2\right)$$
$$\frac{t^2}{\tau^2} = \frac{1}{\left(1 - R^2 v^2\right)}$$

Expressando a razão por meio do coeficiente *A*,

$$\frac{t^2}{\tau^2} = A^2$$

Por essa razão, é fácil ver que o invariante velocidade própria é a unidade ($c$):

$$u^2 = A^2 \frac{1}{A^2} = 1$$

A composição de velocidades no espaço-tempo (1-1) é o produto dos *boosts* na direção do movimento do observador em repouso ($u_0 = 1$) e dos observadores em movimento ($u_1$, $u_2$).

$$u = u_0 u_1 u_2$$

Substituindo os valores das velocidades:

$$\begin{aligned} u &= A_1 \left(1 + R\vec{v}_1\right) A_2 \left(1 + R\vec{v}_2\right) \\ u &= A_1 A_2 \left[1 + R\vec{v}_2 + R\vec{v}_1 + R^2 v_1 v_2\right] \\ u &= A_1 A_2 \left[1 + R^2 v_1 v_2 + R\left(\vec{v}_2 + \vec{v}_1\right)\right] \\ u &= A_1 A_2 \left[1 + R^2 v_1 v_2 + R\left(v_2 + v_1\right) e_x\right] \end{aligned}$$

Evidenciando o fator $1 + R^2 v_1 v_2$ (para os valores que ele seja diferente de zero),

$$u = A_1 A_2 \left(1 + R^2 v_1 v_2\right) \left[1 + R \frac{\left(v_2 + v_1\right)}{1 + R^2 v_1 v_2} e_x\right]$$

Escrevendo usando a definição de paravetor, obtemos:

$$u = A\left(1 + R\vec{v}\right) \qquad \begin{cases} A = A_1 A_2 \left(1 + R^2 v_1 v_2\right) \\ \vec{v} = \dfrac{\left(v_2 + v_1\right)}{1 + R^2 v_1 v_2} e_x \end{cases}$$

A parte escalar é a transformação do fator A e o vetor $\boldsymbol{v}$ é o vetor composição de velocidades. Antes de encerrarmos vamos verificar para quais valores $1 + R^2 v_1 v_2$ é nulo.

$$\begin{aligned} 1 + R^2 v_1 v_2 &\neq 0 \\ -R^2 v_1 v_2 &\neq 1 \end{aligned}$$

Extraindo a raiz quadrada e escrevendo explicitamente a velocidade da luz $c$:

$$\sqrt{-R^2 v_1 v_2} \neq c$$

Na variedade de Minkowski essa restrição está associada ao fato de todo corpo com velocidade diferente (reciprocamente: igual) da luz apresentará velocidade diferente (reciprocamente: igual) da luz em todos os referenciais inerciais.

## Geometria do Espaço-Tempo

Vamos agora obter a métrica do espaço-tempo plano. Em um espaço-tempo ortocrônico, teremos quatro versores: três espaciais e um temporal.

$$B = \{\hat{e}_x, \hat{e}_t\}$$

Definimos a métrica como o produto interno dos versores da base:

$$\begin{cases} \eta_{xx} = \langle \hat{e}_x, \hat{e}_x \rangle = 1 \\ \eta_{tt} = \langle \hat{e}_t, \hat{e}_t \rangle = T \end{cases} \qquad \eta_{ij} = 0, \quad \forall i \neq j$$

Na forma matricial é escrita como:

$$\eta_{ij} = \begin{pmatrix} 1 & 0 \\ 0 & T \end{pmatrix}$$

A diagonalidade da métrica decorre da ortogonalidade dos versores. Não conhecemos os versores de *t,* por essa razão não conhecemos o valor da norma ao quadrado de *t.* Para determinar esse valor, que denotamos por *T*, usaremos a condição de automorfismo:

$$\Lambda^i_j \eta_{ij} \Lambda^j_i = \eta_{ij}$$

$$\begin{bmatrix} A & -vA \\ -R^2vA & A \end{bmatrix} \begin{bmatrix} 1 & 0 \\ 0 & T \end{bmatrix} \begin{bmatrix} A & -R^2vA \\ -vA & A \end{bmatrix} = \begin{bmatrix} 1 & 0 \\ 0 & T \end{bmatrix}$$

$$\begin{bmatrix} A & -vAT \\ -R^2vA & AT \end{bmatrix} \begin{bmatrix} A & -R^2vA \\ -vA & A \end{bmatrix} = \begin{bmatrix} 1 & 0 \\ 0 & T \end{bmatrix}$$

$$\begin{bmatrix} A^2 + v^2A^2T & -R^2vA^2 - vA^2T \\ -R^2vA^2 - vA^2T & R^4v^2A^2 + A^2T \end{bmatrix} = \begin{bmatrix} 1 & 0 \\ 0 & T \end{bmatrix}$$

Desta relação, extraímos três equações lineares em *T*:

$$\begin{cases} A^2 + v^2A^2T = a \\ R^2vA^2 + vA^2T = 0 \\ R^4v^2A^2 + A^2T = T \end{cases}$$

Vamos operar a segunda equação, para obtermos o valor de *T*.

$$\begin{cases} \left(R^2 + T\right)vA^2 = 0 \\ T = -R^2 \end{cases}$$

Vamos usar as duas equações para retirar a prova real:

$$\begin{cases} A^2 - R^2v^2A^2 = 1 \\ A^2\left(1 - R^2v^2\right)1 = 1 \\ A^2\dfrac{1}{A^2} = 1 \\ 1 = 1 \qquad (Q.E.D) \end{cases} \qquad \begin{cases} R^4v^2A^2 - A^2R^2 = -R^2 \\ A^2\left(R^2v^2 - 1\right)R^2 = -R^2 \\ -A^2\dfrac{R^2}{A^2} = -R^2 \\ -R^2 = -R^2 \qquad (Q.E.D) \end{cases}$$

Portanto as componentes da métrica serão:

$$\eta_{ij} = \begin{pmatrix} 1 & 0 \\ 0 & -R^2 \end{pmatrix} \qquad \det\left(\eta_{ij}\right) = -R^2$$

De forma que cada unidade hipercomplexa induz o valor do determinante da métrica:

$$\det \eta_{ij} = \begin{cases} -1, \ espaço-tempo\ de\ Minkowski \\ 0, \ \ espaço-tempo\ de\ Galileu \\ +1, \ espaço-tempo\ de\ Euclides \end{cases}$$

Assim, podemos usar o determinante da métrica do espaço-tempo para determinar a medida do ângulo.

O elemento de linha na variedade espaço-tempo é definida a partir da métrica pela relação:

$$ds^2 = \sum_{i=1}^{2}\sum_{j=1}^{2}\eta_{ij}dx^idx^j$$

Expandindo as somas, obtemos a forma quadrática fundamental:

$$ds^2 = \eta_{xx}dxdx + \eta_{tt}dtdt$$
$$ds^2 = dx^2 - R^2dt^2$$

Determinada a métrica geral do espaço-tempo plano, devemos estudar as expressões das funções de Poincaré para cada número hipercomplexo e verificar como estes números induzem a métrica da variedade.

### 1) Função Parabólica de Poincaré e a Variedade de Galileu

$$P_\varepsilon^+(v) = 1 + \sum_{n=0}^{\infty}\left(\varepsilon^2\right)^n\frac{v^{2n}}{(2n)!} \qquad P_\varepsilon^-(v) = v + \sum_{n=1}^{\infty}\left(\varepsilon^2\right)^n\frac{v^{2n+1}}{(2n+1)!}$$

$$P_\varepsilon^+(v) = 1 + \sum_{n=0}^{\infty}(0)^n\frac{v^{2n}}{(2n)!} \qquad P_\varepsilon^-(v) = v + \sum_{n=1}^{\infty}(0)^n\frac{v^{2n+1}}{(2n+1)!}$$

$$P_\varepsilon^+(v) = 1 + \sum_{n=0}^{\infty}0\cdot\frac{v^{2n}}{(2n)!} \qquad P_\varepsilon^-(v) = v + \sum_{n=1}^{\infty}0\cdot\frac{v^{2n+1}}{(2n+1)!}$$

$$P_\varepsilon^+(v) = 1 \qquad P_\varepsilon^-(v) = v$$

Na forma matricial, teremos:

$$\Lambda_j^i = \begin{bmatrix} 1 & -v \\ -v\varepsilon^2 & 1 \end{bmatrix} \quad \rightarrow \quad G_j^i = \begin{bmatrix} 1 & -v \\ 0 & 1 \end{bmatrix}$$

Essa matriz *G* corresponde a uma rotação parabólica. A métrica desse espaço será dado por:

$$ds^2 = dx^2 - \varepsilon^2 dt^2$$
$$ds^2 = dx^2 - 0dt^2$$
$$ds^2 = dx^2$$

### 2) Função Hiperbólica de Poincaré e a Variedade de Lorentz

$$P_h^+(v) = 1 + \sum_{n=0}^{\infty} \left(h^2\right)^n \frac{v^{2n}}{(2n)!} \qquad P_h^-(v) = v + \sum_{n=1}^{\infty} \left(h^2\right)^n \frac{v^{2n+1}}{(2n+1)!}$$
$$P_h^+(v) = 1 + \sum_{n=0}^{\infty} (1)^n \frac{v^{2n}}{(2n)!} \qquad P_h^-(v) = v + \sum_{n=1}^{\infty} (1)^n \frac{v^{2n+1}}{(2n+1)!}$$
$$P_h^+(v) = \cosh(v) \qquad P_h^-(v) = \sinh(v)$$

Na forma matricial, teremos:

$$\Lambda_j^i = \begin{bmatrix} \cosh(v) & -\sinh(v) \\ -\sinh(v)h^2 & \cosh(v) \end{bmatrix} \quad \rightarrow \quad L_j^i = \begin{bmatrix} \cosh(v) & -\sinh(v) \\ -\sinh(v) & \cosh(v) \end{bmatrix}$$

Essa matriz *L* corresponde a uma rotação hiperbólica. A métrica desse espaço será dado por:

$$ds^2 = dx^2 - h^2 dt^2$$
$$ds^2 = dx^2 - dt^2$$

Se parametrizarmos as coordenadas espaciais como cosseno hiperbólico e a coordenada temporal como seno hiperbólico, deduzimos que nesse espaço, o tempo opera como um eixo ortogonal aos eixos espaciais e as transformações de Lorentz, são rotações hiperbólicas. As assíntotas da hipérbole são retas de 45º, definidas pelo produto da velocidade da luz pelo tempo. Dada a isotropia da velocidade da luz, essas assíntotas definem uma superfície cônica. Os eventos interiores a superfície são os causais, os eventos sobre a superfície são os simultâneos e os eventos fora da superfície são aqueles que as consequências antecedem as causas.

### 3) Função Polar de Poincaré e a Variedade de Euclides

$$P_i^+(v) = 1 + \sum_{n=0}^{\infty} \left(i^2\right)^n \frac{v^{2n}}{(2n)!} \qquad P_i^-(v) = v + \sum_{n=1}^{\infty} \left(i^2\right)^n \frac{v^{2n+1}}{(2n+1)!}$$
$$P_i^+(v) = 1 + \sum_{n=0}^{\infty} (-1)^n \frac{v^{2n}}{(2n)!} \qquad P_i^-(v) = v + \sum_{n=1}^{\infty} (-1)^n \frac{v^{2n+1}}{(2n+1)!}$$
$$P_i^+(v) = \cos(v) \qquad P_i^-(v) = \sin(v)$$

Na forma matricial, teremos:

$$\Lambda^i_j = \begin{bmatrix} \cos(v) & -\sin(v) \\ -\sin(v)i^2 & \cos(v) \end{bmatrix} \rightarrow E^i_j = \begin{bmatrix} \cos(v) & -\sin(v) \\ \sin(v) & \cos(v) \end{bmatrix}$$

Essa matriz *E* corresponde a uma rotação elíptica. A métrica desse espaço será dado por:

$$ds^2 = dx^2 - i^2 dt^2$$
$$ds^2 = dx^2 + dt^2$$

Se parametrizarmos as coordenadas espaciais como cosseno polar e a coordenada temporal como seno polar, deduzimos que nesse espaço, o tempo é uma circunferência (esfera $S^1$) ortogonal aos eixos espaciais e as transformações de Euclides, são rotações esféricas. Esse é o espaço onde o tempo apresenta loops fechados, semelhante as latitudes de um globo. Por estar associado a um número imaginário, esse tempo é denominado de imaginário ou de tempo euclidiano (Windred, 1935).

## Desigualdade Triangular do Espaço-Tempo

Seja $\vec{u}$ e $u_t$ vetores unitários espacial e temporal respectivamente, vamos calcular a norma do bivetor gerado pelo produto exterior destes dois vetores:

$$\vec{u} \wedge \vec{u}_t = b(e_x e_t)$$
$$|\vec{u} \wedge \vec{u}_t|^2 = b^2 |(e_x e_t)|^2$$
$$|\vec{u} \wedge \vec{u}_t|^2 = b^2 |R|^2$$

Assim como para os números hipercomplexos, definimos a norma ao quadrado de *R* pelo seu produto por seu conjugado. Como os vetores são unitários, segue que:

$$|\vec{u} \wedge \vec{u}_t|^2 = b^2 (R\bar{R})$$
$$|\vec{u} \wedge \vec{u}_t|^2 = -b^2 RR$$
$$|\vec{u} \wedge \vec{u}_t|^2 = -b^2 R^2$$

Agora vamos continuar desenvolvendo a equação:

$$|\vec{u} \wedge \vec{u}_t|^2 = (\vec{u} \wedge \vec{u}_t)(\vec{u}_t \wedge \vec{u})$$
$$|\vec{u} \wedge \vec{u}_t|^2 = (\vec{u}\vec{u}_t - \vec{u} \cdot \vec{u}_t)(\vec{u}_t\vec{u} - \vec{u}_t \cdot \vec{u})$$
$$|\vec{u} \wedge \vec{u}_t|^2 = |\vec{u}|^2 |\vec{u}_t|^2 - (\vec{u} \cdot \vec{u}_t)^2$$

Igualando as duas expressões que obtivemos:

$$(\vec{u} \cdot \vec{u}_t)^2 = \left(|\vec{u}|^2 |\vec{u}_t|^2 + b^2 R^2\right)$$

Agora vamos obter os três casos possíveis da desigualdade:

$$(\vec{u}\cdot\vec{u}_t)^2 \begin{cases} \leq |\vec{u}|^2 |\vec{u}_t|^2, \quad se\ R^2 = -1, \\ (Espaço\ de\ Euclides) \\ = |\vec{u}|^2 |\vec{u}_t|^2, \quad se\ R^2 = \ 0, \\ (Espaço\ de\ Galileu) \\ \geq |\vec{u}|^2 |\vec{u}_t|^2, \quad se\ R^2 = +1 \\ (Espaço\ de\ Minkowski) \end{cases}$$

Para quaisquer vetores, a norma ao quadrado de sua soma é dado pela relação fundamental:

$$|\boldsymbol{u}+\boldsymbol{v}|^2 = |\boldsymbol{u}|^2 + |\boldsymbol{v}|^2 + 2(\boldsymbol{u}\cdot\boldsymbol{v})$$

Observe que no caso da soma de um vetor espacial e um temporal, teremos:

$$\left|u^x e_x + u^t e_t\right| = |\vec{u}+\vec{u}_t|$$

que a definição de um quadrivetor do espaço-tempo:

$$|\boldsymbol{u}| = |\vec{u}+\vec{u}_t|$$
$$|\boldsymbol{u}|^2 = |\vec{u}+\vec{u}_t|^2$$

por outro lado, temos que:

$$|\vec{u}+\vec{u}_t|^2 = |\vec{u}|^2 + |\vec{u}_t|^2 + 2(\vec{u}\cdot\vec{u}_t)$$

portanto,

$$|\boldsymbol{u}|^2 = |\vec{u}|^2 + |\vec{u}_t|^2 + 2(\vec{u}\cdot\vec{u}_t)$$

conforme as relações que obtivemos para o produto interno, teremos as seguintes relações de (des)igualdade para cada espaço-tempo:

$$|\boldsymbol{u}| = |\vec{u}+\vec{u}_t| \begin{cases} \leq |\vec{u}| + |\vec{u}_t|, \quad se\ R^2 = -1 \\ (Espaço\ de\ Euclides) \\ = |\vec{u}|, \qquad se\ R^2 = \ 0 \\ (Espaço\ de\ Galileu) \\ \geq |\vec{u}| + |\vec{u}_t|, \quad se\ R^2 = +1 \\ (Espaço\ de\ Minkowski) \end{cases}$$

Isso implica que a passagem do tempo próprio para viajantes acelerados, como no caso do paradoxo dos gêmeos, é uma propriedade geométrica da variedade:

**1) Desigualdade Triangular na Variedade de Euclides**

$$|\boldsymbol{u}| = |\vec{u} + \vec{u}_t| \leq |\vec{u}| + |\vec{u}_t|$$

O espaço Euclidiano opera como um espelho do espaço Lorentziano, isto está ligado a uma propriedade matemática chamada de dualidade. No caso da desigualdade triangular envolvendo viagens no espaço e no tempo, isso implica que o paradoxo dos gêmeos, é o gêmeo inercial que envelhece menos. Nesta variedade, quanto maior a velocidade de um corpo de teste, mais rápido o tempo irá passar, em outras palavras, o tempo sofre uma contração, enquanto o espaço se dilata. Além disso, o efeito Doppler-Fizeau da luz ocorre no sentido contrário: a luz sofre um desvio para o vermelho quando vai de encontro para um corpo e sofre um desvio para o azul quando se afasta do corpo.

**2) Desigualdade Triangular na Variedade de Galileu**

$$|\boldsymbol{u}| = |\vec{u} + \vec{u}_t| = |\vec{u}|$$

Isso significa que a norma de quadrivetor na variedade de Galileu depende apenas de suas componentes espaciais. Em outras palavras, é como se o tempo se contraísse em um ponto, pois seu comprimento é nulo. Fisicamente, isso significa que os eventos mensuráveis na variedade de Galileu se encontram todos na hipersuperfície do presente, de forma que a simultaneidade entre eventos separados seja absoluta. Esse resultado contraria a afirmação de Minkowski de que a variedade de Galileu exige um grupo $G_\infty$, isto é, quando a velocidade da luz é instantânea (infinita). A variedade de Galileu $G_{\varepsilon c}$ é isomórfica a variedade de $G_\infty$, sem exigir uma velocidade da luz instantânea. Nesse caso, a velocidade da luz deixa de ser a mesma para todos os referenciais inerciais e passa a depender do estado de movimento do observador. Como veremos, isso exige modificações nas transformações dos campos elétricos e magnéticos. O fato de uma construção mecânica do eletromagnetismo ser impossível, deriva do fato que o grupo de deslocamento da mecânica racional é $G_\infty$ ou $G_{\varepsilon c}$, e as medidas empíricas sobre o comportamento do campo eletromagnético serem compatíveis com $G_{pc}$.

**3) Desigualdade Triangular na Variedade de Lorentz**

$$|\boldsymbol{u}| = |\vec{u} + \vec{u}_t| \geq |\vec{u}| + |\vec{u}_t|$$

Na variedade Lorentziana a desigualdade triangular revela que qualquer linha do mundo não inercial tem um comprimento menor que uma linha inercial (seguimento de reta). Fisicamente, um observador acelerado sofre uma dilatação do tempo em relação aos observadores inerciais, resultando no famoso paradoxo dos gêmeos. Por se tratar um exemplo amplamente discutido na literatura, não iremos explorar esse exemplo. Ao leitor indicamos o livro *Teoria da Relatividade Especial* (Bohm, 2015), inclusive por sua abordagem do cálculo $K$ de Bondi.

## Por que o Espaço-Tempo é Minkowskiano?

O princípio da relatividade de Poincaré implica o que espaço-tempo é um espaço afim ou cuja medida do ângulo é parabólico. Isso reduz as nove geometrias de Cayley aos espaços Euclideano, Galileano e Minkowskiano. Naturalmente, somos levados a perguntar qual melhor espaço-tempo se adequa a realidade física? Para isso recorremos a experiência, mais precisamente como dois observadores registram um evento no espaço-tempo.

Suponha que durante um instante $T_1$, dois observadores inerciais O' e O se cruzam e sincronizam seus relógios. Chamaremos esse evento de O. Ela combinam que quando seus relógios próprios registrarem o instante $T_2$, eles irão trocar sinais para confirmar se seus relógios ainda estão sincronizados. Porém, como não dispomos de um método para medir a velocidade da luz em um único sentido, imediatamente a recepção do sinal ($T_3$), ele será refletido, retornando ao seu emissor (T). Esse é o processo proposto por Poincaré (1898, 1900, 1904) e Einstein (1905) para realizar a medida do tempo entre referenciais em movimento.

Devido a homogeneidade e a isotropia do espaço, cada observador irá alegar que no tempo $T_2$, o outro observador se deslocou uma distância $x = vT_2$. O sinal emitido irá percorrer uma distância $X$ dada por $x + c(T_3 - T_2)$. Podemos expressar essa equação de forma mais compacta: $X = cT$ se considerarmos que a velocidade de ida e volta é o mesmo. Nestas condições, teremos:

$$\begin{cases} X = cT \\ X' = cT' \end{cases}$$

Se elevarmos ao quadrado, as duas expressões, obtemos o seguinte invariante:

$$\begin{cases} X^2 = c^2T^2 & \leftrightarrow \quad X^2 - c^2T^2 = 0 \\ X'^2 = c^2T'^2 & \leftrightarrow \quad X'^2 - c^2T'^2 = 0 \end{cases}$$

E, portanto:

$$X^2 - c^2T^2 = X'^2 - c^2T'^2 = 0$$

Essa é a forma quadrática fundamental para conexões do tipo-luz em um espaço Minkowskiano. Se assumirmos o postulado da constância da velocidade da luz, como fez Einstein em 1905, somos induzidos a escolher o espaço de Minkowski. É possível, tornar essa condição ainda mais forte, do ponto de vista empírico, utilizando as linhas coordenadas de Plücker e Cayley, pois somente no espaço de Minkowski verifica-se uma total concordância entre as leis do eletromagnetismo e a estrutura geométrica do espaço-tempo.

# Álgebra do Espaço-Tempo Unificada

As funções de Poincaré permitem escrever todos os grupos SO que geram o espaço-tempo como um único grupo que chamaremos de C-Grupo[9], onde o C é uma referência ao formalismo Cliffor que estamos usando. Nesse novo formalismo, um grupo SO (*m, n*) é construído a partir dos parâmetros de Cayley, da seguinte forma:

***m = número de componentes elípticas.***

***n = número de componentes hiperbólicas***

***As componentes parabólicas tem valor nulo***

***O grupo de Galileu,*** associado aos números duais, tem 3 componentes elípticas (espaço) e 1 componente parabólica (tempo), portanto seu grupo é o SO(3,0), que escrevemos, SO(3).

***O grupo de Euclides,*** associado aos números complexos, tem 4 componentes elípticas (três espaciais e uma temporal), portanto seu grupo é o SO(4,0), que escrevemos como SO(4).

***O grupo de Lorentz,*** associado aos números perplexos, tem 3 componentes elípticas (espaço) e 1 hiperbólica (tempo), portanto seu grupo é o SO(3,1), que escrevemos como SO(3,1).

Com o uso do fator *R,* que induz o espaço-tempo, é possível unificar esses grupos em uma super estrutura, o C-Gripo, como provaremos a seguir:

**C-Grupo de Lorentz**

Definimos a matriz s-transformação de Lorentz pela aplicação:

$$\Lambda^R(a)=\begin{pmatrix} P_+^R(a) & -P_-^R(a) \\ -R^2P_-^R(a) & P_+^R(a) \end{pmatrix}$$

Vamos agora provar que as s-transformadas de Lorentz formam um grupo abeliano. Matematicamente, dizemos que um conjunto $G_{GL}$ munido de uma operação interna que chamaremos por produto, $G_{GL}\left(\Lambda^R(a_i),\cdot\right)$, é um grupo se para todo elemento do conjunto verificam-se as quatro primeiras propriedades abaixo:

[9] Nas versões antigas desse trabalho, eu chamei de Super Grupo (S-Grupo), mas achei mais adequado trocar por Cliffor Grupo (C-Grupo).

$$1.\ \Lambda^R(a_3)=\Lambda^R(a_1)\Lambda^R(a_2)\,|\,\Lambda^R(a_3)\in G_{GL}\left(\Lambda^R(a_i),\cdot\right)$$
$$2.\ \Lambda^R(a_1)\left[\Lambda^R(a_2)\Lambda^R(a_3)\right]=\left[\Lambda^R(a_1)\Lambda^R(a_2)\right]\Lambda^R(a_3)$$
$$3.\ \exists\Lambda^R(I)\,|\,\Lambda^R(I)\Lambda^R(a_i)=\Lambda^R(a_i)\Lambda^R(I)=\Lambda^R(a_i)$$
$$4.\ \exists\Lambda^R(a_j)\equiv\left(\Lambda^R\right)^{-1}(a_i)\,|\,\Lambda^R(a_j)\Lambda^R(a_i)=\Lambda^R(a_i)\Lambda^R(a_j)=\Lambda^R(I)$$
$$5.\ \Lambda^R(a_1)\Lambda^R(a_2)=\Lambda^R(a_2)\Lambda^R(a_1)\,|\,\forall\Lambda^R(a_1)\Lambda^R(a_2)\in G_{GL}\left(\Lambda^R(a_i),\cdot\right)$$

Se grupo satisfaz a quinta propriedade é chamado de comutativo ou abeliano. Vamos primeiro verificar a propriedade do fechamento:

$$\Lambda^R(a_3)=\Lambda^R(a_1)\Lambda^R(a_2)\,|\,\Lambda^R(a_3)\in G_{GL}\left(\Lambda^R(a_i),\cdot\right)$$

$$\Lambda^R(a_3)=\begin{pmatrix} P_+^R(a_1) & -P_-^R(a_1) \\ -R^2P_-^R(a_1) & P_+^R(a_1)\end{pmatrix}\cdot\begin{pmatrix} P_+^R(a_2) & -P_-^R(a_2) \\ -R^2P_-^R(a_2) & P_+^R(a_2)\end{pmatrix}$$

$$\Lambda^R(a_3)=\begin{pmatrix} P_+^R(a_1)P_+^R(a_2)+R^2P_-^R(a_1)P_-^R(a_2) & -P_+^R(a_1)P_-^R(a_2)-P_-^R(a_1)P_+^R(a_2) \\ -R^2P_-^R(a_1)P_+^R(a_2)-R^2P_+^R(a_1)P_-^R(a_2) & R^2P_-^R(a_1)P_-^R(a_2)+P_+^R(a_1)P_+^R(a_2)\end{pmatrix}$$

$$\Lambda^R(a_3)=\begin{pmatrix} P_+^R(a_1)P_+^R(a_2)+R^2P_-^R(a_1)P_-^R(a_2) & -\left[P_+^R(a_1)P_-^R(a_2)+P_-^R(a_1)P_+^R(a_2)\right] \\ -R^2\left[P_+^R(a_1)P_-^R(a_2)+P_-^R(a_1)P_+^R(a_2)\right] & P_+^R(a_1)P_+^R(a_2)+R^2P_-^R(a_1)P_-^R(a_2)\end{pmatrix}$$

Usando as regras de soma de arcos, obtemos:

$$\Lambda^R(a_3)=\begin{pmatrix} P_+^R(a_1+a_2) & -P_-^R(a_1+a_2) \\ -R^2P_-^R(a_1+a_2) & P_+^R(a_1+a_2)\end{pmatrix}$$
$$\Lambda^R(a_3)=\begin{pmatrix} P_+^R(a_3) & -P_-^R(a_3) \\ -R^2P_-^R(a_3) & P_+^R(a_3)\end{pmatrix}$$

Observe que o lado direito é a definição da transformação de Lorentz para um ângulo $a_3$, portanto $\Lambda^R(a_3)\in G_{GL}\left(\Lambda^R(a_i),\cdot\right)$. Por esta fórmula podemos concluir que:

$$\Lambda^R(a_3)=\Lambda^R(a_1)\Lambda^R(a_2)=\Lambda^R(a_1+a_2)$$

Vamos usa-la para demonstrar a associatividade:

$$\Lambda^R(a_1)\left[\Lambda^R(a_2)\Lambda^R(a_3)\right]=\left[\Lambda^R(a_1)\Lambda^R(a_2)\right]\Lambda^R(a_3)$$
$$\Lambda^R(a_1)\left[\Lambda^R(a_2+a_3)\right]=\left[\Lambda^R(a_1+a_2)\right]\Lambda^R(a_3)$$
$$\Lambda^R\left(a_1+\left[a_2+a_3\right]\right)=\Lambda^R\left(\left[a_1+a_2\right]+a_3\right)$$

como a soma dos ângulos é associativa, a igualdade é verdadeira.

Agora, vamos provar a comutatividade, assim não precisaremos provar que o elemento neutro e o elemento inverso comutam, já que a comutatividade é assegurada para todos os ângulos.

$$
\begin{aligned}
&\Lambda^R(a_1)\Lambda^R(a_2)=\Lambda^R(a_1+a_2)\\
&\Lambda^R(a_1+a_2)=\Lambda^R(a_2+a_1)\\
&\Lambda^R(a_1)\Lambda^R(a_2)=\Lambda^R(a_2)\Lambda^R(a_1)
\end{aligned}
$$

como a soma de ângulos comuta, então a igualdade está garantida.

Agora vamos determinar quem é o elemento identidade das s-transformações de Lorentz.

$$
\begin{aligned}
&\Lambda^R(I)\Lambda^R(a_i)=\Lambda^R(a_i)\\
&\Lambda^R(I+a_i)=\Lambda^R(a_i)\\
&I+a_i=a_i\Rightarrow I=0
\end{aligned}
$$

como o ângulo zero pertence ao conjunto dos ângulos e é único, portanto existe um único elemento neutro ou identidade, que é expresso pela seguinte matriz:

$$
\Lambda^R(0)=\begin{pmatrix} P_+^R(0) & -P_-^R(0)\\ -R^2P_-^R(0) & P_+^R(0)\end{pmatrix}\quad\Rightarrow\quad \Lambda^R(0)=\begin{pmatrix}1 & 0\\ 0 & 1\end{pmatrix}
$$

Por fim, iremos calcular o elemento inverso:

$$
\exists\Lambda^R(a_j)\equiv\left(\Lambda^R(a_i)\right)^{-1}\mid\Lambda^R(a_j)\Lambda^R(a_i)=\Lambda^R(a_i)\Lambda^R(a_j)=\Lambda^R(I)
$$

$$
\begin{aligned}
&\Lambda^R(a_j)\Lambda^R(a_i)=\Lambda^R(0)\\
&\Lambda^R(a_j+a_i)=\Lambda^R(0)\\
&a_j=-a_i
\end{aligned}
$$

Como o domínio dos ângulos são os números reais, então $-a_i$ é um elemento do conjunto e é único, portanto existe um único elemento inverso. A matriz inversa será dada por:

$$
\begin{aligned}
&\Lambda(-a_i)=\begin{pmatrix} P_+^R(-a_i) & -P_-^R(-a_i)\\ -R^2P_-^R(-a_i) & P_+^R(-a_i)\end{pmatrix}\\
&\Lambda(-a_i)=\begin{pmatrix} P_+^R(a_i) & -\left[-P_-^R(a_i)\right]\\ -R^2\left[-P_-^R(a_i)\right] & P_+^R(a_i)\end{pmatrix}\\
&\Lambda^{-1}(a_i)=\Lambda(-a_i)=\begin{pmatrix} P_+^R(a_i) & P_-^R(a_i)\\ R^2P_-^R(a_i) & P_+^R(a_i)\end{pmatrix}
\end{aligned}
$$

Portanto, provamos que as transformadas de Lorentz formam um grupo abeliano. O uso de funções de Poincaré torna a demonstração extremamente simples e elegante. Agora convém mostrar porque chamamos esse grupo de $\mathbf{SO}(\mathbf{R})$.

Observe que se tomarmos *R* como a unidade imaginária, teremos o grupo de rotações no espaço-tempo euclidiano SO(4):

$$\mathbf{SO}(\mathbf{i})$$

$$\Lambda^{i}(a)=\begin{pmatrix} P_{+}^{i}(a) & -P_{-}^{i}(a) \\ -i^{2}P_{-}^{i}(a) & P_{+}^{i}(a) \end{pmatrix}$$

$$\mathbf{SO}(\mathbf{i})\equiv\mathbf{SO}(\mathbf{4})$$

$$\Lambda^{i}(a)=\begin{pmatrix} \cos(a) & -\sin(a) \\ \sin(a) & \cos(a) \end{pmatrix}$$

Tomando *R* como a unidade dual, teremos o grupo de rotações no espaço de Galileu, SO(3):

$$\mathbf{SO}(\varepsilon)$$

$$\Lambda^{\varepsilon}(a)=\begin{pmatrix} P_{+}^{\varepsilon}(a) & -P_{-}^{\varepsilon}(a) \\ -\varepsilon^{2}P_{-}^{i}(a) & P_{+}^{\varepsilon}(a) \end{pmatrix}$$

$$\mathbf{SO}(\varepsilon)\equiv\mathbf{SO}(\mathbf{3})$$

$$\Lambda^{\varepsilon}(a)=\begin{pmatrix} 1 & -a \\ 0 & 1 \end{pmatrix}$$

Por fim, para obter o grupo de Lorentz SO(1,3), tome $R = p$.

$$\mathbf{SO}(\mathbf{p})$$

$$\Lambda^{p}(a)=\begin{pmatrix} P_{+}^{p}(a) & -P_{-}^{p}(a) \\ -p^{2}P_{-}^{p}(a) & P_{+}^{p}(a) \end{pmatrix}$$

$$\mathbf{SO}(\mathbf{p})\equiv\mathbf{SO}(\mathbf{1,3})$$

$$\Lambda^{p}(a)=\begin{pmatrix} \cosh(a) & -\sinh(a) \\ \sinh(a) & \cosh(a) \end{pmatrix}$$

Portanto, o grupo generalizado de Lorentz $\mathbf{SO}(\mathbf{R})$ permite gerar, por meio da variação do parâmetro *R,* as três principais grupo de rotações que geram as variedades de Galileu, Lorentz e Euclides.

**Geradores Infinitesimais do Espaço-Tempo**

Vamos agora calcular os geradores do espaço-tempo. Usando a equação de Poincaré para calcular os geradores necessários:

$$\text{var } j = \frac{n^2 - n}{2}$$
$$\text{n} = 4 \rightarrow \text{var } j = 6$$

Portanto precisamos de seis parâmetros livres para calcular os geradores do espaço-tempo. Essa é a razão da álgebra de Lie não abeliana do espaço-tempo, que corresponde as linhas de universo serem descritas por 6-vetores. Quem são os nossos seis parâmetros? São as rotações espaciais (*rot*) (três parâmetros) e os boosts de Lorentz (três parâmetros). Portanto as equações com seus parâmetros são (POINCARÉ, 1906):

$$Rot \begin{cases} f_1 = x + y\delta z - z\delta y \\ f_2 = y - x\delta z + z\delta x \\ f_3 = z + x\delta y - y\delta x \end{cases}$$

$$Boosts \begin{cases} f_4 = R^2\left(x + kt\delta y + y\delta kt\right) \\ f_5 = R^2\left(y + kt\delta z + z\delta kt\right) \\ f_0 = R^2\left(z + kt\delta x + x\delta kt\right) \end{cases}$$

Vamos determinar os geradores infinitesimais:

$$X_0 = M_{00}\partial_0 + M_{10}\partial_1 + M_{20}\partial_2 + M_{30}\partial_3$$
$$X_1 = M_{01}\partial_0 + M_{11}\partial_1 + M_{21}\partial_2 + M_{31}\partial_3$$
$$X_2 = M_{02}\partial_0 + M_{12}\partial_1 + M_{22}\partial_2 + M_{32}\partial_3$$
$$X_3 = M_{03}\partial_0 + M_{13}\partial_1 + M_{23}\partial_2 + M_{33}\partial_3$$
$$X_4 = M_{04}\partial_0 + M_{14}\partial_1 + M_{24}\partial_2 + M_{34}\partial_3$$
$$X_5 = M_{05}\partial_0 + M_{15}\partial_1 + M_{25}\partial_2 + M_{35}\partial_3$$

Substituindo os índices das derivadas:

$$X_0 = M_{00}\partial_t + M_{10}\partial_x + M_{20}\partial_y + M_{30}\partial_z$$
$$X_1 = M_{01}\partial_t + M_{11}\partial_x + M_{21}\partial_y + M_{31}\partial_z$$
$$X_2 = M_{02}\partial_t + M_{12}\partial_x + M_{22}\partial_y + M_{32}\partial_z$$
$$X_3 = M_{03}\partial_t + M_{13}\partial_x + M_{23}\partial_y + M_{33}\partial_z$$
$$X_4 = M_{04}\partial_t + M_{14}\partial_x + M_{24}\partial_y + M_{34}\partial_z$$
$$X_5 = M_{05}\partial_t + M_{15}\partial_x + M_{25}\partial_y + M_{35}\partial_z$$

Agora vamos calcular os valores dos coeficientes $M_{ij}$:

| $\mathbf{M_{i1}(ct,x,y,z)=\dfrac{\partial f_i(ct,x,y,z,\delta ct,\delta x,\delta y,\delta z)}{\partial(\delta a_1)}}$ | | |
|---|---|---|
| $M_{01}=\dfrac{\partial(x+y\delta z-z\delta y)}{\partial(\delta kt)}$<br>$M_{01}=0$ | $M_{02}=\dfrac{\partial(y-x\delta z+z\delta x)}{\partial(\delta kt)}$<br>$M_{02}=0$ | $M_{03}=\dfrac{\partial(z+x\delta y-y\delta x)}{\partial(\delta kt)}$<br>$M_{03}=0$ |
| $M_{04}=R^2\dfrac{\partial(x+kt\delta y+y\delta kt)}{\partial(\delta kt)}$<br>$M_{04}=R^2y$ | $M_{05}=R^2\dfrac{\partial(y+ct\delta z+z\delta kt)}{\partial(\delta kt)}$<br>$M_{05}=R^2z$ | $M_{00}=R^2\dfrac{\partial(z+kt\delta x+x\delta kt)}{\partial(\delta kt)}$<br>$M_{00}=R^2x$ |
| $M_{11}=\dfrac{\partial(x+y\delta z-z\delta y)}{\partial(\delta x)}$<br>$M_{11}=0$ | $M_{12}=\dfrac{\partial(y-x\delta z+z\delta x)}{\partial(\delta x)}$<br>$M_{12}=z$ | $M_{13}=\dfrac{\partial(z+x\delta y-y\delta x)}{\partial(\delta x)}$<br>$M_{13}=-y$ |
| $M_{14}=R^2\dfrac{\partial(x+kt\delta y+y\delta kt)}{\partial(\delta x)}$<br>$M_{14}=0$ | $M_{15}=R^2\dfrac{\partial(y+kt\delta z+z\delta kt)}{\partial(\delta x)}$<br>$M_{15}=0$ | $M_{10}=R^2\dfrac{\partial(z+kt\delta x+x\delta kt)}{\partial(\delta x)}$<br>$M_{10}=R^2kt$ |
| $M_{21}=\dfrac{\partial(x+y\delta z-z\delta y)}{\partial(\delta y)}$<br>$M_{21}=-z$ | $M_{22}=\dfrac{\partial(y-x\delta z+z\delta x)}{\partial(\delta y)}$<br>$M_{22}=0$ | $M_{23}=\dfrac{\partial(z+x\delta y-y\delta x)}{\partial(\delta y)}$<br>$M_{23}=x$ |
| $M_{24}=R^2\dfrac{\partial(x+kt\delta y+y\delta kt)}{\partial(\delta y)}$<br>$M_{24}=R^2kt$ | $M_{25}=R^2\dfrac{\partial(y+kt\delta z+z\delta kt)}{\partial(\delta y)}$<br>$M_{25}=0$ | $M_{20}=R^2\dfrac{\partial(z+kt\delta x+x\delta kt)}{\partial(\delta y)}$<br>$M_{20}=0$ |
| $M_{31}=\dfrac{\partial(x+y\delta z-z\delta y)}{\partial(\delta z)}$<br>$M_{31}=y$ | $M_{32}=\dfrac{\partial(y-x\delta z+z\delta x)}{\partial(\delta z)}$<br>$M_{32}=-x$ | $M_{33}=\dfrac{\partial(z+x\delta y-y\delta x)}{\partial(\delta z)}$<br>$M_{33}=0$ |
| $M_{34}=R^2\dfrac{\partial(x+kt\delta y+y\delta kt)}{\partial(\delta z)}$<br>$M_{34}=0$ | $M_{25}=R^2\dfrac{\partial(y+kt\delta z+z\delta kt)}{\partial(\delta z)}$<br>$M_{35}=R^2kt$ | $M_{30}=R^2\dfrac{\partial(z+kt\delta x+x\delta kt)}{\partial(\delta z)}$<br>$M_{30}=0$ |

Substituindo os valores dos coeficientes $M$ nas equações dos geradores infinitesimais:

$$X_0=R^2x\partial_t+R^2kt\partial_x+0\partial_y+0\partial_z \qquad X_0=R^2(x\partial_t+kt\partial_x)$$
$$X_1=0\partial_t+0\partial_x-z\partial_y+y\partial_z \qquad X_1=y\partial_z-z\partial_y$$
$$X_2=0\partial_t+z\partial_x+0\partial_y-x\partial_z \qquad X_2=z\partial_x-x\partial_z$$
$$X_3=0\partial_t-y\partial_x+x\partial_y+0\partial_z \qquad X_3=x\partial_y-y\partial_x$$
$$X_4=R^2y\partial_t+0\partial_x+R^2kt\partial_y+0\partial_z \qquad X_4=R^2(y\partial_t+kt\partial_y)$$
$$X_5=R^2z\partial_t+0\partial_x+0\partial_y+R^2kt\partial_z \qquad X_5=R^2(z\partial_t+kt\partial_z)$$

Os geradores infinitesimais do espaço-tempo são os vetores de Killing de $\mathbf{SO}(\mathbf{R})$.

O conjunto composto pelos elementos $X_1$, $X_2$, $X_3$ são as rotações espaciais ao redor dos eixos *x, y,* e *z*, concomitantemente. Já o conjunto composto pelos elementos $X_0$, $X_5$, $X_6$ são os *boosts* de Lorentz nas direções *x, y, z,* respectivamente.

## 2.1. Constantes da Estrutura do Espaço-Tempo

Vamos agora calcular os tensores da estrutura espaço-tempo por meio dos seus geradores infinitesimais. Deveremos expandir 15 colchetes de Lie, porém como os geradores são funções lineares, os cálculos são simples. Fixando o gerador $X_0$, teremos:

$$[X_0, X_1] = C_{01}^0 e_t + C_{01}^1 e_x + C_{01}^2 e_y + C_{01}^3 e_z$$
$$[X_0, X_1] = R^2(x\partial_t + kt\partial_x)(y\partial_z - z\partial_y) - R^2(y\partial_z - z\partial_y)(x\partial_t + kt\partial_x)$$
$$[X_0, X_1] = C_{01}^0 = C_{01}^1 = C_{01}^2 = C_{01}^3 = 0$$

$$[X_0, X_2] = C_{02}^0 e_t + C_{02}^1 e_x + C_{02}^2 e_y + C_{02}^3 e_z$$
$$[X_0, X_2] = R^2(x\partial_t + kt\partial_x)(z\partial_x - x\partial_z) - R^2(z\partial_x - x\partial_z)(x\partial_t + kt\partial_x)$$
$$[X_0, X_2] = -R^2 kt\partial_x(x\partial_z) - R^2 z\partial_x(x\partial_t)$$
$$[X_0, X_2] = -R^2(z\partial_t + kt\partial_z) = -X_5$$
$$C_{02}^0 = -R^2 z, \;\; C_{02}^3 = -R^2 kt, \;\;\; C_{02}^1 = C_{02}^2 = 0$$

$$[X_0, X_3] = C_{03}^0 e_t + C_{03}^1 e_x + C_{03}^2 e_y + C_{03}^3 e_z$$
$$[X_0, X_3] = R^2(x\partial_t + kt\partial_x)(x\partial_y - y\partial_x) - R^2(x\partial_y - y\partial_x)(x\partial_t + kt\partial_x)$$
$$[X_0, X_3] = R^2 kt\partial_x(x\partial_y) + y\partial_x(x\partial_t)$$
$$[X_0, X_3] = R^2(kt\partial_y + y\partial_t) = X_4$$
$$C_{03}^0 = -R^2 z, \;\; C_{03}^3 = -R^2 kt, \;\;\; C_{03}^1 e_x = C_{03}^2 = 0$$

$$[X_0, X_4] = C_{04}^0 e_t + C_{04}^1 e_x + C_{04}^2 e_y + C_{04}^3 e_z$$
$$[X_0, X_4] = R^4(x\partial_t + kt\partial_x)(y\partial_t + kt\partial_y) - R^4(y\partial_t + kt\partial_y)(x\partial_t + kt\partial_x)$$
$$[X_0, X_4] = R^4 x\partial_t(kt\partial_y) - R^4 y\partial_t(kt\partial_x)$$
$$[X_0, X_4] = R^4(x\partial_y - y\partial_x) = R^4 X_3$$
$$C_{04}^0 = C_{04}^3 = 0, \;\;\;\; C_{04}^1 = -R^4 y, \; C_{04}^2 = R^4 x$$

$$[X_0, X_5] = C_{05}^0 e_t + C_{05}^1 e_x + C_{05}^2 e_y + C_{05}^3 e_z$$
$$[X_0, X_5] = R^4(x\partial_t + kt\partial_x)(z\partial_t + kt\partial_z) - R^4(z\partial_t + kt\partial_z)(x\partial_t + kt\partial_x)$$
$$[X_0, X_5] = R^4 x\partial_t(kt\partial_z) - R^4 z\partial_t(kt\partial_x)$$
$$[X_0, X_5] = R^4(x\partial_z - z\partial_x) = R^4 X_2$$
$$C_{05}^0 = C_{05}^2 = 0, \;\;\;\; C_{05}^1 = -R^4 z, \; C_{05}^3 = R^4 x$$

Veja que a álgebra de Lie desse espaço, corresponde a rotações no espaço-tempo que preservam a forma quadrática. A partir de $X_0$ já geramos $X_2$, $X_3$, $X_4$ e $X_5$. Vamos calcular, os comutadores fixando $X_1, X_2, ..., X_4$.

$$[X_1, X_2] = C_{12}^0 e_t + C_{12}^1 e_x + C_{12}^2 e_y + C_{12}^3 e_z$$
$$[X_1, X_2] = (y\partial_z - z\partial_y)(z\partial_x - x\partial_z) - (z\partial_x - x\partial_z)(y\partial_z - z\partial_y)$$
$$[X_1, X_2] = y\partial_z(z\partial_x) - x\partial_z(z\partial_y)$$
$$[X_1, X_2] = y\partial_x - x\partial_y = -X_3$$
$$[X_1, X_2] = C_{12}^0 = C_{12}^3 e_x = 0, \quad C_{12}^1 = y, \quad C_{12}^2 = -x$$

$$[X_1, X_3] = C_{13}^0 e_t + C_{13}^1 e_x + C_{13}^2 e_y + C_{13}^3 e_z$$
$$[X_1, X_3] = (y\partial_z - z\partial_y)(x\partial_y - y\partial_x) - (x\partial_y - y\partial_x)(y\partial_z - z\partial_y)$$
$$[X_1, X_3] = z\partial_y(y\partial_x) - x\partial_y(y\partial_z)$$
$$[X_1, X_3] = z\partial_x - x\partial_z = X_2$$
$$C_{13}^1 = z, \quad C_{13}^3 = -x, \quad C_{13}^0 = C_{13}^2 = 0$$

$$[X_1, X_4] = C_{14}^0 e_t + C_{14}^1 e_x + C_{14}^2 e_y + C_{14}^3 e_z$$
$$[X_1, X_4] = R^2(y\partial_z - z\partial_y)(y\partial_t + kt\partial_y) - R^2(y\partial_t + kt\partial_y)(y\partial_z - z\partial_y)$$
$$[X_1, X_4] = -R^2 z\partial_y(y\partial_t) - kt\partial_y(y\partial_z)$$
$$[X_1, X_4] = -R^2(z\partial_t + ct\partial_z) = -X_5$$
$$C_{14}^1 = C_{14}^2 = 0, \quad C_{04}^0 = -R^2 z, \; C_{04}^3 = -R^2 kt$$

$$[X_1, X_5] = C_{15}^0 e_t + C_{15}^1 e_x + C_{15}^2 e_y + C_{15}^3 e_z$$
$$[X_1, X_5] = R^2(y\partial_z - z\partial_y)(z\partial_t + kt\partial_z) - R^2(z\partial_t + kt\partial_z)(y\partial_z - z\partial_y)$$
$$[X_1, X_5] = R^2 y\partial_z(z\partial_t) + R^2 kt\partial_z(z\partial_x)$$
$$[X_1, X_5] = R^2(y\partial_t + kt\partial_x) = X_4$$
$$C_{15}^2 = C_{15}^3 = 0, \quad C_{15}^0 = R^2 y, \; C_{15}^3 = R^2 kt$$

$$[X_2, X_3] = C_{23}^0 e_t + C_{23}^1 e_x + C_{23}^2 e_y + C_{23}^3 e_z$$
$$[X_2, X_3] = (z\partial_x - x\partial_z)(x\partial_y - y\partial_x) - (x\partial_y - y\partial_x)(z\partial_x - x\partial_z)$$
$$[X_2, X_3] = z\partial_x(x\partial_y) - y\partial_x(x\partial_z)$$
$$[X_2, X_3] = z\partial_x - y\partial_z = -X_1$$
$$C_{23}^1 = z, \quad C_{23}^3 = y, \quad C_{23}^0 = C_{23}^2 = 0$$

$$[X_2, X_4] = C_{24}^0 e_t + C_{24}^1 e_x + C_{24}^2 e_y + C_{24}^3 e_z$$
$$[X_2, X_4] = R^2(z\partial_x - x\partial_z)(y\partial_t + kt\partial_y) - R^2(y\partial_t + kt\partial_y)(z\partial_x - x\partial_z)$$
$$C_{24}^0 = C_{24}^1 = C_{24}^2 = C_{24}^3 = 0$$

$$[X_2, X_5] = C_{25}^0 e_t + C_{25}^1 e_x + C_{25}^2 e_y + C_{25}^3 e_z$$
$$[X_2, X_5] = R^2 (z\partial_x - x\partial_z)(z\partial_t + kt\partial_z) - R^2 (z\partial_t + kt\partial_z)(z\partial_x - x\partial_z)$$
$$[X_2, X_5] = -R^2 x\partial_z (z\partial_t) - R^2 kt\partial_z (z\partial_x)$$
$$[X_2, X_5] = -R^2 (x\partial_t + kt\partial_x) = -X_0$$
$$C_{25}^2 = C_{25}^3 = 0, \quad C_{25}^0 = -R^2 x, \; C_{25}^1 = -R^2 kt$$

$$[X_3, X_4] = C_{34}^0 e_t + C_{34}^1 e_x + C_{34}^2 e_y + C_{34}^3 e_z$$
$$[X_3, X_4] = R^2 (x\partial_y - y\partial_x)(y\partial_t + kt\partial_y) - R^2 (y\partial_t + kt\partial_y)(x\partial_y - y\partial_x)$$
$$[X_3, X_4] = R^2 x\partial_y (y\partial_t) + R^2 kt\partial_y (y\partial_x)$$
$$[X_3, X_4] = R^2 (x\partial_t + kt\partial_x) = X_0$$
$$C_{34}^2 = C_{34}^3 = 0, \quad C_{34}^0 = R^2 x, \; C_{34}^1 = R^2 kt$$

$$[X_3, X_5] = C_{35}^0 e_t + C_{35}^1 e_x + C_{35}^2 e_y + C_{35}^3 e_z$$
$$[X_3, X_5] = R^2 (x\partial_y - y\partial_x)(z\partial_t + kt\partial_z) - R^2 (z\partial_t + kt\partial_z)(x\partial_y - y\partial_x)$$
$$C_{35}^0 = C_{35}^1 = C_{35}^2 = C_{35}^3 = 0$$

$$[X_4, X_5] = C_{45}^0 e_t + C_{45}^1 e_x + C_{45}^2 e_y + C_{45}^3 e_z$$
$$[X_4, X_5] = R^2 (y\partial_t + kt\partial_y)(z\partial_t + kt\partial_z) - R^2 (z\partial_t + kt\partial_z)(y\partial_t + kt\partial_y)$$
$$[X_4, X_5] = R^2 y\partial_t (kt\partial_z) - R^2 z\partial_t (ct\partial_y)$$
$$[X_4, X_5] = R^2 (y\partial_z - z\partial_x) = R^2 X_1$$
$$C_{45}^0 = C_{45}^2 = 0, \quad C_{45}^1 = -R^2 z, \; C_{45}^3 = R^2 y$$

Portanto, no espaço-tempo Lorentziano e Euclidiano há dois tipos de rotação (espaciais e *boosts*), enquanto no espaço Galileamo só existe uma forma de rotação. Além disso, no espaço-tempo existem três rotações que geram valores nulos.

$$[X_3, X_4] = [X_5, X_2] = X_0 \qquad [X_2, X_1] = X_3$$
$$[X_3, X_2] = X_1 \qquad [X_0, X_4] = R^4 X_3$$
$$[X_4, X_5] = R^2 X_1 \qquad [X_1, X_5] = [X_0, X_3] = X_4$$
$$[X_1, X_3] = X_2 \qquad [X_2, X_0] = [X_4, X_1] = X_5$$
$$[X_0, X_5] = R^4 X_2 \qquad [X_0, X_1] = [X_2, X_4] = [X_3, X_5] = 0$$

Esses permutadores compõe um tensor antissimétrico com 36 componentes, sendo que apenas 12 destas componentes não são nulas, sendo que apenas seis são independentes, que correspondem aos seis geradores do grupo de Lorentz.

**Isomorfismo com o Grupo PSL (2,C)**

Existe uma transformação especial, compatível com o Princípio da Relatividade, definida no corpo dos números complexos, denominada de Transformação de Möbius. Obtemos essa transformação por meio de isomorfismo de grupos de Lie. Observe que o grupo de Poincaré é um grupo do tipo *SO* e, como o grupo $SL(2^{R^4}, \mathbf{R})$ define um mapa de spinores sobre *SO* então o grupo de Lorentz é isomórfico ao grupo de Möbius $PSL(2^{R^4}, \mathbf{R})$. Vamos definir a ação do mapa sobre o espaço-tempo por meio da aplicação:

$$X \mapsto QX\bar{Q}$$
$$X^{\dagger} = \bar{X}^{T} = X$$

onde *X* é uma matriz hermitiana e *Q* uma matriz de determinante unitário, definidas por:

$$X = \begin{pmatrix} Rkt + z & x + y \\ x - y & Rkt - z \end{pmatrix}$$

$$Q = \begin{pmatrix} \alpha & \beta \\ \chi & \delta \end{pmatrix}$$

$$\alpha\delta - \beta\chi = 1$$

As condições impostas sobre *X* e *Q* fazem com que o mapa preserve o determinante:

$$\det X \mapsto \det\left(QX\bar{Q}\right)$$
$$\det X \mapsto \left(\det Q\right)\left(\det X\right)\left(\det \bar{Q}\right)$$
$$\det X \mapsto \det X$$

Essa transformação tem a mesma estrutura da transformação conforme de Möbius de uma superfície de Riemann $\mathbf{R}^2$ e o plano hipercomplexo estendido:

$$w \mapsto \frac{\alpha w + \beta}{\chi w + \delta}$$

$$\alpha\delta - \beta\chi = 1$$

O determinante da matriz *X* deve ser preservado, pois ele define o invariante da forma quadrática fundamental do espaço-tempo:

$$\det X = \left(Rkt + z\right)\left(Rkt - z\right) - \left(x + y\right)\left(x - y\right)$$
$$\det X = R^2k^2t^2 - x^2 - y^2 - z^2$$

O que prova que a aplicação é um mapa entre as transformações de Poincaré e as transformações de Möbius.

**4-Vetores na Variedade Espaço-Tempo**

Como mostramos o espaço-tempo plano é definido pela sua característica anelar $R^2$. Em particular, nossas definições se tornam singulares se $R^2$ for um número nilpotente de segunda ordem. Para tornarmos as nossas definições o mais geral possível e evitar as singularidades, adotaremos a convenção onde a componente temporal assume o papel de quarta coordenada[10] e vamos definir a métrica do espaço-tempo pela seguinte regra:

$$\eta : \left\{ \eta_{3+R^4 \times 3+R^4},\ \eta_{ij} = \eta_{ji} \mid \eta_{\mu\nu} = \delta_{\mu\nu},\ \eta_{4j} = -R^2 \delta_{4j} \right\}$$

Portanto, se $R$ for um número nilpotente de ordem 2, a matriz associada a métrica se torna uma matriz 3x3 que coincide com delta de Kroenecker e a Identidade. Agora podemos estudar a estrutura geral para a construção de 4-vetores de grandezas físicas para podermos estudar como se transformam algumas grandezas mecânicas, eletromagnéticas e ópticas, em variedades do espaço-tempo. Nossos 4-vetores são estruturas algébricas que apresentam quatro componentes:

$$J_i = \left( J_1, J_2, J_3, J_4 \right)$$

Todas as componentes devem ter a mesma dimensão. A componente zero, também chamada de componente *temporal,* é sempre um escalar e, em geral, vem associada com a velocidade da luz no vácuo, pois o eixo $x_0$ é o eixo espacial $kt$. As demais componentes, conhecidas como espaciais, são as componentes de um vetor no espaço. Nestas condições, podemos escrever:

$$J_i = \left( \vec{J}, J_4 \right)$$

Existe uma importante relação entre os vetores covariantes e contravariantes envolvendo o tensor métrico do espaço:

$$J_i = \eta_{ij} J^j$$

Sendo a métrica orientada como ($-R^2$,1,1,1), então as componentes do 4-vetor covariante se relacionam com as contravariantes por meio da lei:

$$\begin{matrix} J_1 = \eta_{11} J^1 \\ J_2 = \eta_{22} J^2 \\ J_3 = \eta_{33} J^3 \\ J_4 = \eta_{44} J^4 \end{matrix} \quad \Rightarrow \quad \begin{matrix} J_1 = J^1 \\ J_2 = J^2 \\ J_3 = J^3 \\ J_4 = -R^2 J^4 \end{matrix} \quad \text{ou} \quad \begin{matrix} J^1 = J_1 \\ J^2 = J_2 \\ J^3 = J_3 \\ J^4 = -R^2 J_4 \end{matrix}$$

[10] Assumiremos o tempo como a quarta coordenada por uma finalidade puramente didática, visto que a convenção não altera os resultados.

Por meio dos 4vetores podemos construir invariantes relativísticos, forma quadráticas, que relacionam as componentes vetoriais e escalares:

$$J_i J^i = J_1 J^1 + J_2 J^2 + J_3 J^3 + J_4 J^4$$

Substituindo os valores do 4vetor contravariante, obtemos:

$$J_i J^i = J_1 J_1 + J_2 J_2 + J_3 J_3 - R^2 J_4 J_4$$
$$J^2 = J_1^2 + J_2^2 + J_3^2 - R^2 J_4^2$$

Que pela definição de norma pode ser escrito da seguinte forma:

$$J^2 = \left\|\vec{J}\right\|^2 - R^2 J_4^2$$
$$J^2 = \vec{J} \cdot \vec{J} - R^2 J_4^2$$

O escalar $J$ é um invariante, isto é, não depende da escolha do referencial. Escolheremos $J$ como sendo a medida efetuada no referencial próprio, quando o ângulo de rotação é zero.

$$J_i'' = \left(J_1 P_+^R(0) - J_4 P_-^R(0), J_2, J_3, J_4 P_+^R(0) - R^2 J_1 P_-^R(0)\right)$$
$$J_i^o = \left(J_1^o, J_2^o, J_3^o, J_4^o\right) \quad (\textit{referencial próprio do corpo})$$

Portanto nosso invariante pode ser expresso pelas relações:

$$\left\|J^o\right\|^2 = \left\|\vec{J}\right\|^2 - R^2 J_4^2$$
$$\vec{J}^o \cdot \vec{J}^o - R^2 J_4^{o2} = \left\|\vec{J}\right\|^2 - R^2 J_4^2$$

Para 4-vetores não-nilpotentes, existe sempre um referencial onde as componentes espaciais são todas nulas. Nessas condições, podemos escrever a relação:

$$R^2 J_4^{o2} = \left\|\vec{J}\right\|^2 - R^2 J_4^2$$

Uma consequência da covariância é que o módulo de um tensor não depende da escolha dos referenciais. Assim, podemos definir a norma de um vetor a partir da característica anelar:

$$\left\|\vec{J}\right\|^2 = R^2 J_4^{o2} + R^2 J_4^2$$
$$\left\|\vec{J}\right\|^2 = R^2\left(J_4^{o2} + J_4^2\right)$$
$$\left\|\vec{J}\right\| = R\left(J_4^{o2} + J_4^2\right)^{1/2}$$

Essa relação permite estabelecer um isomorfismo entre o espaço da norma dos 4-vetores e o espaço das características anelares.

Assim, a covariância de Lorentz para os 4-vetores será:

**COVARIANTE**

$$J_i = (J_1, J_2, J_3, J_4)$$
$$J'_i = \left(J_1 P_+^R(a) - J_4 P_-^R(a), J_2, J_3, J_0 P_+^R(a) - R^2 J_1 P_-^R(a)\right)$$
$$J'_i = \left(\Gamma[J_1 - \beta J_4], J_2, J_3, \Gamma\left[J_4 - R^2 \beta J_1\right]\right)$$

**CONTRAVARIANTE**

$$J^i = \left(J^1, J^2, J^3, J^4\right)$$
$$J^i = \left(J'^1 P_+^R(a) + J'^4 P_-^R(a), J'^2, J'^3, J'^4 P_+^R(a) + R^2 J'^1 P_-^R(a)\right)$$
$$J^i = \left(\Gamma\left[J'^1 + \beta J'^4\right], J'^2, J'^3, \Gamma\left[J'^4 + R^2 \beta J'^1\right]\right)$$

Registre que os p-vetores covariantes são chamados de p-formas ou p-covetores, enquanto os q-vetores contravariantes são chamados de q-vetores.

**C-Grupo de Poincaré**

Antes de prosseguirmos em nosso estudo sobre Teoria da Relatividade Especial, vamos discutir a representação dos Cliffor (C-) Grupos de Poincaré e Lorentz, isto é, as generalizações dos grupos realizadas por meio das funções de Poincaré. Esse capítulo tem como principal fonte o livro Matemática para Físicos com Aplicações (BARCELOS NETO, 2010, p. 157-168). Também iremos abordar o conceito de representação *spinorial*.

Tomemos dois sistemas inerciais de referencial no espaço-tempo de Poincaré-Minkowski. Dado intervalo de universo $ds^2$,

$$ds^2 = \eta_{ij} dx^i dx^j$$

A métrica do espaço-tempo de Poincaré-Minkowski se transforma como um tensor covariante de segunda ordem:

$$\eta_{nm} = \eta_{ij} \frac{\partial x'^i}{\partial x^m} \frac{\partial x'^j}{\partial x^n}$$

Diferenciando a equação em relação a coordenada $x^p$:

$$\eta_{ij} \frac{\partial^2 x'^i}{\partial x^p \partial x^m} \frac{\partial x'^j}{\partial x^n} + \eta_{ij} \frac{\partial x'^i}{\partial x^m} \frac{\partial^2 x'^j}{\partial x^p \partial x^n} = 0$$

O teorema de Schwarz permite permutar as derivadas, assim podemos trocar a ordem livremente, permutando no segundo termo a derivada em $x^m$ com $x^n$ e $x'^i$ e $x'^j$,

$$\eta_{ij}\frac{\partial^2 x'^i}{\partial x^p \partial x^m}\frac{\partial x'^j}{\partial x^n}+\eta_{ij}\frac{\partial^2 x'^i}{\partial x^p \partial x^m}\frac{\partial x'^j}{\partial x^n}=0$$

$$\eta_{ij}\frac{\partial^2 x'^i}{\partial x^p \partial x^m}\frac{\partial x'^j}{\partial x^n}=0$$

Tanto o tensor métrico quanto a matriz de transformação (jacobiano) possuem determinante não-singular, portanto, essa igualdade só é válida se:

$$\frac{\partial^2 x'^i}{\partial x^p \partial x^m}=0$$

Integrando a função em relação a $x^p$ e $x^m$:

$$x'^i=\alpha^i+x^p\left[\Lambda^R\right]^i_p$$

Onde as matrizes são com coeficientes constantes. Qualquer transformação que satisfaça essa relação e forme um grupo é chamado de Grupo de Poincaré ou Grupo Não Homogêneo de Lorentz. Se o coeficiente □$^i$ for nulo, temos o grupo homogêneo de Lorentz. Substituindo essa relação na transformação do tensor métrico:

$$\eta_{nm}=\eta_{ij}\left(\frac{\partial x^p}{\partial x^m}\left[\Lambda^R\right]^i_p\frac{\partial x^p}{\partial x^n}\left[\Lambda^R\right]^j_p\right)$$

As derivadas se transformam como o tensor de Kroenecker:

$$\eta_{nm}=\eta_{ij}\left(\delta^p_m\left[\Lambda^R\right]^i_p\delta^p_n\left[\Lambda^R\right]^j_p\right)$$

$$\eta_{nm}=\eta_{ij}\left(\left[\Lambda^R\right]^i_m\left[\Lambda^R\right]^j_n\right)$$

Em notação absoluta, essa é equação dos automorfismos internos:

$$\eta=\left(\Lambda^R\right)^\dagger\eta\left(\Lambda^R\right)$$

Tomando o determinante:

$$\det\eta=\det\left[\left(\Lambda^R\right)^\dagger\eta\left(\Lambda^R\right)\right]$$

$$\det\eta=\det\left(\Lambda^R\right)^\dagger\det\eta\det\left(\Lambda^R\right)$$

$$\left[\det\left(\Lambda^R\right)\right]^2=1$$

Assim teremos duas soluções possíveis:

$$\left|\det\left(\Lambda^R\right)\right| = 1 \quad \rightarrow \quad \begin{cases} \det\left(\Lambda^R\right) = +1 \\ \det\left(\Lambda^R\right) = -1 \end{cases}$$

Expandindo a transformação da métrica:

$$\eta_{nm} = \eta_{ij}\left(\left[\Lambda^R\right]_m^i \left[\Lambda^R\right]_n^j\right)$$

$$\eta_{nm} = \eta_{00}\left(\left[\Lambda^R\right]_m^0 \left[\Lambda^R\right]_n^0\right) + \eta_{\mu\nu}\left(\left[\Lambda^R\right]_m^\mu \left[\Lambda^R\right]_n^\nu\right)$$

$$\eta_{nm} = -R^2\left(\left[\Lambda^R\right]_m^0 \left[\Lambda^R\right]_n^0\right) + \left(\left[\Lambda^R\right]_m^\mu \left[\Lambda^R\right]_n^\nu\right)$$

Para a coordenada temporal, temos a seguinte transformação:

$$\eta_{00} = -R^2\left(\left[\Lambda^R\right]_0^0 \left[\Lambda^R\right]_0^0\right) + \left(\left[\Lambda^R\right]_0^\mu \left[\Lambda^R\right]_0^\nu\right)$$

$$-R^2 = -R^2\left(\left[\Lambda^R\right]_0^0\right)^2 + \left(\left[\Lambda^R\right]_0^\mu \left[\Lambda^R\right]_0^\nu\right)$$

$$R^2\left(\left[\Lambda^R\right]_0^0\right)^2 - R^2 = \left(\left[\Lambda^R\right]_0^\mu \left[\Lambda^R\right]_0^\nu\right)$$

$$R^2\left\{\left(\left[\Lambda^R\right]_0^0\right)^2 - 1\right\} = \left(\left[\Lambda^R\right]_0^\mu \left[\Lambda^R\right]_0^\nu\right)$$

Aqui há uma relação que nos permite definir a característica anelar da variedade:

$$R^2 = \frac{\left(\left[\Lambda^R\right]_0^\mu \left[\Lambda^R\right]_0^\nu\right)}{\left\{\left(\left[\Lambda^R\right]_0^0\right)^2 - 1\right\}}$$

Se $R$ for um número nilpotente de ordem dois, resulta que:

$$\left(\left[\Lambda^\varepsilon\right]_0^\mu \left[\Lambda^\varepsilon\right]_0^\nu\right) = 0$$

Para os demais números hipercomplexos, teremos:

$$\left(\left[\Lambda^R\right]_0^0\right)^2 = \frac{\left(\left[\Lambda^R\right]_0^\mu \left[\Lambda^R\right]_0^\nu\right)}{R^2} + 1$$

Como o menor valor do produto das matrizes de Lorentz é zero, podemos majorar a expressão acima e concluir que:

$$\left(\left[\Lambda^R\right]_0^0\right)^2 \geq 1$$

Portanto, a matriz temporal de Lorentz admite duas soluções:

$$\left|\left[\Lambda^R\right]_0^0\right| \geq 1 \quad \rightarrow \quad \begin{cases} \left[\Lambda^R\right]_0^0 \geq 1 \\ \left[\Lambda^R\right]_0^0 < 1 \end{cases}$$

Denotando por + e – os valores do determinante e por $\uparrow$ e $\downarrow$ os valores da matriz temporal de Lorentz, teremos quatro conjuntos possíveis:

$$\left\{P_{+\uparrow}^R, P_{-\uparrow}^R, P_{+\downarrow}^R, P_{-\downarrow}^R\right\}$$

Destes conjuntos, podemos formar quatro grupos:

**ORTOCRONO PRÓPRIO**

$$P_{+\uparrow}^R$$

**PRÓPRIO**

$$P_{+\uparrow}^R \cup P_{+\downarrow}^R = P_{+}^R$$

**ORTOCRONO**

$$P_{+\uparrow}^R \cup P_{-\uparrow}^R = P_{\uparrow}^R$$

**GRUPO ANTICRONO**

$$P_{+\uparrow}^R \cup P_{-\downarrow}^R = P_{+}^R$$

**GRUPO ORTOCRONO PRÓPRIO DE LORENTZ**

$$SO(R)$$

$$\left\{\Lambda^R_{\left(3+|R|^2\right)\times\left(3+|R|^2\right)} \mid \left(\Lambda^R\right)_j^i \in \mathbb{R},\ \left(\Lambda^R\right)^\dagger \eta\left(\Lambda^R\right) = \eta,\ \det\left(\Lambda^R\right) = 1,\ \left|\left(\Lambda^R\right)_0^0\right| \geq 1\right\}$$

**C-Transformações Ortocrônicas de Lorentz**

Até o presente momento, trabalhamos apenas com as transformações de Lorentz considerando que o movimento entre os referenciais inerciais fossem longitudinais. Agora, devemos generalizar essas transformações para o movimento inercial arbitrário. Definimos o vetor posição no espaço-tempo de Galileu pela seguinte equação paramétrica:

$$\vec{r}_o = \vec{r} - \vec{v}t$$
$$\vec{r}_o' = \vec{r}' - \vec{v}t$$

Vamos decompor o vetor posição em função de suas componentes longitudinal e transversal a velocidade da partícula em dois referenciais inerciais:

$$\vec{r} = r_{\parallel}\frac{\vec{v}}{v} + \vec{r}_{\perp}$$

$$\vec{r}' = r_{\parallel}'\frac{\vec{v}}{v} + \vec{r}_{\perp}'$$

como a componente longitudinal tem o mesmo sentido da velocidade, o versor da posição longitudinal pode ser definido em função da velocidade. Multiplicando a primeira equação por **v**:

$$\vec{v}\cdot\vec{r} = r_{\parallel}\frac{\vec{v}\cdot\vec{v}}{v} + \vec{v}\cdot\vec{r}_{\perp}$$
$$\vec{v}\cdot\vec{r} = r_{\parallel}v$$

Isolando a componente longitudinal do vetor de posição,

$$r_{\parallel} = \frac{\vec{v}\cdot\vec{r}}{v}$$

Substituindo esse valor na primeira equação,

$$\vec{r} = \frac{(\vec{v}\cdot\vec{r})}{v}\frac{\vec{v}}{\|\vec{v}\|} + \vec{r}_{\perp}$$
$$\vec{r} = \frac{(\vec{v}\cdot\vec{r})}{v^2}\vec{v} + \vec{r}_{\perp}$$

Isolando a componente transversal,

$$\vec{r}_{\perp} = \vec{r} - \frac{(\vec{v}\cdot\vec{r})}{v^2}\vec{v}$$

Com base nas transformações de Lorentz, descobrimos que as componentes transversais se mantém invariantes (LOGUNOV, 2005). Isso permite que escrevamos as seguintes transformações:

$$r'_{\parallel} = \Gamma\left(r_{\parallel} - vt\right)$$
$$r'_{\perp} = r_{\perp}$$
$$t' = \Gamma\left(t - R^2\frac{v}{k^2}r_{\parallel}\right)$$

Substituindo os valores da componente longitudinal e transversal:

$$\frac{(\vec{v}\cdot\vec{r}'_o)}{v} = \Gamma\left(\frac{(\vec{v}\cdot\vec{r}_o)}{v} - vt\right)$$
$$\vec{r}' - \frac{(\vec{v}\cdot\vec{r}'_o)}{v^2}\vec{v} = \vec{r} - \frac{(\vec{v}\cdot\vec{r}_o)}{v^2}\vec{v}$$

Para obtermos as transformações operaremos a segunda equação:

$$\vec{r}' - \left[\frac{(\vec{v}\cdot\vec{r}'_o)}{v}\right]\frac{\vec{v}}{v} = \vec{r} - \frac{(\vec{v}\cdot\vec{r}_o)}{v^2}\vec{v}$$

Substituindo a transformação longitudinal no termo em colchetes:

$$\vec{r}' - \Gamma\left(\frac{(\vec{v}\cdot\vec{r})}{v} - vt\right)\frac{\vec{v}}{v} = \vec{r} - \frac{(\vec{v}\cdot\vec{r})}{v^2}\vec{v}$$

$$\vec{r}' - \Gamma\frac{(\vec{v}\cdot\vec{r})}{v^2}\vec{v} + \Gamma t\vec{v} = \vec{r} - \frac{(\vec{v}\cdot\vec{r})}{v^2}\vec{v}$$

$$\vec{r}' = \vec{r} + \Gamma\frac{(\vec{v}\cdot\vec{r})}{v^2}\vec{v} - \frac{(\vec{v}\cdot\vec{r})}{v^2}\vec{v} + \Gamma t\vec{v}$$

Evidenciando, obtemos a transformação geral de Lorentz da posição e, portanto, as transformações gerais de Lorentz para qualquer variedade espaço-temporal plana são:

$$\vec{r}' = \vec{r} + (\Gamma - 1)\frac{(\vec{v}\cdot\vec{r})}{v^2}\vec{v} + \Gamma t\vec{v}$$

$$t' = \Gamma\left(t - \frac{R^2}{k^2}(\vec{v}\cdot\vec{r})\right)$$

### 2.2. Matrizes Ortocrônicas do S-Grupo de Poincaré

Por meio da Teoria de Grupos estabelecemos que o grupo de Poincaré é um grupo ortocrônico próprio do tipo

$$SO(R) = \left\{\Lambda^R_{(3+R^4)\times(3+R^4)} \mid (\Lambda^R)^i_j \in \mathbb{R},\right.$$
$$\left.(\Lambda^R)^\dagger \eta (\Lambda^R) = \eta,\ \det(\Lambda^R) = 1,\ \left|(\Lambda^R)^0_0\right| \geq 1\right\}$$

que satisfaz a seguinte equação afim:

$$x'^i = \alpha^i + x^p (\Lambda^R)^i_p$$

Agora iremos estudar os subgrupos de Poincaré, as matrizes de transformação e *boost.* Detalhes sobre este capítulo pode ser visto em Barcelos Neto (2010, p. 161-168). Podemos representar a matriz de Lorentz da seguinte forma:

$$(\Lambda^R)^i_p = \begin{pmatrix} P^R & 0 \\ 0 & R_\theta \end{pmatrix}$$

Onde *L* é a matriz de rotações no espaço-tempo e *R* são as matrizes de rotação de SO(2).

$$P^R = \begin{pmatrix} P^R_+ & -P^R_- \\ R^2 P^R_- & P^R_+ \end{pmatrix} \qquad R_\theta = \begin{pmatrix} \cos\theta & -\sin\theta \\ \sin\theta & \cos\theta \end{pmatrix}$$

Se o sistema não apresentar translações (que correspondem a rotações no espaço hipercomplexo), a matriz $P^R$ é a matriz identidade:

$$P^R = \begin{pmatrix} 1 & 0 \\ 0 & 1 \end{pmatrix}$$

Nesse caso, o grupo de Poincaré corresponde ao grupo estacionário de Galileo:

$$\left(\Lambda^R\right)^i_p = \begin{pmatrix} I & 0 \\ 0 & R_\theta \end{pmatrix}$$

Se o sistema não apresentar rotações, *R* é a matriz identidade:

$$R_\theta = \begin{pmatrix} 1 & 0 \\ 0 & 1 \end{pmatrix}$$

E teremos a matriz especial de *boosts* de Lorentz:

$$\left(\Lambda^R\right)^i_p = \begin{pmatrix} P^R & 0 \\ 0 & I \end{pmatrix}$$

Podemos ainda obter uma matriz mais geral de *boosts,* que chamaremos de matriz de Poincaré e denotaremos pela letra $\left(\Upsilon^R\right)^i_p$.

$$\left(\Upsilon^R\right)^i_p = \begin{pmatrix} \Upsilon^0_0 & \Upsilon^0_1 & \Upsilon^0_2 & \Upsilon^0_3 \\ \Upsilon^1_0 & \Upsilon^1_1 & \Upsilon^1_2 & \Upsilon^1_3 \\ \Upsilon^2_0 & \Upsilon^2_1 & \Upsilon^2_2 & \Upsilon^2_3 \\ \Upsilon^3_0 & \Upsilon^3_1 & \Upsilon^3_2 & \Upsilon^3_3 \end{pmatrix}$$

A matriz de transformação de Poincaré deve obedecer a transformação do grupo:

$$x'^i = \alpha^i + x^p \left(\Upsilon^R\right)^i_p \qquad \Rightarrow \qquad \begin{pmatrix} kt' \\ x' \\ y' \\ z' \end{pmatrix} = \begin{pmatrix} \alpha_0 \\ \alpha_1 \\ \alpha_2 \\ \alpha_3 \end{pmatrix} + \begin{pmatrix} \Upsilon^0_0 & \Upsilon^0_1 & \Upsilon^0_2 & \Upsilon^0_3 \\ \Upsilon^1_0 & \Upsilon^1_1 & \Upsilon^1_2 & \Upsilon^1_3 \\ \Upsilon^2_0 & \Upsilon^2_1 & \Upsilon^2_2 & \Upsilon^2_3 \\ \Upsilon^3_0 & \Upsilon^3_1 & \Upsilon^3_2 & \Upsilon^3_3 \end{pmatrix} \begin{pmatrix} kt \\ x \\ y \\ z \end{pmatrix}$$

Efetuando o produto e a soma das matrizes,

$$\begin{pmatrix} kt' \\ x' \\ y' \\ z' \end{pmatrix} = \begin{pmatrix} \alpha_0 + kt\Upsilon^0_0 + x\Upsilon^0_1 + y\Upsilon^0_2 + z\Upsilon^0_3 \\ \alpha_1 + kt\Upsilon^1_0 + x\Upsilon^1_1 + y\Upsilon^1_2 + z\Upsilon^1_3 \\ \alpha_2 + kt\Upsilon^2_0 + x\Upsilon^2_1 + y\Upsilon^2_2 + z\Upsilon^2_3 \\ \alpha_3 + kt\Upsilon^3_0 + x\Upsilon^3_1 + y\Upsilon^3_2 + z\Upsilon^3_3 \end{pmatrix}$$

Tomemos as transformações de coordenadas do espaço-tempo:

$$kt' = \alpha_0 + \Gamma\left(kt - \frac{R^2}{k}(\vec{v}\cdot\vec{r})\right) \qquad \vec{r}' = \vec{\alpha} + \vec{r} + (\Gamma - 1)\frac{(\vec{v}\cdot\vec{r})}{v^2}\vec{v} - \Gamma t\vec{v}$$

Vamos expandir as transformações, começando pela temporal:

$$kt' = \alpha_0 + \Gamma kt - \Gamma \frac{R^2}{k} x v_x - \Gamma \frac{R^2}{k} y v_y - \Gamma \frac{R^2}{k} z v_z$$

Definindo a razão *v/k* como fator beta, nossa equação se torna:

$$kt' = \alpha_0 + \Gamma kt - \Gamma R^2 x \beta_x - \Gamma R^2 y \beta_y - \Gamma R^2 z \beta_z$$

Portanto os coeficientes da primeira linha devem ser:

$$\alpha_0' = \alpha_0, \qquad \left(\Upsilon^R\right)_0^0 = \Gamma, \qquad \left(\Upsilon^R\right)_\mu^0 = -\Gamma R^2 \beta_\mu$$

Agora vamos abrir as equações espaciais:

$$\vec{r}' = \vec{\alpha} + \vec{r} + (\Gamma - 1)\frac{(\vec{v}\cdot\vec{r})}{v^2}\vec{v} - \Gamma t \vec{v}$$

$$x_\mu' = \alpha_\mu + x_\mu + (\Gamma-1)\frac{v_x v_\mu}{v^2} x + (\Gamma-1)\frac{v_y v_\mu}{v^2} y + (\Gamma-1)\frac{v_z v_\mu}{v^2} z - \Gamma t v_\mu$$

$$x_\mu' = \alpha_\mu + x_\mu + (\Gamma-1)\frac{k^2 v_x v_\mu}{k^2 v^2} x + (\Gamma-1)\frac{k^2 v_y v_\mu}{k^2 v^2} y + (\Gamma-1)\frac{k^2 v_z v_\mu}{k^2 v^2} z - \Gamma kt \frac{v_\mu}{k}$$

Usando o fator beta de Lorentz, obtemos:

$$x_\mu' = \alpha_\mu + (-\Gamma\beta_\mu) kt + x_\mu + (\Gamma-1)\frac{\beta_x\beta_\mu}{\beta^2} x + (\Gamma-1)\frac{\beta_y\beta_\mu}{\beta^2} y + (\Gamma-1)\frac{\beta_z\beta_\mu}{\beta^2} z$$

Portanto, as componentes espaciais são:

$$x' = \alpha_x + (-\Gamma\beta_x) kt + \left[1 + (\Gamma-1)\frac{\beta_x^2}{\beta^2}\right] x + (\Gamma-1)\frac{\beta_y\beta_x}{\beta^2} y + (\Gamma-1)\frac{\beta_z\beta_x}{\beta^2} z$$

$$y' = \alpha_y + (-\Gamma\beta_y) kt + (\Gamma-1)\frac{\beta_x\beta_y}{\beta^2} x + \left[1 + (\Gamma-1)\frac{\beta_y^2}{\beta^2}\right] y + (\Gamma-1)\frac{\beta_z\beta_y}{\beta^2} z$$

$$z' = \alpha_z + (-\Gamma\beta_z) kt + (\Gamma-1)\frac{\beta_x\beta_z}{\beta^2} x + (\Gamma-1)\frac{\beta_y\beta_z}{\beta^2} y + \left[1 + (\Gamma-1)\frac{\beta_z^2}{\beta^2}\right] z$$

Portanto as componentes da matriz são:

$$\alpha_0' = \alpha_0 \qquad \vec{\alpha}' = \alpha_\mu$$

$$\left(\Upsilon^R\right)_0^0 = \Gamma \qquad \left(\Upsilon^R\right)_\mu^\mu = 1 + (\Gamma-1)\frac{\beta_\mu^2}{\beta^2}$$

$$\left(\Upsilon^R\right)_\mu^0 = \left(\Upsilon^R\right)_0^\mu = -\Gamma R^2 \beta_\mu \qquad \left(\Upsilon^R\right)_\nu^\mu = \left(\Upsilon^R\right)_\mu^\nu = (\Gamma-1)\frac{\beta_\mu\beta_\nu}{\beta^2}$$

Essa matriz é consistente com a definição do grupo de Poincaré, pois ela deve ser, como esperado, hermitiana:

$$\left(\Upsilon^R\right)^{i\,\dagger}_{j} = \left(\bar{\Upsilon}^R\right)^{j}_{i} = \left(\Upsilon^R\right)^{j}_{i}$$

É fácil verificar que essa matriz é gerada pela seguinte regra:

$$\left(\Upsilon^R\right)^{i}_{j} = \begin{cases} \Gamma & \left(\text{se } i = j = 0\right) \\ \delta_{ij} + \left(\Gamma - 1\right)\dfrac{\beta_i \beta_j}{\beta^2} & \left(\text{se } i \text{ ou } j \neq 0\right) \end{cases} \qquad \beta_0 = -\Gamma \frac{R^2\beta^2}{\left(\Gamma - 1\right)}$$

E a matriz de *boosts* de Poincaré será dada por:

$$\left(\Upsilon^R\right)^{i}_{j} = \begin{pmatrix} \Gamma & -\Gamma R^2\beta_x & -\Gamma R^2\beta_y & -\Gamma R^2\beta_z \\ -\Gamma R^2\beta_x & 1+\left(\Gamma-1\right)\dfrac{\beta_x^2}{\beta^2} & \left(\Gamma-1\right)\dfrac{\beta_x\beta_y}{\beta^2} & \left(\Gamma-1\right)\dfrac{\beta_x\beta_z}{\beta^2} \\ -\Gamma R^2\beta_y & \left(\Gamma-1\right)\dfrac{\beta_x\beta_y}{\beta^2} & 1+\left(\Gamma-1\right)\dfrac{\beta_y^2}{\beta^2} & \left(\Gamma-1\right)\dfrac{\beta_y\beta_z}{\beta^2} \\ -\Gamma R^2\beta_z & \left(\Gamma-1\right)\dfrac{\beta_x\beta_z}{\beta^2} & \left(\Gamma-1\right)\dfrac{\beta_y\beta_z}{\beta^2} & 1+\left(\Gamma-1\right)\dfrac{\beta_z^2}{\beta^2} \end{pmatrix}$$

Por fim, vamos provar que a matriz de Poincaré é ortogonal. Como a matriz de Poincaré é um automorfismo interno da variedade:

$$\left(\Upsilon^R\right)^{i}_{m} \eta_{ij} \left(\Upsilon^R\right)^{j}_{n} = \eta_{mn}$$

Multiplicando pelo conjugado da métrica:

$$\left(\Upsilon^R\right)^{i}_{m} \eta_{ij} \left(\Upsilon^R\right)^{j}_{n} \eta^{nk} = \eta_{mn}\eta^{nk}$$

$$\left(\Upsilon^R\right)_{mj} \left(\Upsilon^R\right)^{jk} = \delta^k_m$$

Multiplicando a equação por $\left(\Upsilon^R\right)^{-1}_{mj} \delta^m_k$

$$\delta^m_k I \left(\Upsilon^R\right)^{jk} = I \left(\Upsilon^R\right)^{-1}_{mj}$$

$$\left(\Upsilon^R\right)^{jm} = \left(\Upsilon^R\right)^{-1}_{mj}$$

Como a matriz é hermitiana, então podemos escrever:

$$\left(\bar{\Upsilon}^R\right)^{jm\,\dagger} = \left(\Upsilon^R\right)^{-1}_{mj}$$

que é a condição de ortogonalidade.

**Representação do C-Grupo de Poincaré**

O C-grupo de Poincaré apresenta uma álgebra de Lie e sua matriz é dada por uma exponencial complexa:

$$\Upsilon = e^{-\frac{i}{2}\omega^{ij}L_{ij}}$$

Onde $\omega^{ij}$ são estruturas antissimétricas que correspondem aos seis parâmetros do grupo e as matrizes $L_{ij}$ são os geradores do grupo. Expandindo o exponencial em série de Taylor:

$$\Upsilon = 1 + \frac{i}{2}\omega^{ij}L_{ij} + O$$

Onde $O$ corresponde aos termos de ordem maior ou igual à 2. Como estamos buscando os geradores infinitesimais o grupo, podemos descartar os termos $O$.

$$\Upsilon = 1 + \frac{i}{2}\omega^{ij}L_{ij}$$

As matrizes geradores desse grupo são dados por:

$$\left(L_{ij}\right)^{m}_{n} = i\left(\delta^{m}_{i}g_{jn} + \delta^{m}_{j}g_{in}\right)$$

Inicialmente vamos introduzir as matrizes auxiliares:

$$A_0 = \begin{pmatrix} 1 & 0 \\ 0 & 1 \end{pmatrix}, \quad A_1 = \begin{pmatrix} 0 & -i \\ -i & 0 \end{pmatrix}, \quad A_2 = \begin{pmatrix} -i & 0 \\ 0 & -i \end{pmatrix}$$

$$A_3 = \begin{pmatrix} 0 & -i \\ 0 & 0 \end{pmatrix}, \quad A_4 = \begin{pmatrix} 0 & i \\ -i & 0 \end{pmatrix}, \quad A_5 = \begin{pmatrix} 0 & 0 \\ 0 & -i \end{pmatrix}$$

Por estas matrizes podemos construir as matrizes de Pauling:

$$\sigma_1 \equiv -A_1^2 = \begin{pmatrix} 0 & 1 \\ 1 & 0 \end{pmatrix}, \quad \sigma_2 \equiv -A_4 = \begin{pmatrix} 0 & -i \\ i & 0 \end{pmatrix},$$

Usando a equação dos geradores, obtemos as matrizes que geram o grupo generalizado de Poincaré:

$$L_{01} = \begin{pmatrix} A_1 & 0 \\ 0 & 0 \end{pmatrix}, \quad L_{02} = \begin{pmatrix} 0 & A_2 \\ A_2 & 0 \end{pmatrix}, \quad L_{03} = \begin{pmatrix} 0 & A_3 \\ A_3^T & 0 \end{pmatrix}$$

$$L_{12} = \begin{pmatrix} 0 & A_4 \\ -A_3 & 0 \end{pmatrix}, \quad L_{13} = \begin{pmatrix} 0 & A_5 \\ -A_5 & 0 \end{pmatrix}, \quad L_{23} = \begin{pmatrix} 0 & 0 \\ 0 & A_4 \end{pmatrix}$$

A álgebra de Lie do grupo de Poincaré é dado por:

$$\left[L_{ij}, L_{kl}\right] = i\left(g_{il}L_{jk} + g_{jk}L_{il} - g_{ik}L_{jl} - g_{jl}L_{ik}\right)$$

Vamos construir os vetores de *boosts K* e *rotações S:*

$$K_i = \left(L_{01}, L_{02}, L_{03}\right), \qquad S_i = \left(L_{12}, L_{13}, L_{23}\right)$$

Que satisfazem as leis de comutação:

$$\left[K_i, K_j\right] = -i\varepsilon_{ijk}S_k,$$
$$\left[S_i, K_j\right] = -i\varepsilon_{ijk}K_k,$$
$$\left[S_i, S_j\right] = +i\varepsilon_{ijk}S_k$$

A primeira relação forma o grupo dos *boosts,* porém esse grupo não apresenta uma álgebra de Lie, pois seus elementos não são todos *boosts.* A terceira relação é o grupo de rotações que por só ter elementos de mesma classe, admite uma álgebra de Lie.

**Spinores e Representação Spinoral**

Um spinor é o equivalente algébrico a um vetor do espaço euclidiano em um espaço complexo. Spinores são elementos que se transformam linearmente quando um espaço euclidiano é submetido a uma rotação infinitesimal. Essa associação dos spinores com as rotações fica evidente em seu próprio nome que deriva da palavra *spin.*que se refere ao momento angular das partículas. Definimos o conceito de representação spinorial as $N(N-1)/2$ matrizes $\Gamma_a$ tais que (BARCELOS NETO, 2010, p. 148-149):

$$\left\{\Gamma_a, \Gamma_b\right\} = \Gamma_a\Gamma_b + \Gamma_b\Gamma_a = 2\delta_{ab}$$

onde o operador $\{\Gamma_a, \Gamma_b\}$ é o anticomutador. O gerador do grupo $M_{ab}$, satisfaz uma álgebra de Lie:

$$M_{ab} = -\frac{i}{4}\left[\Gamma_a, \Gamma_b\right]$$
$$\left[M_{ij}, M_{kl}\right] = i\left(\delta_{il}M_{jk} + \delta_{jk}M_{il} - \delta_{ik}M_{jl} - \delta_{jl}M_{ik}\right)$$

Há duas importantes relações envolvendo comutadores e anticomutadores:

$$\left[AB, C\right] = A\left\{B, C\right\} + \left\{A, C\right\}B$$
$$\left[A, BC\right] = \left\{A, B\right\}C - B\left\{A, C\right\}$$

Para o S-Grupo de Poincaré definiremos os seguintes spinores a partir das matrizes de *boost* e as matrizes de *rotação:*

$$J_i = \frac{1}{2}\left(S_i + iK_i\right), \qquad \bar{J}_i = \frac{1}{2}\left(S_i - iK_i\right)$$

**Linhas Coordenadas do Espaço-Tempo**

Em 1868-1869, os matemáticos J. Plücker e A. Cayley introduziram dentro das álgebras geométricas o conceito de *linhas coordenadas* (WHITTAKER, 1953, p. 34).

*Definição*[11]*:*Sejam ($x_0, x_1, x_2, x_3$) e ($y_0, y_1, y_2, y_3$) coordenadas tétradas de dois pontos de uma linha reta $p$ sobre a variedade espaço-tempo, se escrevermos a sua álgebra de Lie não-abeliana:

$$x_m y_n - x_n y_m = p_{mn}$$

As seis componentes do tensor antissimétrico (ou 6-vetor) $p_{mn}$

$$p_{01},\quad p_{02},\quad p_{03},\quad p_{23},\quad p_{31},\quad p_{12}$$

São chamadas de *linhas coordenadas de p.*

Seguindo a convenção adotada por Whittaker faremos a coordenada temporal, de índice *4,* ser indexada em *0.* Pela teoria elementar das matrizes, sabemos que essa permutação de linhas não altera as propriedades matemáticas e físicas do sistema. Nestas condições, nossas funções de Poincaré, assumem a seguinte forma:

$$\begin{aligned} x_0 &= x_0' P_R^+(v) + x_1' R^2 P_R^-(v) \qquad & x_2 &= x_2' \\ x_1 &= x_1' P_R^+(v) + x_0' P_R^-(v) \qquad & x_3 &= x_3' \end{aligned}$$

Agora vamos calcular as linhas coordenadas do espaço-tempo, assumindo que as transformações em $y_n$ tem a mesma forma que as transformações em $x_m$:

$$\begin{aligned} p_{01} &= x_0 y_1 - x_1 y_0 \\ &= \left[x_0' P_R^+(v) + x_1' R^2 P_R^-(v)\right]\left[y_1' P_R^+(v) + y_0' P_R^-(v)\right] \\ &\quad - \left[x_1' P_R^+(v) + x_0' P_R^-(v)\right]\left[y_0' P_R^+(v) + y_1' R^2 P_R^-(v)\right] \end{aligned}$$

$$\begin{aligned} p_{01} &= x_0 y_1 - x_1 y_0 \\ &= x_0' y_1' \left[P_R^+(v)\right]^2 + x_0' y_0' P_R^+(v) P_R^-(v) \\ &\quad + x_1' y_1' R^2 P_R^+(v) P_R^-(v) + x_1' y_0' R^2 \left[P_R^-(v)\right]^2 \\ &\quad - x_1' y_0' \left[P_R^+(v)\right]^2 + x_1' y_1' R^2 P_R^+(v) P_R^-(v) \\ &\quad - x_0' y_0' P_R^+(v) P_R^-(v) - x_0' y_1' R^2 \left[P_R^-(v)\right]^2 \end{aligned}$$

$$\begin{aligned} p_{01} &= x_0 y_1 - x_1 y_0 \\ &= \left(x_0' y_1' - x_1' y_0'\right)\left[P_R^+(v)\right]^2 + \left(x_1' y_0' - x_0' y_1'\right) R^2 \left[P_R^-(v)\right]^2 \end{aligned}$$

[11] Adaptada de Whittaker (1953, p. 34).

Alterando a posição dos elementos da segunda parcela:

$$\begin{aligned} p_{01} &= x_0 y_1 - x_1 y_0 \\ &= \left(x'_0 y'_1 - x'_1 y'_0\right)\left[P_R^+(v)\right]^2 - \left(x'_0 y'_1 - x'_1 y'_0\right) R^2 \left[P_R^-(v)\right]^2 \end{aligned}$$

$$\begin{aligned} p_{01} &= x_0 y_1 - x_1 y_0 \\ &= \left(x'_0 y'_1 - x'_1 y'_0\right)\left(\left[P_R^+(v)\right]^2 - R^2 \left[P_R^-(v)\right]^2\right) \end{aligned}$$

Usando a relação fundamental da trigonometria hipercomplexa:

$$p_{01} = x_0 y_1 - x_1 y_0 = x'_0 y'_1 - x'_1 y'_0 = p'_{01}$$

Portanto, concluímos que a transformação da linha $p_{01}$ é dada por:

$$p_{01} = p'_{01}$$

Agora vamos calcular a transformação de $p_{0a}$, onde o índice $a$ varia de 2 à 3.

$$\begin{aligned} p_{0a} &= x_0 y_a - x_a y_0 \\ &= \left[x'_0 P_R^+(v) + x'_1 R^2 P_R^-(v)\right]\left[y'_a\right] - \left[x'_a\right]\left[y'_0 P_R^+(v) + y'_1 R^2 P_R^-(v)\right] \end{aligned}$$

$$\begin{aligned} p_{0a} &= x_0 y_a - x_a y_0 \\ &= x'_0 y'_a P_R^+(v) + x'_1 y'_a R^2 P_R^-(v) - x'_a y'_0 P_R^+(v) - x'_a y'_1 R^2 P_R^-(v) \\ &= \left(x'_0 y'_a - x'_a y'_0\right) P_R^+(v) + \left(x'_1 y'_a - x'_a y'_1\right) R^2 P_R^-(v) \end{aligned}$$

Que resulta nas transformações:

$$p_{0a} = p'_{0a}\, P_R^+(v) + p'_{1a}\, R^2 P_R^-(v)$$

Substituindo os valores do índice de $a$ e levando em consideração a antissimetria,

$$p_{02} = p'_{02}\, P_R^+(v) + p'_{12}\, R^2 P_R^-(v)$$

$$p_{03} = p'_{03}\, P_R^+(v) - p'_{31}\, R^2 P_R^-(v)$$

Calculemos a transformação de $p_{23}$:

$$\begin{aligned} p_{23} &= x_2 y_3 - x_2 y_3 \\ &= x'_2 y'_3 - x'_3 y'_2 = p'_{23} \end{aligned}$$

Portanto, $p_{23}$ é também um invariante:

$$p_{23} = p'_{23}$$

Por fim, determinaremos os valores de $p_{1a}$, $a$ varia de 2 à 3:

$$
\begin{aligned}
p_{1a} &= x_1 y_a - x_a y_1 \\
&= \left[x_1' P_R^+(v) + x_0' P_R^-(v)\right]\left[y_a'\right] - \left[x_a'\right]\left[y_1' P_R^+(v) + y_0' P_R^-(v)\right]
\end{aligned}
$$

$$
\begin{aligned}
p_{1a} &= x_1 y_a - x_a y_1 \\
&= x_1' y_a' P_R^+(v) + x_0' y_a' P_R^-(v) - x_a' y_1' P_R^+(v) - x_a' y_0' P_R^-(v) \\
&= \left(x_1' y_a' - x_a' y_1'\right) P_R^+(v) + \left(x_0' y_a' - x_a' y_0'\right) P_R^-(v)
\end{aligned}
$$

Que resulta nas transformações:

$$p_{1a} = p_{1a}' \, P_R^+(v) + p_{0a}' \, P_R^-(v)$$

Substituindo os valores do índice de *a* e levando em consideração a antissimetria,

$$p_{31} = p_{31}' \, P_R^+(v) + p_{03}' \, P_R^-(v)$$

$$p_{12} = p_{12}' \, P_R^+(v) + p_{02}' \, P_R^-(v)$$

Portanto as transformações das componentes do 6-vetor (ou tensor antissimétrico) são:

$$
\begin{aligned}
p_{01} &= p_{01}' \\
p_{02} &= p_{02}' \, P_R^+(v) + p_{12}' \, R^2 P_R^-(v) \\
p_{03} &= p_{03}' \, P_R^+(v) - p_{31}' \, R^2 P_R^-(v) \\
p_{23} &= p_{23}' \\
p_{31} &= p_{31}' \, P_R^+(v) - p_{03}' \, P_R^-(v) \\
p_{12} &= p_{12}' \, P_R^+(v) + p_{02}' \, P_R^-(v)
\end{aligned}
$$

Observe que as linhas coordenadas com as coordenadas temporais dependem da assinatura e as linhas compostas apenas pelas coordenadas especiais são as mesmas para todas as variedades.

## Teoria Eletromagnética

Nessa seção exploraremos a conexão entre as propriedades geométricas do espaço-tempo e o eletromagnetismo. Como observamos anteriormente, as linhas coordenadas do espaço-tempo Lorentziano coincidem com as componentes do campo Eletromagnético. Nós generalizaremos esse fato para todas variedades, a partir de um postulado, cujo enunciado é dado pelo seguinte princípio:

*"As componentes do campo eletromagnético são as linhas coordenadas do espaço-tempo. Em outras palavras, o campo eletromagnético é definido pela geometria do espaço-tempo afim"*

## O Campo Eletromagnético como as Linhas Coordenadas do Espaço-Tempo

As linhas coordenadas estão relacionadas as componentes do campo eletromagnético. Se tomarmos a assinatura da variedade Lorentziana, $R^2 = 1$, então as linhas coordenadas correspondem as transformações do campo elétrico e do campo magnético:

$$p_{01} = E_x, \qquad p_{02} = E_y, \qquad p_{03} = E_z$$
$$p_{23} = B_x, \qquad p_{31} = B_y, \qquad p_{12} = B_z$$

Portanto, as linhas coordenadas do espaço-tempo lorentziano representam o campo eletromagnético. Isso não é surpreendente, já que desde a construção da teoria eletromagnética de Maxwell-Hertz, os campos elétricos e magnéticos eram associados a propriedades do éter lumífero. A teoria da relatividade rejeita a substancialidade do éter, e propõe uma estrutura mais sofisticada denominada por H. Minkowski de espaço-tempo. Essa estrutura geométrica herda a operacionalidade do éter, portanto, as vibrações mecânicas do éter se transformam em linhas coordenadas do espaço-tempo.

Motivados por essa relação entre o campo eletromagnético e as linhas coordenadas do espaço-tempo, nós iremos postular o seguinte princípio:

*As componentes do campo eletromagnético correspondem as linhas coordenadas do espaço-tempo, conforme a seguinte regra:*

$$p_{01} = E_x, \quad p_{02} = E_y, \quad p_{03} = E_z,$$
$$p_{23} = B_x, \quad p_{31} = B_y, \quad p_{12} = B_z$$

Portanto, as transformações das componentes do campo elétrico e magnético para as variedades espaço-temporais planas serão:

$$\begin{aligned} E_x &= E'_x & B_x &= B'_x \\ E_y &= E'_y P_R^+(v) + B'_z R^2 P_R^-(v) & B_y &= B'_y P_R^+(v) - E'_z P_R^-(v) \\ E_z &= E'_z P_R^+(v) - B'_y R^2 P_R^-(v) & B_z &= B'_z P_R^+(v) + E'_y P_R^-(v) \end{aligned}$$

A partir dessas transformações é fácil ver porque é impossível estabelecer uma construção mecânica do eletromagnetismo. A variedade elementar da mecânica racional é a de Galileu (LANGEVIN, 1922), portanto ela tem como assinatura um número dual (nilpotente). Neste caso as transformações devem ser:

$$\begin{aligned} E_x &= E'_x & B_x &= B'_x \\ E_y &= E'_y P_\varepsilon^+(v) + B'_z \varepsilon^2 P_\varepsilon^-(v) & B_y &= B'_y P_\varepsilon^+(v) - E'_z P_\varepsilon^-(v) \\ E_z &= E'_z P_\varepsilon^+(v) - B'_y \varepsilon^2 P_\varepsilon^-(v) & B_z &= B'_z P_\varepsilon^+(v) + E'_y P_\varepsilon^-(v) \end{aligned}$$

E substituindo os valores,

$$\begin{aligned} E_x &= E'_x & B_x &= B'_x \\ E_y &= E'_y & B_y &= B'_y - \beta E'_z \\ E_z &= E'_z & B_z &= B'_z + \beta E'_y \end{aligned}$$

Estas devem ser as transformações do campo elétrico e do campo magnético para que os fenômenos eletromagnéticos admitam uma descrição mecânica. É possível mostrar que essa condição exige que o campo elétrico seja irrotacional. Isso significa que em um espaço-tempo galileano não é possível ocorrer o fenômeno da indução de Faraday. Por essa razão, é necessário substituir a variedade galileana, por uma variedade euclidiana ou lorentziana. Nas próximas seções, mostraremos a forma generalizada das equações de Maxwell e se elas nos permitem decidir por vias experimentais qual a variedade mais adequada.

Na seção anterior, verificamos que os campos eletromagnéticos correspondem as linhas coordenadas da variedade espaço-tempo. Essa observação sugere que a própria teoria eletromagnética tenha aspectos topológicos. Nessa seção, propomos uma forma topológica para as equações de Maxwell a partir do estudo das linhas coordenadas da variedade que definem a álgebra de Lie generalizada da Variedade.

Para obtermos as equações do eletromagnetismo válidas em qualquer variedade espaço-temporal, não podemos assumir que as equações de Maxwell, modificadas por Lorentz, sejam as mesmas. Para achar as novas equações introduziremos dois postulados:

**1)** *As equações devem ser covariantes de Poincaré.*

**2)** *As componentes do campo elétrico e magnético devem ser as linhas coordenadas da variedade*:

$$p_{mn} = \begin{pmatrix} 0 & E_x & E_y & E_z \\ -E_x & 0 & B_z & -B_y \\ -E_y & -B_z & 0 & -B_x \\ -E_z & B_y & B_x & 0 \end{pmatrix}$$

$$p'_{mn} = \begin{pmatrix} 0 & E_x & \Gamma\left(E_y + R^2 \mathrm{B} B_z\right) & \Gamma\left(E_z - R^2 \mathrm{B} B_y\right) \\ -E_x & 0 & -\Gamma\left(B_z + \mathrm{B} E_y\right) & \Gamma\left(B_y + \mathrm{B} E_z\right) \\ -\Gamma\left(E_y + R^2 \mathrm{B} B_z\right) & \Gamma\left(B_z - \mathrm{B} E_y\right) & 0 & -B_x \\ -\Gamma\left(E_z - R^2 \mathrm{B} B_y\right) & -\Gamma\left(B_y + \mathrm{B} E_z\right) & B_x & 0 \end{pmatrix}$$

Nós poderíamos introduzir um terceiro postulado que afirmaria que devemos buscar a forma que menos modifique as equações de Maxwell. Embora adotemos essa premissa, faremos por uma questão de simplicidade, não porque se impõe ao nosso espírito que soluções mais sofisticadas devam ser rejeitadas. De fato, convidamos ao leitor explorar outras possibilidades. O primeiro postulado é uma condição natural imposta pelo natureza das variedades que estamos analisando: espaço-temporais. O segundo postulado é observado na variedade lorentziana e se os efeitos associados a propagação das ondas eletromagnéticas dependem das qualidades topológicas da variedade, podemos inferir que a teoria eletromagnética é induzida pela assinatura da métrica do espaço-tempo. Nós procuraremos equações modificadas de Maxwell no vácuo da forma:

$$\nabla \cdot \vec{E} = 0 \qquad \nabla \times \vec{E} = -a\frac{\partial \vec{B}}{\partial t}$$

$$\nabla \cdot \vec{B} = 0 \qquad \nabla \times \vec{B} = -b\frac{\partial \vec{E}}{\partial t}$$

onde *a* e *b* são constantes a determinar que podem ser funções da característica-*R*.

Tomemos as transformações de Poincaré e a lei de transformação das derivadas parciais:

$$x' = \Gamma\left(x - vt\right) \qquad \partial_x = \Gamma\left(\partial_{x'} - \mathrm{B}R^2\partial_{t'}\right)$$
$$t' = \Gamma\left(t - \mathrm{B}R^2 x\right) \qquad \partial_t = \Gamma\left(\partial_{t'} - v\partial_{x'}\right)$$

Primeiro vamos determinar o coeficiente *a.* Para isso usaremos apenas a primeira componente da lei de Faraday e a lei de Gauss para o campo magnético:

$$\partial_x B_x + \partial_y B_y + \partial_z B_z = 0$$
$$a\partial_t B_x = \partial_y E_z - \partial_z E_y$$

Substituindo as transformações de *x* e de *t*, teremos:

$$\Gamma\left(\partial_{x'} B_x - \mathrm{B}R^2\partial_{t'} B_x\right) + \partial_y B_y + \partial_z B_z = 0$$
$$a\Gamma\left(\partial_{t'} B_x - v\partial_{x'} B_x\right) = \partial_y E_z - \partial_z E_y$$

E, após distribuir:

$$a\partial_t B_x = \partial_y E_z - \partial_z E_y$$
$$a\left(\Gamma\partial_{t'} B_x - v\Gamma\partial_{x'} B_x\right) = \partial_y E_z - \partial_z E_y$$

Substituindo o valor da derivada espacial da componente *x* do campo magnético na primeira componente da equação de Faraday, obtemos:

$$a\left(\Gamma\partial_{t'} B_x - v\Gamma\mathrm{B}R^2\partial_{t'} B_x + v\partial_y B_y + v\partial_z B_z\right) = \partial_y E_z - \partial_z E_y$$
$$a\left(\Gamma\partial_{t'} B_x - \Gamma\mathrm{B}^2R^2\partial_{t'} B_x + v\partial_y B_y + v\partial_z B_z\right) = \partial_y E_z - \partial_z E_y$$
$$a\Gamma\left(1 - \mathrm{B}^2R^2\right)\partial_{t'} B_x + a\mathrm{B}\partial_y B_y + a\mathrm{B}\partial_z B_z = \partial_y E_z - \partial_z E_y$$
$$a\frac{\Gamma}{\Gamma^2}\partial_{t'} B_x = \partial_y\left(E_z - a\mathrm{B}B_y\right) - \partial_z\left(E_y + a\mathrm{B}B_z\right)$$
$$a\partial_{t'} B_x = \partial_y\Gamma\left(E_z - a\mathrm{B}B_y\right) - \partial_z\Gamma\left(E_y + a\mathrm{B}B_z\right)$$

No sistema *S'* as equações devem apresentar a mesma forma que no sistema *S:*

$$\nabla' \cdot \vec{E}' = 0 \qquad \nabla' \times \vec{E}' = -a\frac{\partial \vec{B}'}{\partial t'}$$

$$\nabla' \cdot \vec{B}' = 0 \qquad \nabla' \times \vec{B}' = -b\frac{\partial \vec{E}'}{\partial t'}$$

Por inspeção, obtemos parte das transformações dos campos:

$$B_{x'} = B_x$$
$$E_{z'} = \Gamma\left(E_z - a\mathrm{B}B_y\right)$$
$$E_{y'} = \Gamma\left(E_y + a\mathrm{B}B_z\right)$$

Para obtermos o valor da constante a, basta compararmos as linhas coordenadas com as componentes do campo elétrico:

$$\Gamma\left(E_z - a\mathrm{B}B_y\right) = \Gamma\left(E_z - R^2\mathrm{B}B_y\right)$$

Portanto,

$$a = R^2$$

Agora vamos determinar o valor da constante *b*. para isso usaremos apenas a primeira componente da lei de Ampére e a lei de Gauss para o campo elétrico:

$$\partial_x E_x + \partial_y E_y + \partial_z E_z = 0$$
$$b\partial_t E_x = \partial_y B_z - \partial_z B_y$$

Substituindo as transformações de *x* e de *t*, teremos:

$$\begin{cases}\Gamma\left(\partial_{x'}E_x - \mathrm{B}R^2\partial_{t'}E_x\right) + \partial_y E_y + \partial_z E_z = 0\\ b\Gamma\left(\partial_{t'}E_x - v\partial_{x'}E_x\right) = \partial_y B_z - \partial_z B_y\end{cases} \qquad \begin{cases}\Gamma\partial_{x'}E_x = \Gamma\mathrm{B}R^2\partial_{t'}E_x + \partial_y E_y + \partial_z E_z\\ b\left(\Gamma\partial_{t'}E_x - v\Gamma\partial_{x'}E_x\right) = \partial_y B_z - \partial_z B_y\end{cases}$$

Substituindo o valor da derivada espacial da componente *x* do campo magnético na primeira componente da equação de Faraday, obtemos:

$$b\left(\Gamma\partial_{t'}E_x - v\Gamma\mathrm{B}R^2\partial_{t'}E_x + v\partial_y E_y + v\partial_z E_z\right) = \partial_y B_z - \partial_z B_y$$

$$b\left(\Gamma\partial_{t'}E_x - \Gamma\mathrm{B}^2R^2\partial_{t'}E_x + v\partial_y E_y + v\partial_z E_z\right) = \partial_y B_z - \partial_z B_y$$

$$b\Gamma\left(1 - \mathrm{B}^2R^2\right)\partial_{t'}E_x + b\mathrm{B}\partial_y E_y + b\mathrm{B}\partial_z E_z = \partial_y B_z - \partial_z B_y$$

$$b\frac{\Gamma}{\Gamma^2}\partial_{t'}E_x = \partial_y\left(B_z - b\mathrm{B}E_y\right) - \partial_z\left(B_y + b\mathrm{B}E_z\right)$$
$$b\partial_{t'}E_x = \partial_y\Gamma\left(B_z - b\mathrm{B}E_y\right) - \partial_z\Gamma\left(B_y + b\mathrm{B}E_z\right)$$

No sistema *S'* as equações devem apresentar à mesma forma que no sistema *S* (covariância de Poincaré):

$$\nabla' \cdot \vec{E}' = 0 \qquad \nabla' \times \vec{E}' = -\frac{a}{k}\frac{\partial \vec{B}'}{\partial t'}$$

$$\nabla' \cdot \vec{B}' = 0 \qquad \nabla' \times \vec{B}' = -\frac{b}{k}\frac{\partial \vec{E}'}{\partial t'}$$

Por inspeção, obtemos parte das transformações dos campos:

$$E_{x'} = E_x$$
$$-B_{z'} = \Gamma\left(B_z - b\mathrm{B}E_y\right)$$
$$-B_{y'} = \Gamma\left(B_y + b\mathrm{B}E_z\right)$$

Para obtermos o valor da constante a, basta compararmos as linhas coordenadas com as componentes do campo elétrico:

$$\Gamma\left(B_z - b\mathrm{B}E_y\right) = -\Gamma\left(B_z - \mathrm{B}E_y\right)$$

Portanto,

$$b = -1$$

E desta forma, as equações de Maxwell no vácuo para variedades espaço-temporais arbitrárias serão:

$$\nabla' \cdot \vec{E}' = 0 \qquad \nabla' \times \vec{E}' = -R^2\frac{\partial \vec{B}'}{\partial t'}$$

$$\nabla' \cdot \vec{B}' = 0 \qquad \nabla' \times \vec{B}' = \frac{\partial \vec{E}'}{\partial t'}$$

Para a variedade galileana, $R^2 = 0$, recuperamos o resultado anterior, que o campo elétrico é irrotacional. Para variedade lorentziana, $R^2 = +1$, obtemos a lei de Faraday-Lenz. Por fim, para a variedade euclidiana, $R^2 = -1$, a lei de Faraday-anti-Lenz[12]

$$Galileana \quad \mapsto \quad \nabla \times \vec{E} = 0$$

$$Lorentziana \quad \mapsto \quad \nabla \times \vec{E} = -\frac{\partial \vec{B}}{\partial t}$$

$$Euclidiana \quad \mapsto \quad \nabla \times \vec{E} = +\frac{\partial \vec{B}}{\partial t}$$

[12] Anti-Lenz porque o sentido da corrente induzida produz um campo magnético no mesmo sentido do fluxo magnético que lhe deu origem.

**Unificando as Equações de Maxwell**[13]

Por meio da álgebra geométrica é possível escrever as equações de Maxwell como única equação. Para este fim, introduziremos o conceito de pseudo-escalar $I$ que permite relacionar o produto vetorial de Gibbs-Heaviside com o produto exterior de Grassmann.

$$\vec{u} \wedge \vec{v} = I\left(\vec{u} \times \vec{v}\right), \qquad onde, \quad I^2 = -1$$

Como o produto exterior de dois vetores produz um bivetor e o produto vetorial, um pseudo-vetor[14], então essas grandezas se associam por meio da equação:

$$\widehat{B} = I\vec{B}$$

Para os operadores diferenciais definimos os produtos de Clifford por meio das relações:

$$\begin{aligned} \nabla\vec{u} &= \operatorname{div}\vec{u} + \nabla \wedge \vec{u} \\ \nabla\vec{u} &= \operatorname{div}\vec{u} + I \operatorname{rot}\vec{u} \end{aligned} \qquad \nabla f = \operatorname{grad} f$$

Vamos definir um novo operador diferencial, denominado de multivetor gradiente:

$$\widehat{\nabla} = \nabla + \frac{1}{c}\frac{\partial}{\partial t}$$

Por meio do multivetor gradiente, **as quatro equações de Maxwell-Heaviside** podem ser escritas como **uma única equação**:

$$\widehat{\nabla}\left(\vec{E} + \widehat{B}\right) = 4\pi\left(\rho - \frac{\vec{J}}{c}\right)$$

onde $\vec{E}$, é o vetor campo elétrico, $\widehat{B}$, o bivetor campo magnético, $\rho$ é a densidade de carga, $\vec{J}$ é a corrente de deslocamento e $c$ é a velocidade da luz no vácuo.

Para realizarmos a demonstração faremos a substituição direta e, por fim, faremos a comparação das quantidades multivetoriais. Inicialmente, escrevamos o multivetor gradiente por extenso:

$$\left(\nabla + \frac{1}{c}\frac{\partial}{\partial t}\right)\left(\vec{E} + \widehat{B}\right) = 4\pi\left(\rho - \frac{\vec{J}}{c}\right)$$

Distribuindo os operadores do lado direito:

$$\nabla\vec{E} + \frac{1}{c}\frac{\partial\vec{E}}{\partial t} + \nabla\widehat{B} + \frac{1}{c}\frac{\partial\widehat{B}}{\partial t} = 4\pi\left(\rho - \frac{\vec{J}}{c}\right)$$

[13] Essa seção é uma sumarização do trabalho: *Uma mini-introdução à concisa álgebra geométrica do eletromagnetismo* (Ferreira, 2006)
[14] Para detalhes ver: *Da força ao tensor: evolução do conceito físico e da representação matemática do campo eletromagnético* (Silva, 2002).

Aplicando os produtos de Clifford, obtemos:

$$\mathrm{div}\vec{E} + \nabla \wedge \vec{E} + \frac{1}{c}\frac{\partial \vec{E}}{\partial t} + \mathrm{div}\widehat{B} + \nabla \wedge \widehat{B} + \frac{1}{c}\frac{\partial \widehat{B}}{\partial t} = 4\pi\left(\rho - \frac{\vec{J}}{c}\right)$$

Explorando a relação entre produto exterior e o produto vetorial,

$$\mathrm{div}\vec{E} + I\left(\mathrm{rot}\,\vec{E}\right) + \frac{1}{c}\frac{\partial \vec{E}}{\partial t} + \mathrm{div}\widehat{B} + I\left(\mathrm{rot}\,\widehat{B}\right) + \frac{1}{c}\frac{\partial \widehat{B}}{\partial t} = 4\pi\left(\rho - \frac{\vec{J}}{c}\right)$$

Substituindo a relação entre o bivetor e o pseudovetor campo magnético:

$$\mathrm{div}\vec{E} + I\left(\nabla \times \vec{E}\right) + \frac{1}{c}\frac{\partial \vec{E}}{\partial t} + I\mathrm{div}\vec{B} + I^2\left(\nabla \times \vec{B}\right) + I\frac{1}{c}\frac{\partial \vec{B}}{\partial t} = 4\pi\left(\rho - \frac{\vec{J}}{c}\right)$$

$$\mathrm{div}\vec{E} + I\left(\mathrm{rot}\,\vec{E}\right) + \frac{1}{c}\frac{\partial \vec{E}}{\partial t} + I\left(\mathrm{div}\vec{B}\right) - \mathrm{rot}\,\vec{B} + I\frac{1}{c}\frac{\partial \vec{B}}{\partial t} = 4\pi\left(\rho - \frac{\vec{J}}{c}\right)$$

Vamos organizar a equação em blocos ordenados: escalar, pseudo-escalar, vetor, pseudo-vetor:

$$\underbrace{\left[\mathrm{div}\vec{E}\right]}_{\textbf{escalar}} + \underbrace{I\left[\left(\mathrm{div}\vec{B}\right)\right]}_{\textbf{pseudo - escalar}} + \underbrace{\left[\frac{1}{c}\frac{\partial \vec{E}}{\partial t} - \mathrm{rot}\,\vec{B}\right]}_{\textbf{vetor}} + \underbrace{I\left[\left(\mathrm{rot}\,\vec{E}\right) + \frac{1}{c}\frac{\partial \vec{B}}{\partial t}\right]}_{\textbf{pseudo - vetor}} = 4\pi\left(\underbrace{\rho}_{\textbf{escalar}} - \underbrace{\frac{\vec{J}}{c}}_{\textbf{vetor}}\right)$$

Igualando cada bloco, conforme sua natureza algébrica:

$$\mathrm{div}\vec{E} = 4\pi\rho \qquad \mathrm{div}\vec{B} = 0$$

$$\mathrm{rot}\,\vec{B} = \frac{4\pi}{c}\vec{J} + \frac{1}{c}\frac{\partial \vec{E}}{\partial t} \qquad \mathrm{rot}\,\vec{E} = -\frac{1}{c}\frac{\partial \vec{B}}{\partial t}$$

que são as 4 equações de Maxwell-Heaviside (Q.E.D.).

**Unificando as Equações de Maxwell em Variedades Arbitrárias**

Por meio da álgebra geométrica vimos que é possível escrever as equações de Maxwell como única equação. O procedimento que aplicamos foi válido apenas para uma variedade Lorentziana. Para estendermos as nossas definições para as demais variedades, devemos modificar o multivetor gradiente para acomodar o bivetor temporal *R.* Isso pode ser feito utilizando o conceito de inversão graduada. Para os fins dessa seção, redefiniremos o multivetor gradiente da seguinte forma:

$$\widehat{\nabla}_R = \nabla + \left(R\right)^{p(p-1)}\frac{1}{c}\frac{\partial}{\partial t}$$

onde $p$ representa a gradação do p-vetor.

Por meio dos multivetores gradientes, **as quatro equações de Maxwell-Heaviside** podem ser escritas como **uma única equação**:

$$\widehat{\nabla}_R\left(\vec{E}+\widehat{B}\right)=4\pi\left(\rho-\frac{\vec{J}}{c}\right)$$

onde $\vec{E}$, é o 1-vetor campo elétrico, $\widehat{B}$, o 2-vetor campo magnético, $\rho$ é o escalar densidade de carga, $\vec{J}$ é o 1-vetor corrente de deslocamento e $c$ é a velocidade da luz no vácuo.

Para realizarmos a demonstração faremos a substituição direta e, por fim, faremos a comparação das quantidades escalares, pseudo-escalares, vetoriais e pseudo-vetoriais. Inicialmente, escrevamos o multivetor gradiente por extenso:

$$\left(\nabla+\left(R\right)^{p(p-1)}\frac{1}{c}\frac{\partial}{\partial t}\right)\left(\vec{E}+\widehat{B}\right)=4\pi\left(\rho-\frac{\vec{J}}{c}\right)$$

Distribuindo os operadores do lado direito:

$$\nabla\vec{E}+\left(R\right)^{p(p-1)}\frac{1}{c}\frac{\partial\vec{E}}{\partial t}+\nabla\widehat{B}+\left(R\right)^{p(p-1)}\frac{1}{c}\frac{\partial\widehat{B}}{\partial t}=4\pi\left(\rho-\frac{\vec{J}}{c}\right)$$

Substituindo as gradações de *p,* obtemos:

$$\nabla\vec{E}+\left(R\right)^{\cancel{1(1-1)}_{=0}}\frac{1}{c}\frac{\partial\vec{E}}{\partial t}+\nabla\widehat{B}+\left(R\right)^{\cancel{2(2-1)}_{=2}}\frac{1}{c}\frac{\partial\widehat{B}}{\partial t}=4\pi\left(\rho-\frac{\vec{J}}{c}\right)$$

$$\nabla\vec{E}+\frac{1}{c}\frac{\partial\vec{E}}{\partial t}+\nabla\widehat{B}+R^2\frac{1}{c}\frac{\partial\widehat{B}}{\partial t}=4\pi\left(\rho-\frac{\vec{J}}{c}\right)$$

Aplicando os produtos de Clifford e as relações com os pseudo-escalares, obtemos:

$$\text{div}\vec{E}+I\left(\text{rot}\,\vec{E}\right)+\frac{1}{c}\frac{\partial\vec{E}}{\partial t}+I\left(\text{div}\vec{B}\right)-\text{rot}\,\vec{B}+I\left(R^2\frac{1}{c}\frac{\partial\vec{B}}{\partial t}\right)=4\pi\left(\rho-\frac{\vec{J}}{c}\right)$$

Pondo a equação em blocos ordenados: escalar, pseudo-escalar, vetor, pseudo-vetor:

$$\underbrace{\left[\text{div}\vec{E}\right]}_{\textbf{escalar}}+I\underbrace{\left[\left(\text{div}\vec{B}\right)\right]}_{\textbf{pseudo - escalar}}+\underbrace{\left[\frac{1}{c}\frac{\partial\vec{E}}{\partial t}-\text{rot}\,\vec{B}\right]}_{\textbf{vetor}}+I\underbrace{\left[\left(\text{rot}\,\vec{E}\right)+R^2\frac{1}{c}\frac{\partial\vec{B}}{\partial t}\right]}_{\textbf{pseudo - vetor}}=4\pi\left(\underbrace{\rho}_{\textbf{escalar}}-\underbrace{\frac{\vec{J}}{c}}_{\textbf{vetor}}\right)$$

Igualando cada bloco, conforme sua natureza algébrica:

$$\text{div}\vec{E}=4\pi\rho \qquad \text{div}\vec{B}=0$$

$$\text{rot}\,\vec{B}=\frac{4\pi}{c}\vec{J}+\frac{1}{c}\frac{\partial\vec{E}}{\partial t} \qquad \text{rot}\,\vec{E}=-R^2\frac{1}{c}\frac{\partial\vec{B}}{\partial t}$$

que são as 4 equações generalizadas de Maxwell-Heaviside (Q.E.D.).

**Unificando as Equações de Maxwell em Variedades Arbitrárias: Nova Abordagem**

Recentemente, eu descobri uma nova abordagem para se unificar as equações de Maxwell em variedades arbitrárias usando a álgebra multivetorial. Esse novo método é muito mais simples e inteligível que o anterior. Sua maior vantagem é que ele define quais grandezas eletromagnéticas são induzidas pelo fator *R*. Nesse novo formalismo, escrevemos as equações de Maxwell, como a seguinte equação multivetorial:

$$\widehat{\Box}(R)\widehat{F}(R)=4\pi\widehat{J}(R)$$

que doravante chamarei de equação de Capiberibe, cujas componentes são dadas por:

$$\widehat{\Box}(R)=\vec{\nabla}+R\frac{1}{c}\frac{\partial}{\partial t} \quad \text{(multi-operador gradiente } R\text{)}$$
$$\widehat{F}(R)=\vec{E}+R\widehat{B} \quad \text{(multivetor eletromagnético } R\text{)}$$
$$\widehat{J}(R)=\rho-R\frac{1}{c}\vec{J} \quad \text{(paravetor densidade de carga/corrente } R\text{)}$$

Para realizarmos a demonstração novamente faremos a substituição direta:

$$\left(\vec{\nabla}+R\frac{\partial}{\partial t}\right)\left(\vec{E}+R\widehat{B}\right)=4\pi\left(\rho-\frac{R}{c}\vec{J}\right)$$

$$\vec{\nabla}\vec{E}+R\vec{\nabla}\widehat{B}+R\frac{1}{c}\frac{\partial\vec{E}}{\partial t}+R^2\frac{1}{c}\frac{\partial\widehat{B}}{\partial t}=4\pi\rho-4\pi\frac{R}{c}\vec{J}$$

$$\nabla\vec{E}+\frac{1}{c}\frac{\partial\vec{E}}{\partial t}+\nabla\widehat{B}+R^2\frac{1}{c}\frac{\partial\widehat{B}}{\partial t}=4\pi\left(\rho-\frac{\vec{J}}{c}\right)$$

Aplicando os produtos de Clifford e as relações com os pseudo-escalares, obtemos:

$$\text{div}\vec{E}+I\left(\text{rot}\,\vec{E}\right)+R\frac{1}{c}\frac{\partial\vec{E}}{\partial t}+IR\left(\text{div}\vec{B}\right)-R\left(\text{rot}\,\vec{B}\right)+I\left(R^2\frac{1}{c}\frac{\partial\vec{B}}{\partial t}\right)=4\pi\rho-4\pi R\frac{1}{c}\vec{J}$$

Pondo a equação em blocos ordenados: escalar, pseudo-escalar, vetor, pseudo-vetor:

$$\underbrace{\left[\text{div}\vec{E}\right]}_{\textbf{escalar}}+I\underbrace{\left[R\left(\text{div}\vec{B}\right)\right]}_{\textbf{pseudo - escalar}}+\underbrace{\left[R\frac{1}{c}\frac{\partial\vec{E}}{\partial t}-R\left(\text{rot}\,\vec{B}\right)\right]}_{\textbf{vetor}}+I\underbrace{\left[\left(\text{rot}\,\vec{E}\right)+R^2\frac{1}{c}\frac{\partial\vec{B}}{\partial t}\right]}_{\textbf{pseudo - vetor}}=\underbrace{4\pi\rho}_{\textbf{escalar}}-\underbrace{4\pi R\frac{\vec{J}}{c}}_{\textbf{vetor}}$$

Igualando cada bloco, conforme sua natureza algébrica e realizando as simplificações:

$$\text{div}\vec{E}=4\pi\rho \qquad \text{div}\vec{B}=0$$
$$\text{rot}\,\vec{B}=\frac{4\pi}{c}\vec{J}+\frac{1}{c}\frac{\partial\vec{E}}{\partial t} \qquad \text{rot}\,\vec{E}=-R^2\frac{1}{c}\frac{\partial\vec{B}}{\partial t}$$

que são as 4 equações generalizadas de Maxwell-Heaviside (Q.E.D.).

**Gauge de Poincaré**

No eletromagnetismo clássico podemos associar ao campo elétrico um escalar, denominado de potencial escalar elétrico $\phi$, e ao campo magnético, um vetor, denominado de potencial vetor magnético **A.** Estes dois potenciais são usados para criar um 4-vetor denominado de 4-potencial eletromagnético.

$$A_i = \left(\phi, \vec{A}\right), \; A_i' = \left(\phi', \vec{A}'\right)$$

No referencial *S'* as componentes do 4-potencial se transformam como:

$$\phi' = \Gamma\left(\phi - R^2 \mathrm{B} A_x\right) \qquad A_y' = A_y$$
$$A_x' = \Gamma\left(A_x - \mathrm{B}\phi\right) \qquad A_z' = A_z$$

No referencial próprio, não há um campo magnético, portanto a partícula terá apenas um escalar potencial elétrico:

$$A_i^o = \left(\phi^o, 0, 0, 0\right)$$

Portanto as equações para construção de nosso invariante são:

$$J_0^o = \phi^o$$
$$J_0 = \phi$$
$$\left\|\vec{J}\right\| = \left\|\vec{A}\right\|$$

Usando a regra dos invariantes relativísticos, obtemos:

$$R^2 \phi^{o2} = R^2 \phi^2 - \left\|\vec{A}\right\|^2$$

Para qualquer referencial inercial é válida a relação:

$$R^2 \phi'^2 - \left\|\vec{A}'\right\|^2 = R^2 \phi^2 - \left\|\vec{A}\right\|^2$$
$$R^2 \left(\phi'^2 - \phi^2\right) = \left\|\vec{A}'\right\|^2 - \left\|\vec{A}\right\|^2$$

Os potenciais elétrico e magnético são os geradores dos campos elétrico e magnético. Para provar essas relações vamos usar as seguintes identidades vetoriais:

$$\nabla \cdot \left(\nabla \times \vec{A}\right) = 0, \qquad \nabla \times \left(\nabla \varphi\right) = 0$$

E as equações de Maxwell na forma vetorial:

$$\nabla \cdot \vec{E} = \rho \qquad \nabla \times \vec{E} = -R^2 \frac{\partial \vec{B}}{\partial t}$$

$$\nabla \cdot \vec{B} = 0 \qquad \nabla \times \vec{B} = \vec{j} + \frac{\partial \vec{E}}{\partial t}$$

Como o divergente do campo magnético é sempre nulo isso implica, pelas identidades vetoriais, que o campo magnético é gerado pelo rotacional do vetor potencial magnético:

$$\vec{B} = \nabla \times \vec{A}$$

Na ausência de um campo magnético, uma carga $q$ está sujeita a uma força elétrica dada por:

$$\vec{f}_e = -q\nabla\varphi$$
$$\vec{E} = -\nabla\varphi$$

Se considerarmos que a partícula se desloca em uma campo eletromagnético, devemos acrescentar ao campo elétrico um vetor **V** a ser determinado:

$$\vec{E} = -\nabla\varphi + \vec{V}$$

Para determinarmos a forma desse vetor, vamos substituir a lei de formação do campo elétrico na terceira de equação de Maxwell.

$$\nabla \times \left(-\nabla\varphi + \vec{V}\right) = -R^2 \frac{\partial \vec{B}}{\partial t}$$

Distribuindo o produto vetorial sobre os vetores e substituindo o campo magnético:

$$-\nabla \times \left(\nabla\varphi\right) + \nabla \times \vec{V} = -R^2 \frac{\partial\left(\nabla \times \vec{A}\right)}{\partial t}$$

Pela identidade vetorial, a primeira parcela do lado esquerdo é zero, além disso, a derivada temporal comuta com o rotacional. Assim, podemos escrever nossa equação da seguinte forma:

$$\nabla \times \vec{V} = \nabla \times \left(-R^2 \frac{\partial \vec{A}}{\partial t}\right)$$

Portanto, o vetor **V** será dado por:

$$\vec{V} = -R^2 \frac{\partial \vec{A}}{\partial t}$$

E a regra de formação dos campos elétrico e magnético são:

$$\begin{cases} \vec{E} = -\nabla\phi - R^2 \dfrac{\partial \vec{A}}{\partial t}, \\ \vec{B} = \nabla \times \vec{A} \end{cases}$$

No sistema *S'* esses vetores terão coordenadas definidas por:

$$\begin{cases} \vec{E}' = -\nabla'\phi' - R^2 \dfrac{\partial \vec{A}'}{\partial t'}, \\ \vec{B}' = \nabla' \times \vec{A}' \end{cases}$$

Essas são as transformações do gauge de Poincaré que é válido para qualquer variedade espaço-temporal. Por meio dessa transformação, podemos calcular as transformações do campo elétrico e do campo magnético. Comecemos pelo campo elétrico, para isso escreveremos as equações das componentes do campo elétrico no referencial *S'* e as do campo magnético no referencial *S*.

$$E_i = -\left(\partial_i\phi + R^2\partial_t A_i\right) \qquad B_y = \left(\partial_z A_x - \partial_x A_z\right)$$
$$B_x = \left(\partial_y A_z - \partial_z A_y\right) \qquad B_z = \left(\partial_x A_y - \partial_y A_x\right)$$

Começaremos estudando a componente $x$ do campo elétrico. Aplicando as transformações do 4-Gradiente e do 4-Potencial,

$$E'_x = -\Gamma\left(\partial_x\phi' + BR^2\partial_t\phi' + R^2\partial_t A'_x + \mathrm{B}R^2\partial_x A'_x\right)$$
$$E'_x = -\Gamma^2\left(\partial_x\left(\phi - BR^2A_x\right) + BR^2\partial_t\left(\phi - BR^2A_x\right) + R^2\partial_t\left(A_x - B\phi\right) + BR^2\partial_x\left(A_x - B\phi\right)\right)$$
$$E'_x = -\Gamma^2\left[\partial_x\left(1 - B^2R^2\right)\phi + \partial_x\left(BR^2 - BR^2\right)A_x + \partial_t\left(BR^2 - BR^2\right)\phi + R^2\partial_t\left(1 - B^2R^2\right)A_x\right]$$

Usando o fator de Poincaré e realizando as simplificações algébricas:

$$E'_x = -\frac{\Gamma^2}{\Gamma^2}\left[\partial_x\phi + R^2\partial_t A_x\right]$$
$$E'_x = -\left[\partial_x\phi + R^2\partial_t A_x\right]$$
$$E'_x = E_x$$

Para a componente $y$, teremos a relação entre o sistema *S'* e *S*:

$$E'_y = -\left(\partial'_y\phi' + R^2\partial'_t A'_y\right)$$
$$E'_y = -\Gamma\left(\partial_y\left(\phi - \mathrm{B}R^2A_x\right) + R^2\partial_t A_y + \mathrm{B}R^2\partial_x A_y\right)$$
$$E'_y = -\Gamma\left(\partial_y\phi + \frac{R^2}{k}\partial_t A_y + \mathrm{B}R^2\partial_x A_y - \mathrm{B}R^2\partial_y A_x\right)$$
$$E'_y = -\Gamma\left(\partial_y\phi + R^2\partial_t A_y + \mathrm{B}R^2\left(\partial_x A_y - \partial_y A_x\right)\right)$$
$$E'_y = \Gamma\left[-\left(\partial_y\phi + R^2\partial_t A_y\right) - \mathrm{B}R^2\left(\partial_x A_y - \partial_y A_x\right)\right]$$

A primeira parcela dentro do colchetes é a componente $y$ do campo elétrico e a segunda parcela é a componente $z$ do campo magnético, ambas no referencial $S$.

$$E'_y = \Gamma\left(E_y - \mathrm{B}R^2 B_z\right)$$

E, analogamente, para componente $z$, teremos:

$$\begin{aligned}
E'_z &= -\left(\partial'_z \phi' + R^2 \partial'_t A'_z\right) \\
E'_y &= -\Gamma\left(\partial_z\left(\phi - \mathrm{B}R^2 A_x\right) + R^2 \partial_t A_z + \mathrm{B}R^2 \partial_x A_z\right) \\
E'_y &= \Gamma\left[-\left(\partial_z \phi + R^2 \partial_t A_z\right) + \mathrm{B}R^2\left(\partial_y A_z - \partial_x A_z\right)\right]
\end{aligned}$$

A primeira parcela dentro do colchetes é a componente $z$ do campo elétrico e a segunda é a componente $y$ do campo magnético no referencial $S$.

$$E'_z = \Gamma\left(E_z + \mathrm{B}R^2 B_y\right)$$

Para o campo magnético, usaremos o conjunto de equações:

$$E_i = -\left(\partial_i \phi + R^2 \partial_t A_i\right)$$

$$\begin{aligned}
B'_x &= \left(\partial'_y A'_z - \partial'_z A'_y\right) & B_x &= \left(\partial_y A_z - \partial_z A_y\right) \\
B'_y &= \left(\partial'_z A'_x - \partial'_x A'_z\right) & B_y &= \left(\partial_z A_x - \partial_x A_z\right) \\
B'_z &= \left(\partial'_x A'_y - \partial'_y A'_x\right) & B_z &= \left(\partial_x A_y - \partial_y A_x\right)
\end{aligned}$$

Para a componente $x$ do campo magnético, usando o potencial, obtemos:

$$B'_x = \left(\partial_y A_z - \partial_z A_y\right)$$

O termo em parêntesis é a componente $B_x$, portanto:

$$B'_x = B_x$$

Para a componente $y$, teremos:

$$\begin{aligned}
B'_y &= \left(\partial'_z A'_x - \partial'_x A'_z\right) \\
B'_y &= \Gamma\left(\partial_z\left(A_x - \mathrm{B}\phi\right) - \partial_x A_z - R^2 \mathrm{B} \partial_t A_z\right) \\
B'_y &= \Gamma\left(\partial_z A_x - \mathrm{B}\partial_z \phi - \partial_x A_z - R^2 \mathrm{B} \partial_t A_z\right) \\
B'_y &= \Gamma\left[\left(\partial_z A_x - \partial_x A_z\right) - \mathrm{B}\left(\partial_z \phi + R^2 \partial_t A_z\right)\right]
\end{aligned}$$

A primeira parcela no colchetes é a componente $y$ do campo magnético e a segunda parcela é a componente $z$ do campo elétrico:

$$B'_y = \Gamma\left(B_y + \mathrm{B}E_z\right)$$

Por derradeiro, a componente do *z* se transforma pela regra:

$$B'_z = \left(\partial'_x A'_y - \partial'_y A'_x\right) \qquad B'_y = \Gamma\left(\partial_x A_y - \partial_y A_x + \mathrm{B}R^2\partial_t A_y + \mathrm{B}\partial_y \phi\right)$$

$$B'_z = \Gamma\left(\partial_x A_y + \mathrm{B}R^2\partial_t A_y - \partial_y\left(A_x - \mathrm{B}\phi\right)\right) \qquad B'_y = \Gamma\left[\left(\partial_x A_y - \partial_y A_x\right) + \mathrm{B}\left(\partial_y \phi + R^2\partial_t A_y\right)\right]$$

A primeira parcela no colchetes é a componente *z* do campo magnético e a segunda parcela é a componente *y* do campo elétrico com o sinal invertido:

$$B'_z = \Gamma\left(B_z - \mathrm{B}E_y\right)$$

Portanto, deduzimos sem qualquer dificuldade e ambiguidade, as transformações do campo elétrico e do campo magnético. Esse método é ainda mais simples que o método empregado por Lorentz em 1904, Poincaré em 1905-1906 e Einstein em 1905. Observe que nossa formulação difere de outras notações, pois estamos adotando mesmo sistema de medidas adotado por Albert Einstein, conhecido como sistema de coordenadas hertzianos. As convenções adotadas não alteram o significado físico das equações.

**Oscilações Eletromagnéticas**

Para provarmos que nossa formulação é consistente com a métrica do espaço-tempo arbitrário, vamos calcular as equações de propagação do campo elétrico e do campo magnético. Tomemos as equações de Maxwell modificadas:

$$\nabla \cdot \vec{E} = 0 \qquad \nabla \times \vec{E} = -R^2 \frac{\partial \vec{B}}{\partial t}$$

$$\nabla \cdot \vec{B} = 0 \qquad \nabla \times \vec{B} = \frac{\partial \vec{E}}{\partial t}$$

Aplicando o rotacional sobre o rotacional do campo elétrico,

$$\nabla \times \left(\nabla \times \vec{E}\right) = R^2 \nabla \times \left(\frac{\partial \vec{B}}{\partial t}\right)$$

Como os operadores comutam, podemos reescrever a equação:

$$\nabla \times \left(\nabla \times \vec{E}\right) = R^2 \frac{\partial}{\partial t}\left(\nabla \times \vec{B}\right)$$

Substituindo o valor do rotacional do campo magnético:

$$\nabla \times \left(\nabla \times \vec{E}\right) = R^2 \frac{\partial}{\partial t}\left(-\frac{\partial \vec{E}}{\partial t}\right)$$

$$\nabla \times \left( \nabla \times \vec{E} \right) = -R^2 \frac{\partial^2 \vec{E}}{\partial t^2}$$

Usando a identidade de Laplace para o duplo rotacional:

$$\nabla \left( \nabla \cdot \vec{E} \right) - \nabla^2 \vec{E} = -R^2 \frac{\partial^2 \vec{E}}{\partial t^2}$$

Como a divergência do campo elétrico no vácuo é zero,

$$\nabla^2 \vec{E} - R^2 \frac{\partial^2 \vec{E}}{\partial t^2} = 0$$

Aplicando o rotacional sobre o rotacional do campo magnético,

$$\nabla \times \left( \nabla \times \vec{B} \right) = -\nabla \times \left( \frac{\partial \vec{E}}{\partial t} \right)$$

Como os operadores comutam, podemos reescrever a equação:

$$\nabla \times \left( \nabla \times \vec{B} \right) = -\frac{\partial}{\partial t} \left( \nabla \times \vec{E} \right)$$

Substituindo o valor do rotacional do campo elétrico:

$$\nabla \times \left( \nabla \times \vec{B} \right) = \frac{\partial}{\partial t} \left( -R^2 \frac{\partial \vec{E}}{\partial t} \right)$$

$$\nabla \times \left( \nabla \times \vec{B} \right) = -R^2 \frac{\partial^2 \vec{B}}{\partial t^2}$$

Usando a identidade de Laplace para o duplo rotacional:

$$\nabla \left( \nabla \cdot \vec{B} \right) - \nabla^2 \vec{B} = -R^2 \frac{\partial^2 \vec{B}}{\partial t^2}$$

Como a divergência do campo elétrico no vácuo é zero,

$$\nabla^2 \vec{B} - R^2 \frac{\partial^2 \vec{B}}{\partial t^2} = 0$$

Que coincide com as formas topológicas da luz que calculamos anteriormente, por um processo diferente. Portanto, as modificações que empregamos são consistentes com forma da luz. Desta forma, o campo elétrico e magnético e as formas de propagação da radiação no vácuo são propriedades topológicas da variedade.

**Oscilações Eletromagnéticas em Variedades Galileanas e Euclidianas**

Vamos agora verificar como as equações de Maxwell se comportam nas variedades espaço-temporais Galileana e Euclidiana. A variedade Lorentziana corresponde a teoria eletromagnética usual e dispensa análise. Mais uma vez, escrevamos as equações de Maxwell no vácuo:

$$\nabla \cdot \vec{E} = 0 \qquad \nabla \times \vec{E} = -R^2 \frac{\partial \vec{B}}{\partial t}$$

$$\nabla \cdot \vec{B} = 0 \qquad \nabla \times \vec{B} = \frac{\partial \vec{E}}{\partial t}$$

A única equação que é afetada pela característica-R da variedade é a lei de Faraday. Para uma variedade galileana, $R^2$ é nilpotente, portanto é zero. A equação de Faraday assume a forma:

$$\nabla \times \vec{E} = 0$$

Isso significa que o campo elétrico é irrotacional em todos os pontos e por isso o campo elétrico é apenas uma função do potencial elétrico. Essa equação também indica que não existe indução elétrica por meio da variação de um campo magnético. Além disso, os campos elétricos e magnéticos e a forma da luz, não seriam de ondas esféricas, mas harmônicos esféricos que satisfariam a equação de Laplace-Beltrami:

$$\nabla^2 \vec{E} = 0 \qquad \nabla^2 \vec{B} = 0 \qquad \nabla^2 \Psi = 0$$

$$\vec{E}(r,\theta,\varphi) = R(\vec{r})Y_l^m(\theta,\varphi) \qquad \vec{B}(r,\theta,\varphi) = R(\vec{r})Y_l^m(\theta,\varphi) \qquad \Psi(r,\theta,\varphi) = R(\vec{r})Y_l^m(\theta,\varphi)$$

Na variedade de Galileu também não podemos associar a velocidade de propagação desses harmônicos esféricos com a velocidade da luz, pois a constante *k* de velocidade não está presente.

Para uma variedade euclidiana, $R^2$ é a unidade negativa. A equação de Faraday assume a forma:

$$\nabla \times \vec{E} = \frac{\partial \vec{B}}{\partial t}$$

Isso significa que o campo elétrico sofre uma rotação no sentido oposto, em relação a variedade lorentziana. Essa equação também indica que a indução por meio da variação de um campo magnético ocorre no sentido contrário do usual. Além disso, os campos elétricos e magnéticos e a forma da luz, não seriam de ondas esféricas, mas harmônicos esféricos associados perturbações periódicas no tempo que satisfazem a equação de Laplace-Beltrami temporal:

$$\nabla^2 \Psi + \frac{\partial^2 \Psi}{\partial t^2} = 0$$

$$\Psi(r,\theta,\varphi,t) = R(\vec{r})Y_l^m(\theta,\varphi)\left(Ae^{ikt}\right)$$

Na variedade de Euclides também podemos associar a velocidade de propagação desses harmônicos esféricos com a velocidade da luz, embora o valor de propagação da velocidade da luz possa ser diferente de *c*. O caráter negativo da dimensão de tempo, inverte a orientação da indução e do rotacional do campo elétrico. Essas propriedades podem de alguma forma estar ligada as exóticas propriedades dos meta-materiais. Caso essa hipótese se verifique, poderíamos supor que o meta-material atua localmente sobre o tempo fazendo que ele apresenta um caráter dimensional negativo e fechado.

**A Forma da Luz**

Até o presente momento temos caracterizado as equações gerais do espaço-tempo e quais características particulares a unidade hipercomplexa induz a sua forma. Agora vamos determinar a equação diferencial que rege o comportamento da luz e a sua dependência com fator *R*. Como cada variedade tem uma natureza geométrica única, a forma da luz também deverá ser induzida por *R*. Como foi previsto por Maxwell e confirmado por Hertz, as luz se comporta como uma onda eletromagnética que satisfaz a equação de D'Alambert:

$$\nabla^2\varphi - \frac{\partial^2\varphi}{\partial^2 t} = 0$$

Desta forma: podemos afirmar que se $\varphi$ é um ente observável associado à radiação eletromagnética, então este deve satisfazer a seguinte relação:

$$\Delta\varphi = 0$$

onde $\Delta$ é o operador laplaciano generalizado.

Assim, nosso objetivo será determinar um laplaciano geral para, então, impormos que o fator *R* seja tal que o laplaciano generalizado corresponda ao operador d'alambertiano.

Tomemos o vetor nabla generalizado:

$$\nabla^i = \left(\frac{\partial}{\partial x}, \frac{\partial}{\partial y}, \frac{\partial}{\partial z}, \frac{\partial}{\partial t}\right)$$

Definimos o laplaciano generalizado, pela expressão:

$$\Delta = \sum_{i=1}^{4}\sum_{j=1}^{4} \nabla^i \eta_{ij} \nabla^j$$

Expandindo as duas somas,

$$\Delta = \nabla^1\eta_{11}\nabla^1 + \nabla^2\eta_{22}\nabla^2 + \nabla^3\eta_{33}\nabla^3 + \nabla^4\eta_{44}\nabla^4$$
$$\Delta = \nabla^1\nabla^1 + \nabla^2\nabla^2 + \nabla^3\nabla^3 - R^2\nabla^4\nabla^4$$
$$\Delta = \left(\nabla^1\right)^2 + \left(\nabla^2\right)^2 + \left(\nabla^3\right)^2 - R^2\left(\nabla^4\right)^2$$

Ou de forma compacta,

$$\Delta = \nabla^2 - R^2 \frac{\partial^2}{\partial t^2}$$

Aplicando o operador laplaciano generalizado sobre o potencial da radiação eletromagnética, teremos:

$$\Delta\varphi \equiv \nabla^2\varphi - R^2 \frac{\partial^2 \varphi}{\partial t^2}$$

Como dito anteriormente, cada unidade hipercomplexa irá designar uma forma para radiação eletromagnética, a saber:

**1) Se *R²* = 0**
**(Variedade de Galileu)**

$$\nabla^2\varphi = 0$$

**2) Se *R²* = *-1***
**(Variedade de Euclides)**

$$\nabla^2\varphi + \frac{\partial^2\varphi}{\partial t^2} = 0$$

**3) Se *R²* = *+1***
**(Variedade de Lorentz)**

$$\nabla^2\varphi - \frac{\partial^2\varphi}{\partial t^2} = 0$$

**Álgebra Geométrica Unificada do Espaço-Tempo**

Embora tenhamos definido um operador nabla capaz de unificar as equações de Maxwell-Heaviside para todas as variedades espaço-temporais, esse operador não permite construir potenciais para a variedade. Recentemente, descobrimos uma modificação das equações multivetoriais de Maxwell que preservam os resultados anteriores e permitem definir de forma unívoca uma teoria do potencial para a variedade. Vamos introduzir o multivetor gradiente da seguinte forma:

$$\hat{\nabla}(R) = \nabla + R\frac{1}{c}\frac{\partial}{\partial t}$$

Por meio dos multivetores gradientes, **as quatro equações de Maxwell-Heaviside** podem ser escritas como **uma única equação**:

$$\hat{\nabla}(R)\left(\vec{E} + \hat{B}\right) = 4\pi\left(\rho - R\frac{\vec{J}}{c}\right)$$

onde $\vec{E}$, é o 1-vetor campo elétrico, $\hat{B}$, o 2-vetor campo magnético, $\rho$ é o escalar densidade de carga, $\vec{J}$ é o 1-vetor corrente de deslocamento e $c$ é a velocidade da luz no vácuo.

Para realizarmos a demonstração faremos a substituição direta e, por fim, faremos a comparação das quantidades escalares, pseudo-escalares, vetoriais e pseudo-vetoriais. Inicialmente, escrevamos o multivetor gradiente por extenso:

$$\left(\nabla + R\frac{1}{c}\frac{\partial}{\partial t}\right)\left(\vec{E} + R\widehat{B}\right) = 4\pi\left(\rho - R\frac{\vec{J}}{c}\right)$$

Distribuindo os operadores do lado direito:

$$\nabla\vec{E} + R\frac{1}{c}\frac{\partial\vec{E}}{\partial t} + R\nabla\widehat{B} + R^2\frac{1}{c}\frac{\partial\widehat{B}}{\partial t} = 4\pi\left(\rho - R\frac{\vec{J}}{c}\right)$$

Aplicando os produtos de Clifford e as relações com os pseudo-escalares, obtemos:

$$\text{div}\vec{E} + I\left(\text{rot}\,\vec{E}\right) + R\frac{1}{c}\frac{\partial\vec{E}}{\partial t} + IR\left(\text{div}\vec{B}\right) - R\text{rot}\,\vec{B} + I\left(R^2\frac{1}{c}\frac{\partial\vec{B}}{\partial t}\right) = 4\pi\left(\rho - R\frac{\vec{J}}{c}\right)$$

Pondo a equação em blocos ordenados: escalar, pseudo-escalar, vetor, pseudo-vetor:

$$\underbrace{\left[\text{div}\vec{E}\right]}_{\textbf{escalar}} + I\underbrace{\left[R\left(\text{div}\vec{B}\right)\right]}_{\textbf{pseudo - escalar}} + R\underbrace{\left[\frac{1}{c}\frac{\partial\vec{E}}{\partial t} - \text{rot}\,\vec{B}\right]}_{\textbf{vetor}} + I\underbrace{\left[\left(\text{rot}\,\vec{E}\right) + R^2\frac{1}{c}\frac{\partial\vec{B}}{\partial t}\right]}_{\textbf{pseudo - vetor}} = 4\pi\left(\underbrace{\rho}_{\textbf{escalar}} - \underbrace{R\frac{\vec{J}}{c}}_{\textbf{vetor}}\right)$$

Igualando cada bloco, conforme sua natureza algébrica:

$$\text{div}\vec{E} = 4\pi\rho \qquad \text{div}\vec{B} = 0$$

$$\text{rot}\,\vec{B} = \frac{4\pi}{c}\vec{J} + \frac{1}{c}\frac{\partial\vec{E}}{\partial t} \qquad \text{rot}\,\vec{E} = -R^2\frac{1}{c}\frac{\partial\vec{B}}{\partial t}$$

que são as 4 equações generalizadas de Maxwell-Heaviside (Q.E.D.).

Agora vamos introduzir a involução (ou conjugado) do operador nabla generalizado:

$$\widehat{\nabla}\left(R^*\right) \equiv \widehat{\nabla}\left(R\right)^* = \nabla - R\frac{1}{c}\frac{\partial}{\partial t}$$

Multiplicando o multivetor nabla pelo seu conjugado, obtemos a expressão do laplaciano generalizado:

$$\Delta\left(R\right) \equiv \widehat{\nabla}\left(R\right)\widehat{\nabla}\left(R^*\right) = \nabla^2 - R^2\frac{1}{c^2}\frac{\partial^2}{\partial t^2}$$

Então para um k-vetor $\varphi_k$, o potencial desse k-vetor será dado por:

**1) Se $R^2 = 0$**
**(Variedade de Galileu)**

$$\nabla^2\varphi_k = 0$$

**2) Se $R^2 = -1$**
**(Variedade de Euclides)**

$$\nabla^2\varphi_k + \frac{\partial^2\varphi_k}{\partial t^2} = 0$$

**3) Se $R^2 = +1$**
**(Variedade de Lorentz)**

$$\nabla^2\varphi_k - \frac{\partial^2\varphi_k}{\partial t^2} = 0$$

que concorda com os resultados que obtivemos anteriormente

**Determinação Empírica do Espaço-Tempo**

A nossa abordagem, porém, tem a vantagem de permitir por meio dos fenômenos eletromagnéticos identificar qual variedade plana se adequa a descrição dos fenômenos físicos. Isso ocorre porque as linhas coordenadas da variedade correspondem as componentes do tensor eletromagnético. Portanto, os fenômenos eletromagnéticos, em particular, os fenômenos elétricos, são propriedades intrínsecas da variedade.

De nossa análise da teoria do eletromagnetismo, obtivemos ao menos duas formas de identificar a variedade, a saber:

**1. A rotacionalidade do campo elétrico.**

Na variedade euclidiana, o campo elétrico é irrotacional, portanto o fenômeno de indução elétrica não pode ser observado. Por outro lado, na variedade euclidiana, um fluxo magnético variável induz uma corrente elétrico, mas no sentido inverso da Lei de Lenz. Somente a variedade lorentziana prevê uma indução elétrica que satisfaz a lei de Lenz.

Historicamente, Emil Lenz estabeleceu essa lei qualitativa em 1834 a partir observações empíricas das correntes induzidas por fluxos magnéticos variáveis. Desta forma, mesmo antes da formulação da Teoria da Relatividade Especial, já podíamos determinar o tipo de variedade que melhor corresponde a uma região infinitesimal do espaço-tempo, sem precisar recorrer ao segundo postulado, a constância da velocidade da luz, ou ao argumento de inteligibilidade de Minkowski. Registre que não para identificação da variedade não é preciso estabelecer a intensidade da corrente induzida, apenas a orientação. A lei de Lenz é suficiente.

**2. A Forma da luz.**

Das três variedades, a única que a luz apresenta a forma de uma onda esférica que oscila no vácuo (ou no éter), em concordância com as experiências de Hertz, é a variedade lorentziana. Desta forma, a experiência nos conduz, ao menos por enquanto, a rejeitar as variedades de Galileu e Euclides. Como as experiências de Hertz datam do século XIX, e eram aceitas, sem restrições, no começo do século XX, se Einstein tivesse seguido essa abordagem, ele poderia ter construído uma relatividade com embasamento mais sólido e recorrendo a um único princípio norteador. Observe que a partir do estudo da variedade de Lorentz, induzida pela unidade perplexa, podemos deduzir como teoremas a invariância da velocidade da luz e a constância da velocidade da luz. Desta forma, as experiência de Quirino Majorana realizadas em 1919, com fontes de radiação em alto movimento, se tornam testes experimentais que confirmam uma das previsões da teoria e aumentam seu conteúdo empírico.

**Orientação do Tempo e a Entropia**

O C-Grupo de Poincaré permite compreender a orientação do tempo em qualquer variedade do tipo espaço-tempo plana. No espaço-tempo de Galileu ($G^{3+0}$), cuja variedade tem característica anelar nilpotente, não podemos definir a componente zero da transformação generalizada de Lorentz, pois para essa variedade, verifica-se que:

$$0 = \frac{\left( \left[ \Lambda^{\varepsilon} \right]_0^{\mu} \left[ \Lambda^{\varepsilon} \right]_0^{\nu} \right)}{\left\{ \left( \left[ \Lambda^{\varepsilon} \right]_0^{0} \right)^2 - 1 \right\}}$$

Portanto, não existe uma orientação do tempo no espaço-tempo de Galileu. Essa é razão para as equações da mecânica serem preservadas tanto no sentido futuro do tempo quanto no sentido passado. A variedade de Galileu é simétrica no tempo.

No espaço-tempo de Euclides ($E^{3-1}$), cuja variedade tem característica anelar imaginária, se a componente zero da matriz generalizada de Lorentz for maior que a unidade, temos um tempo negativo, portanto um eixo fechado, orientado no sentido anti-horário. Caso a componente zero menor que a unidade, o caráter anticrônico faz com que o tempo esteja orientado no sentido horário. Por derradeiro, no espaço-tempo de Lorentz ($M^{3+1}$), cuja variedade tem característica anelar perplexa, se a componente zero da matriz generalizada de Lorentz for maior que a unidade, temos um tempo positivo, portanto um eixo aberto, orientado no sentido crescente (futuro). Caso a componente zero menor que a unidade, o caráter anticrônico faz com que o tempo esteja orientado no sentido decrescente (passado). Tanto no espaço-tempo de Euclides quanto no de Lorentz há uma antissimetria no tempo, determinado pela componente zero da matriz de Lorentz.

O formalismo adotado nesse trabalho permite explorar a relação entre o tempo e a entropia. Se determinarmos que a variação da entropia é uma função da componente zero da Matriz de Lorentz, mesmo em um universo cíclico (euclidiano), a entropia continua crescente na fase de retorno, pois a componente zero apenas determina o sentido de rotação do tempo. Assim, podemos escrever que:

$$Se \ \left( \Lambda^R \right)_0^0 \geq 1 \quad \rightarrow \quad dS \geq 0$$

$$Se \ \left( \Lambda^R \right)_0^0 < 1 \quad \rightarrow \quad dS < 0$$

Para demonstrar essa relação, recordemos que na formulação geral, temos a seguinte correspondência:

$$\gamma \mapsto \left( \Lambda^R \right)_0^0 \qquad c \mapsto k(R)$$

Sendo a segunda lei da Termodinâmica é um invariante relativístico (MARTINS, 2012), para tornar nossas equações covariantes precisamos realizar uma pequena alteração na primeira Lei da Termodinâmica.

$$dE = KdQ - dW$$

onde $K$ é uma constante adimensional a ser determinada. Para uma variedade do tipo espaço-tempo, a transformação da energia será dado por:

$$dE = \left|\left(\Lambda^R\right)_0^0\right| \left[\frac{v^2}{k^2} d\left(P_o V_o\right) + dE_o\right]$$

Para deduzir a transformação relativística do trabalho termodinâmico, temos que considerar que a velocidade de uma haste rígida em seu referencial próprio não varia, embora seu momento $G$ sofra um aumento. O diferencial da equação do momento da barra será dada por:

$$-dW = -PdV + \frac{d\vec{G}}{dt} d\vec{r}$$

Pela convenção adotada, como há entrada de energia na barra, o trabalho deve ser negativo para que a variação da energia seja positiva. A força aplicada sobre a barra tende a reduzir seu volume, portanto o volume final tende a ser menor que o inicial.

$$dW = PdV - \frac{d\vec{G}}{dt} d\vec{r}$$

$$dW = PdV - d\vec{G} \frac{d\vec{r}}{dt}$$

$$dW = PdV - d\vec{G} \cdot \vec{v}$$

Usando a relação entalpia-momento (MARTINS, 2012),

$$d\vec{G} = \frac{dH}{k^2} \vec{v}$$

Substituindo esse resultado na relação do trabalho:

$$dW = PdV - \frac{dH}{k^2} v^2$$

$$dW = PdV - \frac{v^2}{k^2} dH$$

$$dW = PdV - \frac{v^2}{k^2} d\left(PV + E\right)$$

$$dW = \frac{P_o dV_o}{\left|\left(\Lambda^R\right)_0^0\right|} - \left|\left(\Lambda^R\right)_0^0\right| \frac{v^2}{k^2} d\left(P_o V_o + E_o\right)$$

Portanto a transformação do trabalho termodinâmico será:

$$dW = \frac{P_o dV_o}{\left|\left(\Lambda^R\right)_0^0\right|} - \left|\left(\Lambda^R\right)_0^0\right| \frac{v^2}{k^2} d\left(P_o V_o\right) - dE_o$$

$$dW = PdV - \frac{v^2}{k^2} d\left(PV\right) + E$$

Agora podemos determinar a transformação do calor. Da primeira lei da termodinâmica podemos escrever o diferencial do calor como:

$$KdQ = dE + dW$$

Substituindo os diferenciais de energia e trabalho, obtemos:

$$KdQ = \frac{P_o dV_o}{\left|\left(\Lambda^R\right)_0^0\right|} - \left|\left(\Lambda^R\right)_0^0\right| \frac{v^2}{k^2} d\left(P_o V_o\right)$$

$$-\left|\left(\Lambda^R\right)_0^0\right| \frac{v^2}{k^2} dE_o + \left|\left(\Lambda^R\right)_0^0\right| \left[\frac{v^2}{k^2} d\left(P_o V_o\right) + dE_o\right]$$

$$KdQ = \frac{P_o dV_o}{\left|\left(\Lambda^R\right)_0^0\right|} + \left|\left(\Lambda^R\right)_0^0\right| dE_o - \left|\left(\Lambda^R\right)_0^0\right| \frac{v^2}{k^2} dE_o$$

$$KdQ = \frac{P_o dV_o}{\left|\left(\Lambda^R\right)_0^0\right|} + \left|\left(\Lambda^R\right)_0^0\right| dE_o \left(1 - \frac{v^2}{k^2}\right)$$

$$KdQ = \frac{1}{\left|\left(\Lambda^R\right)_0^0\right|} \left(P_o dV_o + dE_o\right)$$

O termo em parêntesis é a o calor no referencial próprio, portanto o calor se transforma como:

$$KdQ = \frac{dQ_o}{\left|\left(\Lambda^R\right)_0^0\right|}$$

Como a variação da entropia é um invariante relativístico, a desigualdade de Clausius pode ser escrita da seguinte forma:

$$dS \geq \int \frac{dQ}{T}$$

$$dS_o \geq \int \frac{dQ_o}{\left|\left(\Lambda^R\right)_0^0\right| KT_o}$$

Aqui há uma questão conceitual importante envolvendo o fator K, a saber: a escolha de K define se a temperatura se transforma conforme Planck, Ott, Avramov ou conforme outras relações.

a) *Se assumirmos que K é igual a unidade, a temperatura se transforma de acordo com a análise de Planck (1907).*

$$K = 1 \quad \rightarrow \quad dQ = \frac{dQ_o}{\left|\left(\Lambda^R\right)_0^0\right|} \quad \rightarrow \quad T = \frac{T_o}{\left|\left(\Lambda^R\right)_0^0\right|}$$

b) *Se assumirmos que K é o inverso da componente zero da matriz de Lorentz, a temperatura é um invariante relativístico, como sugere o físico russo I. Avramov (2003).*

$$K = \frac{1}{\left|\left(\Lambda^R\right)_0^0\right|} \quad \rightarrow \quad dQ = dQ_o \quad \rightarrow \quad T = T_o$$

c) *Se assumirmos que K é o inverso ao quadrado da componente zero da matriz de Lorentz, a se transforma de acordo com a análise de Ott.*

$$K = \frac{1}{\left[\left|\left(\Lambda^R\right)_0^0\right|\right]^2} \quad \rightarrow \quad dQ = \left|\left(\Lambda^R\right)_0^0\right| dQ_o \quad \rightarrow \quad T = \left|\left(\Lambda^R\right)_0^0\right| T_o$$

Observe que na formulação de Ott, o calor deve se transformar com a mesma lei que obtivemos para a energia.

d) *Para o caso mais geral, teremos que:*

$$K = \left|\left(\Lambda^R\right)_0^0\right|^n \quad \rightarrow \quad dQ = \frac{dQ_o}{\left|\left(\Lambda^R\right)_0^0\right|^{n+1}} \quad \rightarrow \quad T = \frac{T_o}{\left|\left(\Lambda^R\right)_0^0\right|^{n+1}}$$

Usualmente, assumimos as transformações de Ott como verdadeiras, portanto, a desigualdade Clausius será:

$$dS_o \geq \int \left|\left(\Lambda^R\right)_0^0\right| \frac{dQ_o}{T_o}$$

Da desigualdade de $\left|\left(\Lambda^R\right)_0^0\right|$, deduzimos que:

$$\text{Se } \left|\left(\Lambda^R\right)_0^0\right| \geq 1, \text{ então } \; dS_o \geq \int \frac{dQ_o}{T_o}$$

$$\text{Se } \left|\left(\Lambda^R\right)_0^0\right| < 1, \text{ então } \; dS_o < \int \frac{dQ_o}{T_o}$$

No espaço-tempo de Galileu como não podemos determinar a componente zero da matriz de Lorentz, não existe uma justificativa física para relacionarmos a orientação do tempo com a entropia.

# Cálculo-K Generalizado

Nessa seção apresentaremos uma síntese das ideias de Bondi, seguindo a abordagem a apresentada por David Bohm (2015, p. 175-190) e generaliza-la para variedades espaço-temporais planas arbitrárias, por meio das funções de Poincaré. Para tornar mais simples as deduções, usaremos os diagramas convencionais de Minkowski, porém os resultados são válidas para espaços euclidianos e galielanos, pois a variedade de Lorentz é homeomórfica as variedades de Galileu e Euclides. Este homeomorfismo será definido como uma aplicação linear que preserva as coordenadas *t* e *x,* mas transforma as transformações de Lorentz em funções de Poincaré:

$$R : \mathbb{H} \to \mathfrak{R}$$
$$R\left(xP_{+}^{p}(\theta) \pm ktP_{-}^{p}(\theta)\right) = x^{\mu}P_{+}^{R}(\theta) \pm ktP_{-}^{R}(\theta)$$
$$R\left(ktP_{+}^{p}(\theta) \pm xP_{-}^{p}(\theta)\right) = ktP_{+}^{R}(\theta) \pm xRP_{-}^{R}(\theta)$$

Vamos construir um diagrama de Minkowski. Tomemos dois segmentos de reta ortogonais OA e OB que representam, respectivamente, o eixo *ct* e o eixo *x.* Cada ponto nesse diagrama representa um evento que é representado por suas coordenadas especiais e temporal. Para um observador estacionário *S'*, todos os eventos se encontram na linha AO, que denominamos de linha de mundo de *S'.* A linha OB representa todos os fenômenos simultâneos ao observador *S'.* Suponha que no evento O seja disparado uma onda esférica luminosa de raio *ct.* Para o observador a posição desse raio no eixo OB, devido ao princípio da isotropia, a linha de mundo desse raio deverá ser descrito, pela seguinte função: $x = \pm ct,$ que correspondem, respectivamente, aos eixos OC (+*ct*) e OD (–*ct*). Para obtermos a inclinação da reta, tomemos o arco-tangente das retas OC e OA:

$$\delta = \arctan\left(\frac{OC}{OA}\right)$$
$$\delta = \arctan\left(\frac{ct}{ct}\right)$$
$$\delta = \pi/4 \quad (45^{\circ})$$

Portanto os raios OC e OD formam ângulos de 45 graus com os eixos OA e OB. Como observa Bohm (2015, p. 177) "é claro que em três dimensões há muitas direções possíveis para um raio de luz, de modo que todo o conjunto de raios de luz através de O é representado por um cone. As linhas OC e OD correspondem então à intersecção deste "cone de luz" com o plano x-ct." Vamos supor um observador *S* se desloca com velocidade constante *v* em relação ao observador *S'.* Do ponto de vista geométrico, o observador *S* equivale a uma rotação hiperbólica dos eixos OA e OB com um ângulo □. Se denotarmos por OE e por OF os eixos *ct'* e *x'*, respectivamente, o diagrama de Minkowski, na perspectiva de *S',* apresentará a seguinte representação:

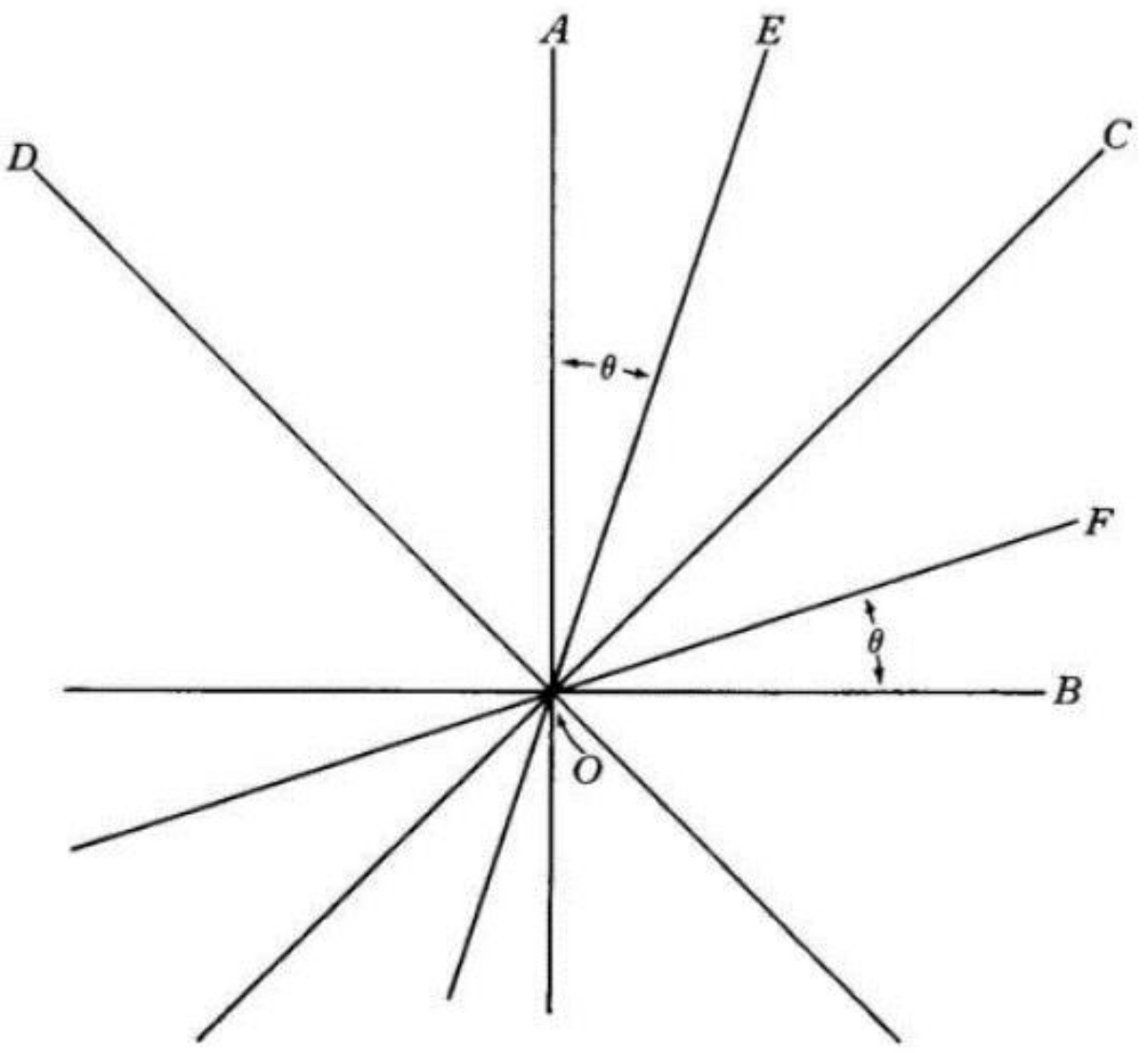

E as transformações será dada por:

$$OE = OAP_{+}^{R}(\theta) - OB\left(RP_{-}^{R}(\theta)\right)$$
$$OF = OBP_{+}^{R}(\theta) - OAP_{-}^{R}(\theta)$$

Se um evento for simultâneo no referencial *S* isso implica que o intervalo *OE* deve ser nulo.

$$0 = OAP_{+}^{R}(\theta) - OBP_{-}^{R}(\theta)$$
$$OAP_{+}^{R}(\theta) = OBR^{2}P_{-}^{R}(\theta)$$
$$OA = OBP_{\odot}^{R}(\theta)$$

Portanto os eventos simultâneos de *S* se localizam na reta *OF* e por isso no referencial *S’,* estes eventos não serão simultâneos. Se tomarmos a perspectiva do referencial *S’,* o diagrama de Minkowski assume o seguinte aspecto:

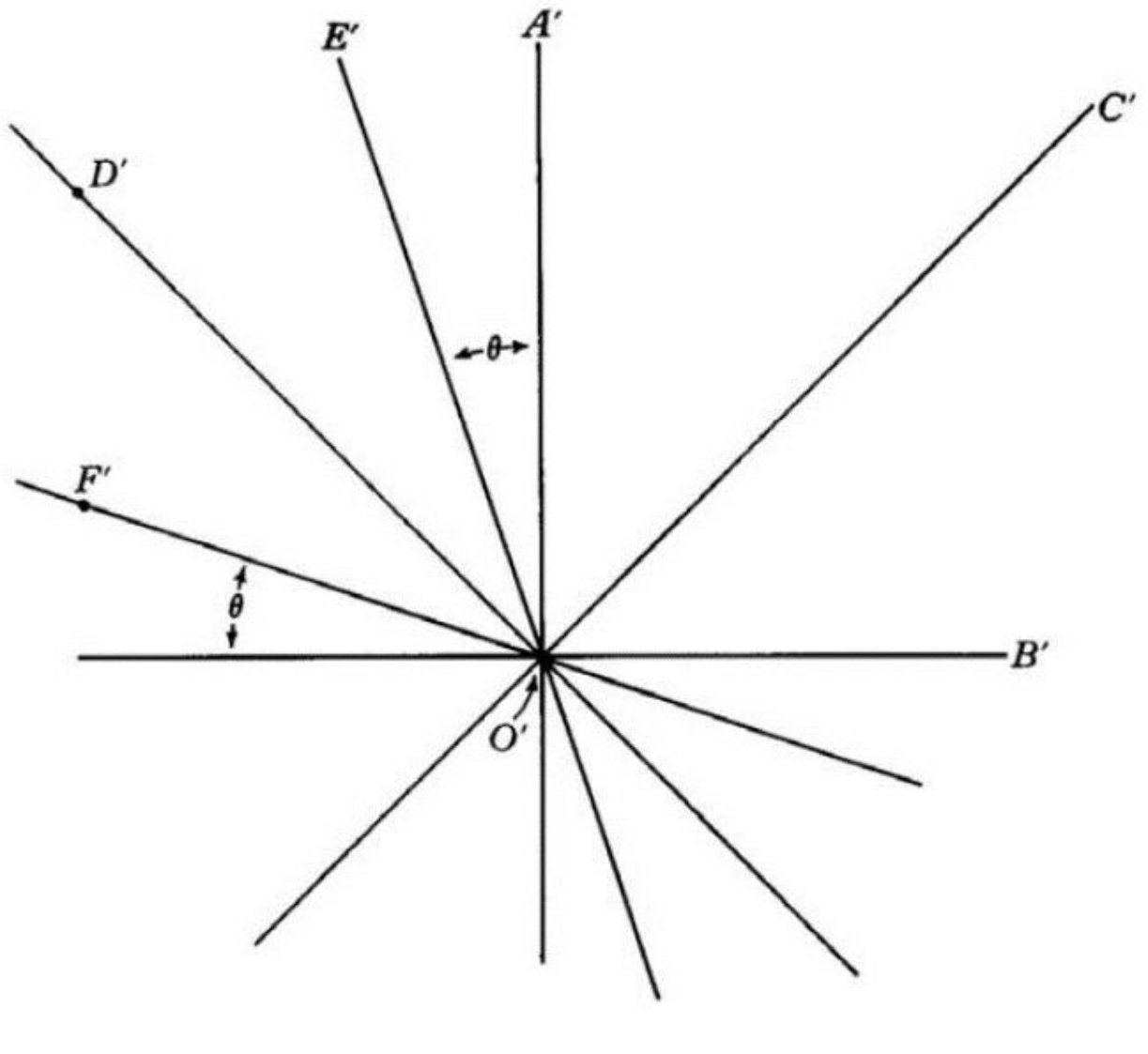

Suponha que os observadores *S* e *S'* portam relógios idênticos e síncronos. Vamos supor que em intervalos constantes, o observador estacionário *S'* envia sinais $N_1$, $N_2$, ..., $N_n$ para o observador *S*. Estes sinais viajam à velocidade da luz e alcançam o observador *S* nos eventos $N'_1$, $N'_2$, ..., $N'_n$.

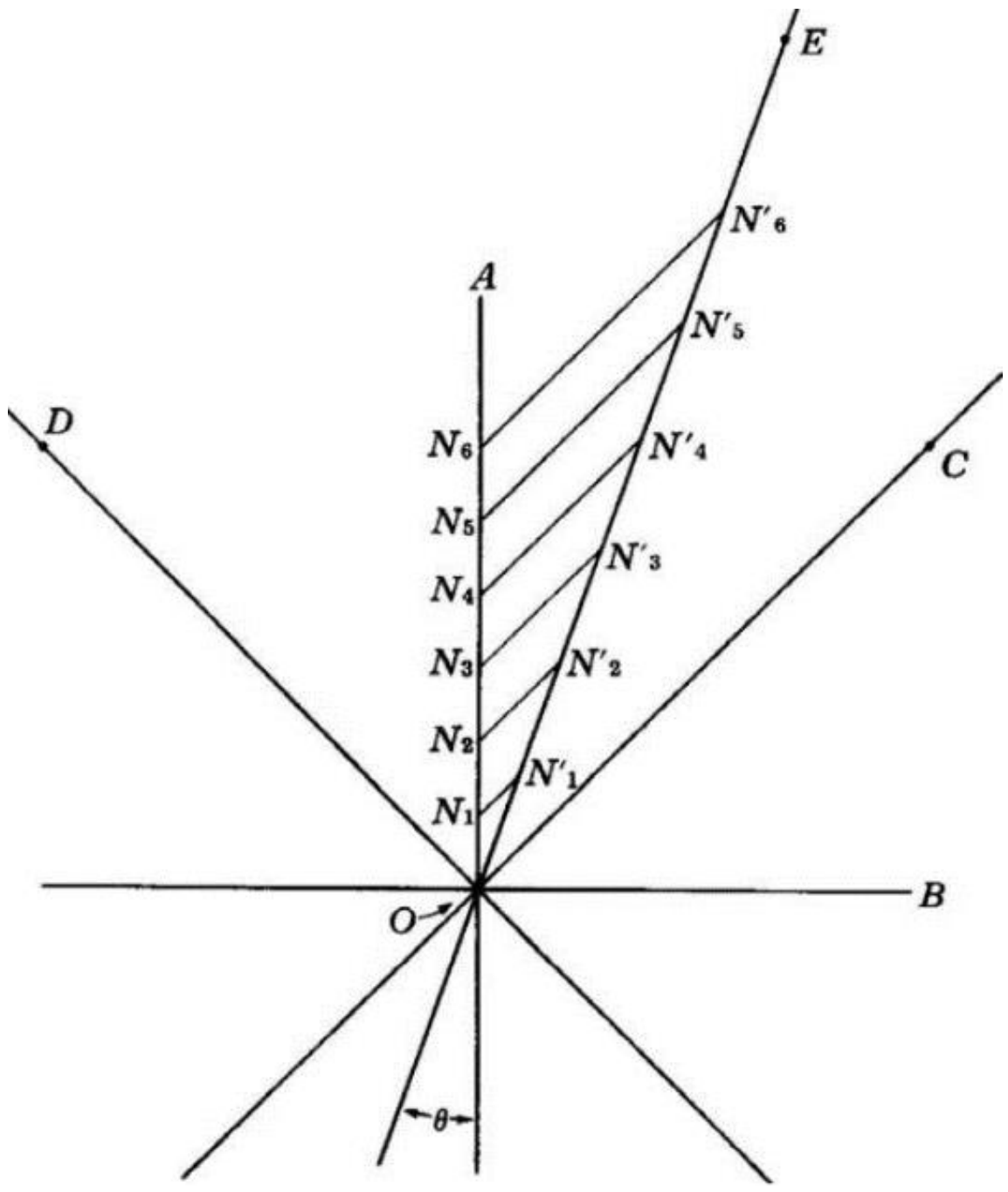

Se o observador *S'* envia sinais em intervalos regulares $T_o$, o observador *S* receberá estes sinais em intervalos *T* devido ao efeito Doppler-Fizeau. Como já observamos, essa é uma consequência da própria natureza ondulatória da luz e não do princípio da relatividade. De qualquer forma, podemos definir uma constante *K* que é a razão entre os dois períodos.

$$K = \frac{T}{T_o}$$

Suponha que o pulso é recebido pelo observador em *S,* ele é imediatamente refletido para o observador *S'.* Assim, podemos dizer que o referencial *S* emite sinais $M_1$, $M_2$, ..., $M_n$ em intervalos regulares $T_o$ e que são recebidos em $M'_1$, $M'_2$, ..., $M'_n$ em intervalos *T'.* Para este referencial podemos definir uma constante *K,*

$$K' = \frac{T'}{T_o}$$

Observe que entre os eventos $N_i$ e $N'_i$ e os eventos $M_j$ e $M'_j$, traçamos linhas $N_iN'_i$ e $M_jM'_j$. Como estas linhas representam as linhas de mundo de raios luminosos trocados entre os referenciais *S'* e *S,* as linhas $N_iN'$ devem ser paralelas ao eixo OC e as linhas $M_jM'_j$, paralelas a OD. Bohm (2015, p. 180) assinala que: “os caminhos dos sinais de rádio, com uma inclinação de 45°, indicam que em ambos os sistemas a velocidade da luz tem o mesmo valor, c. É assim que incorporamos no diagrama de Minkowski o fato observado de que a velocidade da luz é invariante, a mesma para todos os observadores.”.

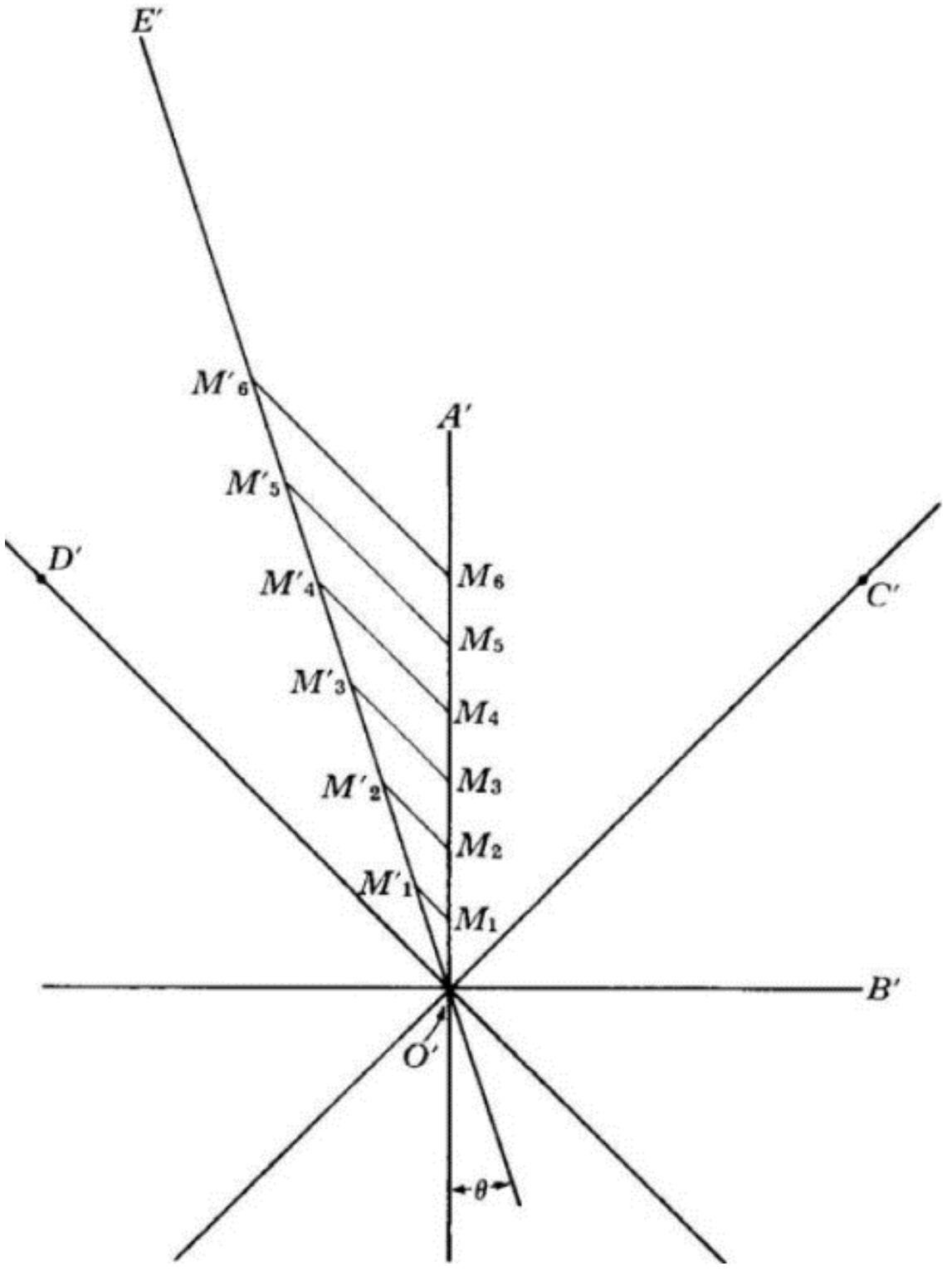

Se o espaço a propagação da velocidade da luz é isotrópica e não existe um referencial privilegiado, isto é, os referenciais *S'* e *S* são equivalentes, como impõe o princípio da relatividade, a razão dos períodos não deve depender do referencial adotado,

$$K = K'$$

Devemos nos lembrar, no entanto, que o exposto é verdadeiro apenas em uma teoria relativista, na qual a luz tem a mesma velocidade em cada sistema de referência. Assim, na mecânica newtoniana, os raios de luz seriam representados como linhas a 45 ° dos eixos apenas em um sistema em repouso no éter, de modo que o raciocínio pelo qual mostramos a igualdade de K e K' não seria insustentável. (BOHM, 2015, p. 182).

Após essas considerações, vamos introduzir o cálculo K. Suponha que na posição O, os observadores em *S'* e *S* troquem sinais luminosos e sincronizem seus relógios. Como nessa posição, ambos ocupam o praticamente o mesmo espaço, a troca de sinais luminosos será praticamente instantânea. Nesse momento, os observadores ajustam seus relógios para marcar o tempo zero.

$$t = t' = 0$$

No instante $T_o$, que corresponde ao evento *N,* o observador *S'* emite um sinal para o observador em *S.* Esse sinal é recebido no tempo *T,* que corresponde à $T = KT_o$, no evento *N'.* O pulso é imediatamente refletido e atinge o observador em *S'* no instante $T_1$, que corresponde à $T_1 = KT$, no evento *N''*. Substituindo o valor de *T*, obteremos: $T_1 = KT^2_o$.

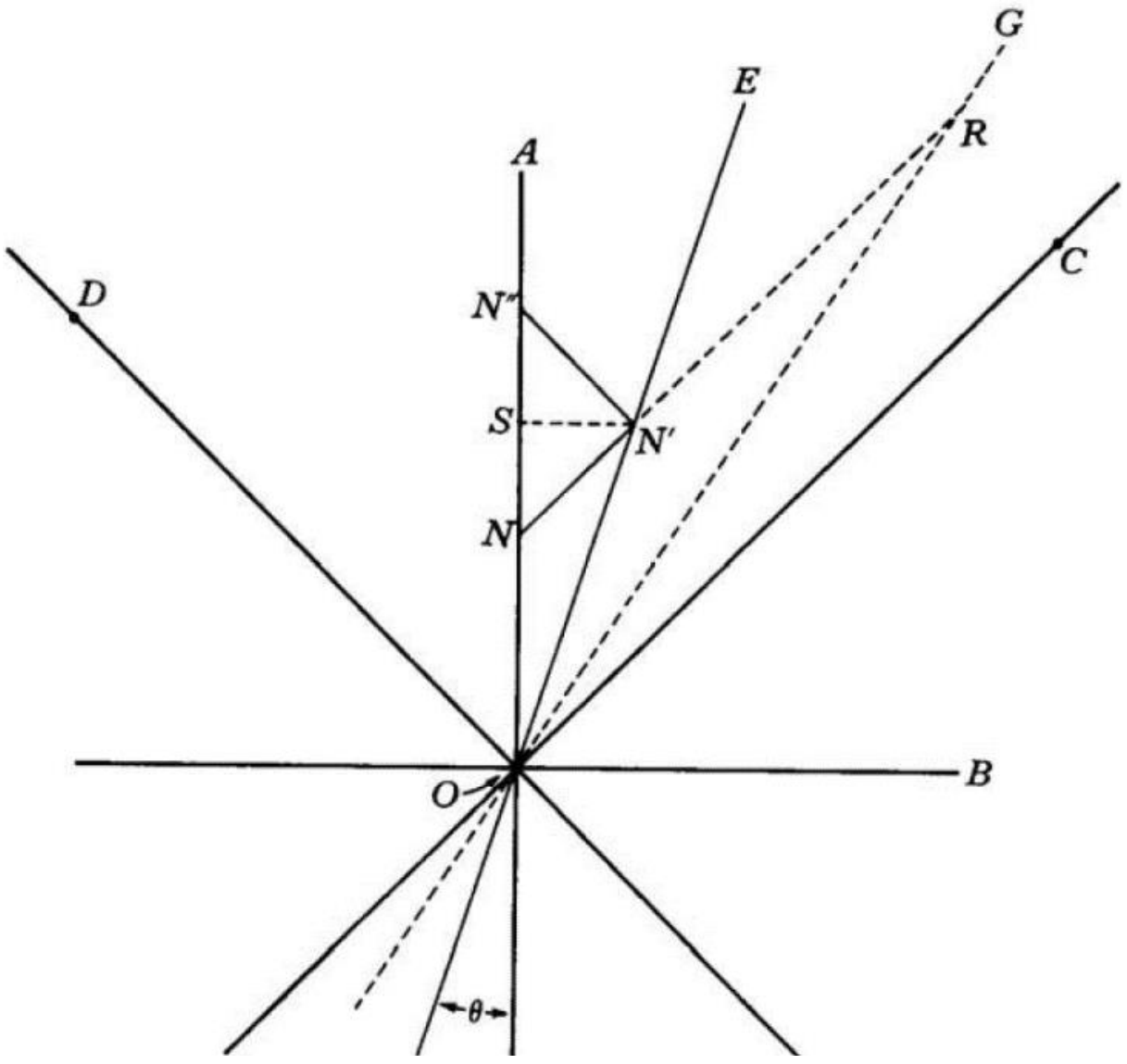

Observe que no diagrama de Minkowski, o evento *S* corresponde ao ponto médio da linha NN''. As linhas N'N'' e NN' formam um ângulo de 45º com a linha SN'. As linhas SN e SN'' formam um ângulo de 90º com a linha SN'. Isso implica que os triângulos SNN' e SN'N'' são isócesles. Portanto, a medida de NN' e de SN' e SN'' e N'N'' são iguais. Nestas condições, podemos escrever as seguintes relações:

$$SN = SN' = SN'' = \frac{NN''}{2}$$

Do triângulo retângulo OSN*'*, podemos concluir que o ângulo entre as linhas ON' e OS é □. As retas SN' e OS se relacionam pela tangente de Poincaré desse ângulo (registre que estamos em um "plano hipercomplexo").

$$SN = SN' = OSP^{R}_{\odot}\left(\theta\right) = \frac{NN''}{2}$$

É imediato que o seguimento OS pode ser escrito como a soma de suas partes:

$$OS = ON + NS = ON + \frac{NN''}{2}$$

O evento *N* corresponde a emissão do sinal em $T_o$. Portanto, o período entre a sincronização dos relógios e a emissão do sinal por *S'*, será:

$$ON = T_o$$

De forma equivalente, o período entre a sincronização dos relógios e a emissão do sinal pelo observador *S*, será:

$$ON' = T$$

A diferença entre a emissão e o retorno do sinal em *S'*, $T_1 - T_o$, será o intervalo NN”:

$$NN'' = T_1 - T_o$$
$$NN'' = \left(K^2 - 1\right)T_o$$

Usando as duas equações envolvendo OS, podemos determinar o valor de *K*.

$$OS = ON + \frac{NN''}{2}$$

$$OSP_{\odot}^{R}\left(\theta\right) = \frac{NN''}{2}$$

Multiplicando a primeira equação pela tangente de Poincaré,

$$OSP_{\odot}^{R}\left(\theta\right) = \left(ON + \frac{NN''}{2}\right)P_{\odot}^{R}\left(\theta\right)$$

Substituindo esse valor na segunda equação:

$$\frac{NN''}{2} = \left(ON + \frac{NN''}{2}\right)P_{\odot}^{R}\left(\theta\right)$$

Isolando ON, obtemos a relação:

$$ON = \frac{\left(1 - P_{\odot}^{R}\left(\theta\right)\right)NN''}{2P_{\odot}^{R}\left(\theta\right)}$$

Substituindo os valores dos segmentos,

$$T_o = \frac{\left(1 - P_{\odot}^{R}\left(\theta\right)\right)\left(K^2 - 1\right)T_o}{2P_{\odot}^{R}\left(\theta\right)}$$
$$\left(1 - P_{\odot}^{R}\left(\theta\right)\right)\left(K^2 - 1\right) = 2P_{\odot}^{R}\left(\theta\right)$$
$$K^2 - K^2P_{\odot}^{R}\left(\theta\right) - 1 + P_{\odot}^{R}\left(\theta\right) = 2P_{\odot}^{R}\left(\theta\right)$$
$$K^2\left(1 - P_{\odot}^{R}\left(\theta\right)\right) = 1 + P_{\odot}^{R}\left(\theta\right)$$
$$K^2 = \frac{1 + P_{\odot}^{R}\left(\theta\right)}{1 - P_{\odot}^{R}\left(\theta\right)}$$

Extraindo a raiz quadrada, concluímos o cálculo de *K:*

$$K = \sqrt{\frac{1 + P_{\odot}^{R}\left(\theta\right)}{1 - P_{\odot}^{R}\left(\theta\right)}}$$

A expressão acima pode ser escrita da seguinte forma:

$$K = \sqrt{\frac{k + Rv}{k - Rv}}$$

O fator *K* corresponde ao efeito Doppler relativístico. Isso não é nenhuma surpresa, visto que como o referencial *S* se desloca em relação à *S'* com velocidade constante, a constância da velocidade da luz impõe que os pulsos sofram uma transformação de suas frequências. Vamos usar o cálculo K para achar a transformação do período. A coordenada *t* corresponde ao seguimento OS.

$$OS = ON + NS = ON + \frac{NN''}{2}$$

$$t = T_o + \frac{\left(K^2 - 1\right)T_o}{2}$$

$$t = \frac{\left(K^2 + 1\right)T_o}{2}$$

No sistema *S,* o tempo corresponde ao eixo *ON'*

$$t' = ON' = T = KT_o$$

Dividindo as *t* por *t'*:

$$\frac{t}{t'} = \frac{\left(K^2 + 1\right)}{2K}$$

Substituindo o valor de $K^2$:

$$\frac{t}{t'} = \left(\frac{1 + P_{\odot}^{R}(\theta)}{1 - P_{\odot}^{R}(\theta)} + 1\right)\frac{1}{2K}$$

$$\frac{t}{t'} = \left(\frac{1}{1 - P_{\odot}^{R}(\theta)}\right)\sqrt{\frac{1 - P_{\odot}^{R}(\theta)}{1 + P_{\odot}^{R}(\theta)}}$$

$$\frac{t}{t'} = \sqrt{\frac{1}{\left(1 - P_{\odot}^{R2}(\theta)\right)}}$$

Usando as relações de Poincaré, obtemos: a fórmula da dilatação do tempo:

$$t = t'\sqrt{P_{+}^{R2}(\theta)}$$

$$t = t'P_{+}^{R}(\theta) = \Gamma t'$$

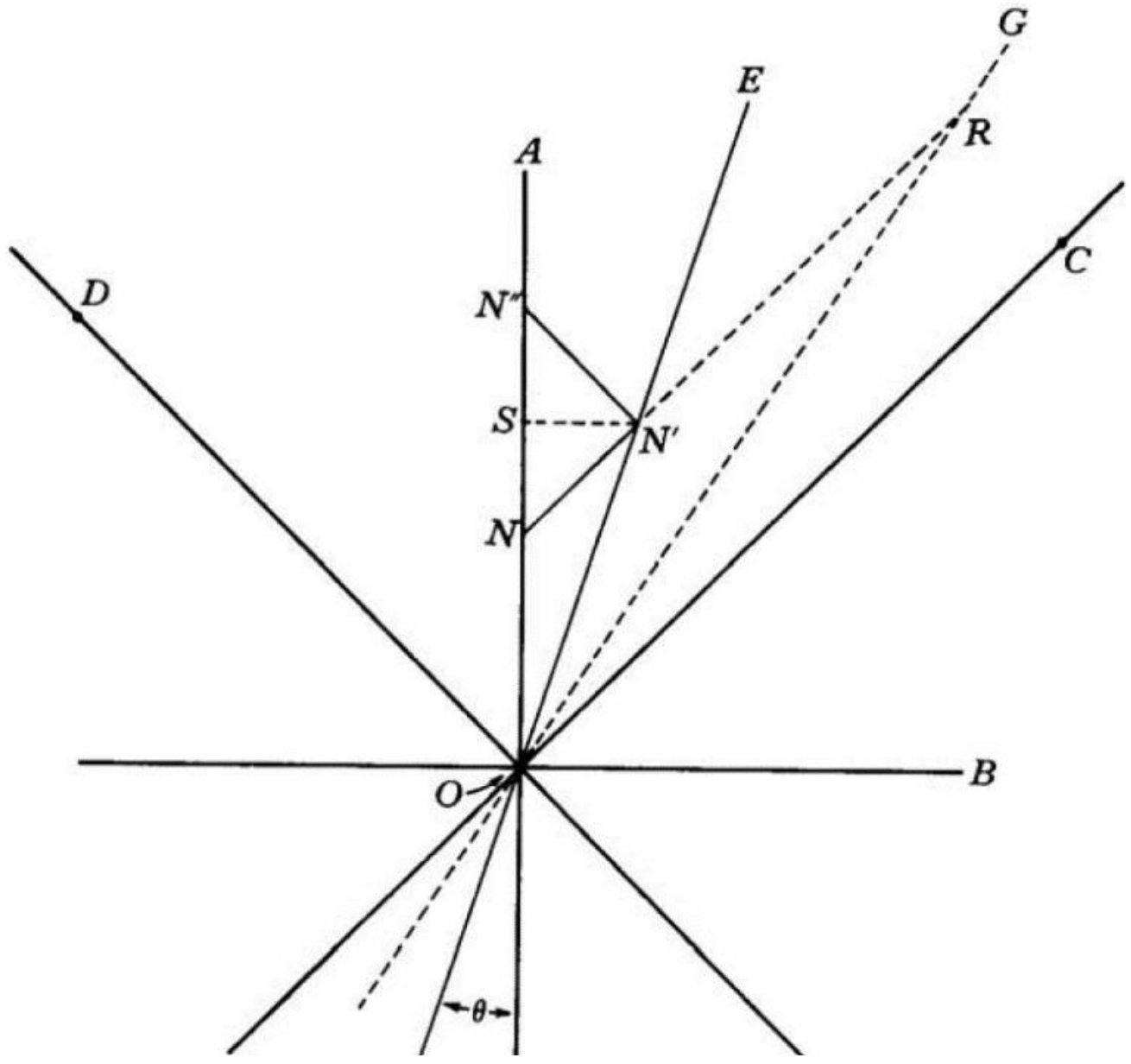

Agora estudaremos a composição das velocidades relativísticas usando fator *K*. Para isso vamos assumir a existência de um terceiro observador *S''* descrita pela linha de mundo OG e que se desloca em relação à *S'* com velocidade constante *w*. No instante $T_o$ o ocorre um evento N: o observador *S'* emite um sinal na direção do observador OG que é recebido no evento R no tempo $T_2$. Esses eventos se relacionam pela equação:

$$T_2 = K(w)T_o$$

Por outro lado, consideremos que o observador *S'* emita no evento *N* um sinal para o observador *S,* que desloca com velocidade constante *v.* Este sinal é recebido por *S*' no evento N'. Portanto, o tempo medido pelo observador *S,* será:

$$T_1 = K(v)T_o$$

Assim que o observador *S* recebe o sinal de *S',* no evento N', ele retransmite esse sinal para o observador *S'',* que se desloca com velocidade constante *u.* O sinal é recebido no instante $T_2$ e marca o evento *R.*

$$T_2 = K(u)T_1$$

Usando as três relações que obtivemos, podemos escrever as equações:

$$T_2 = K(w)T_o = K(u)T_1$$
$$K(w)T_o = K(u)K(v)T_o$$
$$K(w) = K(u)K(v)$$

Essa propriedade do cálculo *K* permite demonstrar que eles apresentam uma estrutura de grupo, assim como as transformações de Lorentz. Portanto, existe um importante grupo associado ao cálculo

K que é o grupo de dos fatores K ou grupo de Bondi. Sem mais delongas, voltemos ao cálculo da composição da velocidade:

$$K(w) = K(u)K(v)$$
$$K^2(w) = K^2(u)K^2(v)$$

Abrindo as funções *K* quadráticas e as tangentes hiperbólicas:

$$\left(\frac{1+P_{\odot}^{R}w}{1-P_{\odot}^{R}w}\right) = \left(\frac{1+P_{\odot}^{R}u}{1-P_{\odot}^{R}u}\right)\left(\frac{1+P_{\odot}^{R}v}{1-P_{\odot}^{R}v}\right)$$
$$\left(\frac{k+Rw}{k-Rw}\right) = \left(\frac{k+Ru}{k-Ru}\right)\left(\frac{k+Rv}{k-Rv}\right)$$
$$\left(\frac{k+Rw}{k-Rw}\right) = \left(\frac{k^2+uRk+vRk+R^2vu}{k^2-uRk-vRk+R^2vu}\right)$$

Vamos multiplicar os fatores em cruz para evidenciar a velocidade resultante *w.*

$$(k+Rw)\left(k^2-uRk-vRk+R^2vu\right) = (k-Rw)\left(k^2+uRk+vRk+R^2vu\right)$$
$$\left(k^3-uRk^2-vRk^2+vukR^2+wk^2R-uwkR^2-vwkR^2+vuwR^4\right) =$$
$$\left(k^3+uRk^2+vRk^2+vukR^2-wk^2R-uwkR^2-vwkR^2-vuwR^4\right)$$

Realizando as implicações algébricas, chegamos a equação:

$$2wRk^2 + 2wvuR^4 = 2uRk^2 + 2vRk^2$$
$$w\left(k^2+vuR^2\right)R^2 = (u+v)R^2k^2$$

$$w = \frac{(u+v)k^2}{\left(k^2+R^2vu\right)}$$

Evidenciando a velocidade da luz no denominador e simplificando com o numerador, obtemos a regra de composição de velocidades

$$w = \frac{u+v}{1+R^2\frac{vu}{k^2}}$$

Bohm (2015, p. 186-187), faz uma importante observação sobre processos de medida:

Como a velocidade da luz é a mesma para todos os observadores, não precisamos de padrões separados de tempo e distância. Por esta razão, é suficiente que todos os observadores tenham relógios equivalentemente construídos. Não é necessário assumir além disso que eles têm bastões de medida padrão. Isso torna as fundações lógicas do procedimento de medição muito simples, porque é possível usar os períodos de vibrações de átomos ou moléculas como relógios padrão, que podem depender de funcionar de maneira equivalente para todos os observadores.

Por fim, vamos deduzir as transformadas de Lorentz do tempo usando o método *K:* Para isso construiremos uma nova linha de mundo representado pela linha SP, que inicialmente se encontra fora do cone de luz, mas em um dado instante intercepta a linha OC e passa a fazer parte da região de vínculos casuais dos observadores *S* e *S'*. Em um instante $T_1$, o observador em *S'* inicia um evento *M. S'* emite um pulso para o observador *S,* que é recebido no evento *N*. Instantaneamente, o observador *S* emite um sinal para um observador *S"* que registra esse evento *P,* e reflete o sinal que atinge o *S* no evento *Q* e *S'* no evento *R,* no instante $T_2$.

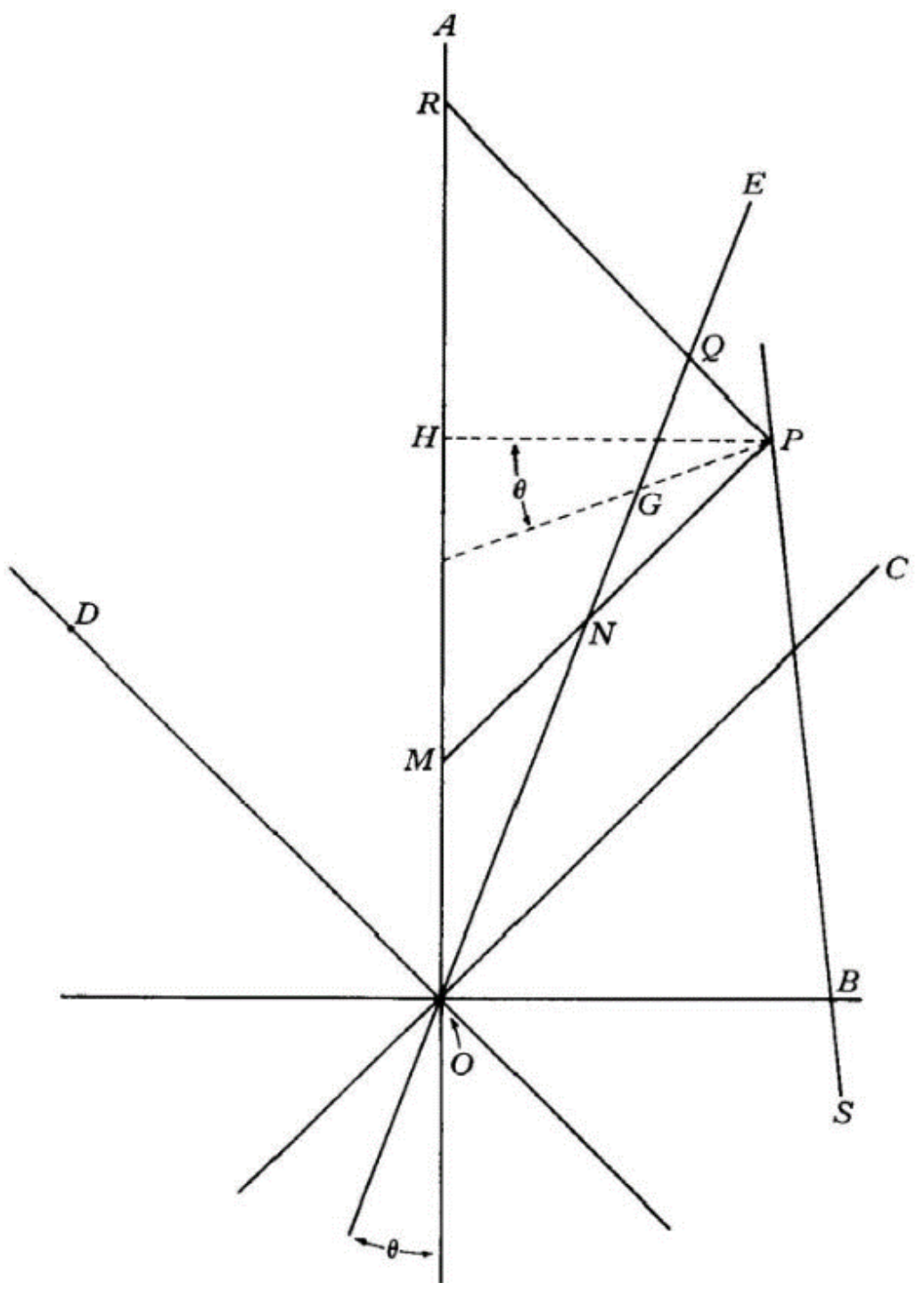

Pela simetria do problema, o seguimento MR corresponde a duração $T_2$. Porém esse seguimento é a soma dos seguimentos MP e PR. Porém, pelo princípio da reflexão, estes dois seguimentos devem ter o mesmo comprimento:

$$MR = MP + PR$$
$$MP = PR$$

$$MP = \frac{MR}{2}$$

O pulso é emitido no evento M, no tempo $T_1$ e retorna no instante $T_2$, portanto o seguimento MR tem "comprimento" $T_2 - T_1$

$$MR = T_2 - T_1 \qquad\qquad MP = \frac{T_2 - T_1}{2}$$

Queremos determinar em qual instante ocorre o evento *P,* segundo o observador no referencial *S'*. Pela geometria elementar, temos que:

$$MP = P - M$$
$$P = M + PM$$

O evento M ocorre no instante $T_1$, substituindo na equação:

$$P = T_1 + \frac{T_2 - T_1}{2}$$

$$P = \frac{T_2 + T_1}{2}$$

Se multiplicarmos o segmento *MP* por *k*, obtemos o "tempo próprio"*:*

$$\tau = k\frac{T_2 - T_1}{2}$$

Se multiplicarmos o ponto *P* por *k*, obtemos o "espaço próprio":

$$s = k\frac{T_2 + T_1}{2}$$

Portanto existe uma relação simples entre os períodos e as medidas de comprimento e tempo:

$$s - \tau = kT_1, \qquad s + \tau = kT_2$$

O princípio da relatividade nos impõe que as mesmas medidas devem ser realizadas pelo observador em *S*:

$$\tau' = k\frac{T_2' - T_1'}{2}, \qquad s' = k\frac{T_2' + T_1'}{2}$$
$$s' + \tau' = kT_2', \qquad s' - \tau' = kT_1'$$

mas, segundo o cálculo-*K,*

$$T_1' = \mathrm{K}T_1, \qquad T_2 = \mathrm{K}T_2'$$

que nos conduz a relação:

$$T_1T_2' = T_1T_2$$

Substituindo as relações entre os tempos e comprimentos:

$$\frac{(s' + \tau')(s' - \tau')}{k^2} = \frac{(s + \tau)(s' - \tau')}{k^2}$$

$$\left(\tau'^2 - s'^2\right) = \left(\tau^2 - s^2\right)$$

Essa é a forma "quadrática própria". As coordenadas próprias e locais de tempo e espaço se relacionam por meio das relações:

$$\begin{cases} s = x \\ R\tau = kt \end{cases}$$

Que é a forma quadrática do espaço-tempo. Das relações entre os dois sistemas inerciais, temos as seguintes relações:

$$T_1 = \frac{T_1'}{\mathrm{K}} \qquad\qquad T_2 = \mathrm{K}\, T_2'$$

Vamos agora obter a transformação de Lorentz, substituindo a relação *x*:

$$s = k\frac{KT_2' - \left(T_1'/K\right)}{2}$$

$$s = \frac{K^2\left(kT_2'\right) - \left(kT_1'\right)}{2K}$$

$$s = \frac{K^2\left(\tau' + s'\right) - \left(\tau' - s'\right)}{2K}$$

$$s = \frac{\left(K^2 - 1\right)\tau' + \left(K^2 + 1\right)s'}{2K}$$

Vamos calcular os o valor dos termos nos parêntesis:

$$\left(K^2 - 1\right) = \frac{1 + P_{\odot}^{R}(\theta)}{1 - P_{\odot}^{R}(\theta)} - 1 \qquad\qquad \left(K^2 + 1\right) = \frac{1 + P_{\odot}^{R}(\theta)}{1 - P_{\odot}^{R}(\theta)} + 1$$

$$\left(K^2 - 1\right) = \frac{2P_{\odot}^{R}(\theta)}{1 - P_{\odot}^{R}(\theta)} \qquad\qquad \left(K^2 + 1\right) = \frac{2}{1 - P_{\odot}^{R}(\theta)}$$

Substituindo os valores do fator *K:*

$$s = \frac{2s' + \left(2\tau' P_{\odot}^{R}(\theta)\right)}{2K\left(1 - P_{\odot}^{R}(\theta)\right)}$$

Agora vamos calcular o fator no denominador:

$$K\left(1 - P_{\odot}^{R}(\theta)\right) = \left(\sqrt{\frac{1 + P_{\odot}^{R}(\theta)}{1 - P_{\odot}^{R}(\theta)}}\right)\left(1 - P_{\odot}^{R}(\theta)\right)$$

$$K\left(1 - P_{\odot}^{R}(\theta)\right) = \left(\sqrt{\left(1 + P_{\odot}^{R}(\theta)\right)\left(1 - P_{\odot}^{R}(\theta)\right)}\right)$$

$$K\left(1-P_{\odot}^{R}(\theta)\right)=\left(\sqrt{\left(1-P_{\odot}^{R2}(\theta)\right)}\right)$$

$$K\left(1-P_{\odot}^{R}(\theta)\right)=\frac{1}{P_{+}^{R}(\theta)}$$

Substituindo na equação,

$$\begin{aligned}
&\tau=P_{+}^{R}(\theta)\left(\tau'+s'P_{\odot}^{R}(\theta)\right)\\
&R\tau=R\tau'P_{+}^{R}(\theta)+R^{2}s'P_{-}^{R}(\theta)\\
&kt=kt'P_{+}^{R}(\theta)+R^{2}x'P_{-}^{R}(\theta)\\
&t=t'P_{+}^{R}(\theta)+R^{2}\frac{x'}{k}P_{-}^{R}(\theta)
\end{aligned}$$

Essa é a transformação da coordenada *t.* Vamos obter a transformação do espaço.

$$s=k\frac{KT_{2}'+\left(T_{1}'/K\right)}{2}$$

$$s=\frac{\left(K^{2}+1\right)s'+\left(K^{2}-1\right)\tau'}{2K}$$

$$\begin{aligned}
&s=P_{+}^{R}(\theta)\left(s'+P_{\odot}^{R}(\theta)\right)\\
&s=s'P_{+}^{R}(\theta)+R\tau'P_{-}^{R}(\theta)\\
&x=x'P_{+}^{R}(\theta)+kt'P_{-}^{R}(\theta)
\end{aligned}$$

Para encerrarmos este tópico sobre cálculo *K,* recorremos as reflexões de Bohm (2015, p. 190):

É evidente que o cálculo K nos fornece uma maneira muito direta de obter muitas das relações que foram historicamente derivadas primeiro com base na transformação de Lorentz. A vantagem do cálculo de K é que torna muito evidente a conexão entre essas relações e os princípios e fatos básicos subjacentes à teoria. De fato, partindo do princípio da relatividade e da invariância da velocidade da luz, vimos que a própria transformação de Lorentz se segue simplesmente de certas características geométricas e estruturais dos padrões de certos conjuntos de eventos físicos. No entanto, por mais elegante e direto que seja, o cálculo de K ainda não foi desenvolvido o suficiente para substituir a transformação de Lorentz em todas as diferentes relações que são significativas na teoria da relatividade. Assim, a situação atual é que a abordagem da transformação de Lorentz e a abordagem do cálculo do K se complementam, no sentido de que cada uma delas oferece percepções que não são prontamente obtidas na outra. Assim, o cálculo *K* de Bondi está generalizado para qualquer variedade Espaço-Tempo e podemos usa-la com a mesma eficiência no espaço euclidiano e galileano.

# Anti-Matéria como a Matéria em um Espaço-Tempo Euclidiano

O estudo da variedade espaço-tempo euclidiano é um assunto pouco abordado nas comunicações de físicas, uma vez que na natureza não se tem observado vizinhanças infinitamente pequenas onde a covariância de Lorentz deve ser substituída pela covariância de Euclides. A variedade euclidaiana suscitou algum interesse entre os pesquisadores de curvas fechadas no tempo, pois o tempo euclidiano (ou imaginário) é localmente fechado. Vilenki (1986), James Hartle e Stephen Hawking (1983), dedicaram seus esforços no programa de emergência do tempo imaginário para a cosmologia. Deltete e Guy (1996, p. 185) apresentaram alguns problemas que inviabilizam as pesquisas com tempo imaginário na cosmologia:

> Todos esses modelos conjeturam uma junção entre regiões (ou eras) de tempo imaginário e regiões (ou eras) de tempo real; portanto, para facilitar a referência, chamaremos o último problema de 'problema de junção'. Dizer que o "problema de união" é problemático, no entanto, não significa que os proponentes de tais modelos se preocupem muito com o significado do tempo imaginário ou com a transição do tempo imaginário para o tempo real. Eles não; mas eles deveriam. Pois, como argumentaremos neste ensaio, várias tentativas de interpretar a transição de uma maneira logicamente consistente e fisicamente significativa fracassam. Além disso, como não parece haver nenhuma maneira de resolver o 'problema de junção' de uma maneira logicamente consistente e fisicamente significativa, concluímos que a noção de 'emergir do tempo imaginário' é incoerente. Uma consequência dessa conclusão parece ser que toda a classe de modelos cosmológicos que apelam ao tempo imaginário é assim refutada.

Nesta seção buscamos entender melhor as propriedades de uma variedade plana que emerge a partir de um tempo imaginário ou euclidiano por meio do estudo da inércia da energia para esse espaço-tempo. Nosso estudo sugere que o Mar de Dirac, um desdobramento da previsão da equação de Dirac sobre os estados de energia, é uma variedade euclidiana. Como a variedade euclidiana permite viagens hiper-luminais e curvas fechadas no tempo, é por isso que a propulsão de Alcubierre exige em uma variedade localmente minkowskiana energia ou inércia negativa (ALCUBIERRE, 1994). Nossa hipótese é que essa exigência implique que o espaço-tempo ao redor dos viajantes de Alcubierre seja euclidiano. Também apontamos que os efeitos residuais do mar de Dirac, como efeito Kasimir, são qualidades naturais da variedade de Euclides.

### A Inércia da Energia em Variedades Euclidianas

Nas variedades minkowskianas, a energia contribui para o conteúdo inercial de um sistema fechado, somando a sua massa total (LANGEVIN, 1913, 1922). Nas variedades galileanas, a energia não apresenta inércia. Podemos previamente conjecturar que na variedade euclidiana a energia também contribua para o conteúdo inercial de um sistema fechado, mas subtraindo sua massa total. Para verificarmos essa hipótese, vamos operar as transformações de Euclides que obtivemos anteriormente:

$$\begin{cases} w = \dfrac{u+v}{1-\dfrac{uv}{c^2}} \\ \Gamma = \dfrac{1}{\sqrt{1+\dfrac{v^2}{c^2}}} \end{cases}$$

Nós precisamos diferenciar a composição das velocidades, mas para tornar os cálculos mais simples, iremos usar um pequeno "truque matemático", que consiste em aplicar *ln* nos dois lados da equação:

$$\ln w = \ln\left(\frac{u+v}{1-\dfrac{uv}{c^2}}\right)$$

$$\ln w = \ln(u+v) - \ln\left(1-\frac{uv}{c^2}\right)$$

Diferenciando a composição das velocidades em relação à *u*:

$$\frac{d(\ln w)}{du} = \frac{d}{du}\left[\ln(u+v) - \ln\left(1-\frac{uv}{c^2}\right)\right]$$

$$\frac{1}{w}\frac{dw}{du} = \frac{1}{(u+v)}\frac{d(u+v)}{du} - \frac{1}{\left(1-\dfrac{uv}{c^2}\right)}\frac{d}{du}\left(1-\frac{uv}{c^2}\right)$$

$$\frac{1}{w}\frac{dw}{du} = \frac{1}{(u+v)} + \frac{v}{c^2\left(1-\dfrac{uv}{c^2}\right)}$$

$$\frac{1}{w}\frac{dw}{du} = \frac{1}{(u+v)} + \frac{v}{(c^2-uv)}$$

$$\frac{1}{w}\frac{dw}{du} = \frac{(c^2-uv)+(u+v)v}{(u+v)(c^2-uv)}$$

$$\frac{1}{w}\frac{dw}{du} = \frac{(c^2-uv)+(v^2+uv)}{(u+v)(c^2-uv)}$$

$$\frac{1}{w}\frac{dw}{du} = \frac{(c^2+v^2)}{(u+v)(c^2-uv)}$$

Evidenciando $c^2$ no numerador e no denominador, obtemos:

$$\frac{1}{w}\frac{dw}{du} = \frac{c^2\left(1+\frac{v^2}{c^2}\right)}{c^2\left(u+v\right)\left(1-\frac{uv}{c^2}\right)}$$

$$\frac{dw}{du} = \frac{1}{\Gamma^2}\frac{w}{\left(u+v\right)\left(1-\frac{uv}{c^2}\right)}$$

Substituindo o valor de *w*:

$$\frac{dw}{du} = \frac{1}{\Gamma^2}\frac{\left(u+v\right)}{\left(u+v\right)\left(1-\frac{uv}{c^2}\right)\left(1-\frac{uv}{c^2}\right)}$$

Portanto o diferencial de *w* será

$$dw = \frac{1}{\Gamma^2}\frac{du}{\left(1-\frac{uv}{c^2}\right)^2}$$

No instante em que P está momentaneamente se movendo com as coordenadas K (ou seja, quando $u = 0$, então P está em repouso em K e w = v), temos

$$dw = \frac{1}{\Gamma^2}du$$

Tomemos o tempo próprio, para *y* e *z*, fixos:

$$d\tau^2 = dx^2 + c^2dt^2$$

Evidenciando *dt,* obtemos:

$$c^2d\tau^2 = c^2 + \frac{dx^2}{dt^2}$$

$$c^2d\tau^2 = \left(c^2 + v^2\right)dt^2$$

$$d\tau^2 = \left(1+\frac{v^2}{c^2}\right)dt^2$$

$$d\tau = \sqrt{1+\frac{v^2}{c^2}}\,dt$$

$$d\tau = \frac{dt}{\Gamma}$$

$$dt = \Gamma d\tau$$

Divdindo *dw* por *dt,*

$$\frac{dw}{dt}=\frac{1}{\Gamma^2}\frac{du}{dt}$$
$$\frac{dw}{dt}=\frac{1}{\Gamma^2}\frac{du}{(\Gamma d\tau)}$$

Levando em consideração que o lado esquerdo é a aceleração no referencial em *movimento* e a derivada do lado direito, a aceleração no referencial *estacionário*, obtemos:

$$a=\frac{a_0}{\Gamma^3}$$

$$a_0=\Gamma^3 a$$

Agora devemos estudar a transformação das forças longitudinais[15]. Segundo Brown (2012):

> Por simetria, uma força *F* exercida ao longo do eixo do movimento entre uma partícula em repouso em k em uma partícula idêntica *P* em repouso em *K* deve ser de magnitude igual e oposta em relação aos dois quadros de referência. Além disso, por definição, uma força de magnitude F aplicada a uma partícula de "massa em repouso" $m_o$ resultará em uma aceleração $a_0 = F/m_o$ em termos de coordenadas inerciais nas quais a partícula está momentaneamente em repouso.

Como as forças longitudinais são invariantes,

$$F=F_0$$

Expressando a força $F_o$ como o produto de sua massa inercial própria pela sua aceleração no referencial próprio:

$$F=m_0 a_0$$

Substituindo a aceleração no referencial próprio pela aceleração no referencial em movimento:

$$F=m_0\Gamma^3 a$$

Usando a definição de força como a variação da quantidade de movimento:

$$F=\frac{d(mv)}{dt}$$

Podemos escrever a expressão da força da seguinte forma:

$$F=\frac{d(m_0\Gamma v)}{dt}$$

Portanto, podemos concluir que a transformação da massa será:

[15] Para uma derivação mais rigorosa ver: Martins (2012, p. 104-105).

$$m = m_0 \Gamma$$

Como variedade euclidiana, à medida que a velocidade aumenta, o fator gama diminui, então, diferente da variedade minkowskiana, à medida que a velocidade do corpo aumenta, a sua inércia diminui. Como explicar esse fato? Na variedade minkowskina atribuímos uma inércia a energia (LANGEVIN, 1913, 1922). Um aumento de velocidade do corpo, corresponde a um aumento de sua energia cinética, e como a energia apresenta inércia, a massa total do corpo também deve aumentar na mesma proporção. Na variedade euclidiana a energia deve apresentar uma inércia negativa, por isso quanto maior a energia transferida ao corpo, menor será sua massa. Para verificarmos esse fato, vamos calcular a relação massa-energia. Novamente iremos tomar o logaritmo da relação e depois deriva-la em relação ao tempo:

$$\ln(m) = \ln(m_0 \Gamma)$$

$$\ln(m) = \ln(m_0) + \ln(\Gamma)$$

$$\ln(m) = \ln(m_0) + \frac{1}{2}\ln\left(\frac{c^2}{c^2 + v^2}\right)$$

$$\ln(m) = \ln(m_0) + \frac{1}{2}\ln(c^2) - \frac{1}{2}\ln(c^2 + v^2)$$

Derivando a função em relação à *t:*

$$\frac{m'}{m} = -\frac{1}{2}\frac{2vv'}{(c^2 + v^2)}$$

$$\frac{m'}{m} = -\frac{vv'}{(c^2 + v^2)}$$

Evidenciando $c^2$ no denominador:

$$\frac{m'}{m} = -\frac{vv'}{c^2\left(1 + \frac{v^2}{c^2}\right)}$$

$$\frac{m'}{m} = -\frac{v'\Gamma^2 v}{c^2}$$

$$m' = -\frac{m\Gamma^2 v' v}{c^2}$$

Escrevendo a derivada da massa na notação diferencial e usando a transformação da massa,

$$\frac{dm}{dt} = -\frac{m_0\Gamma^3 v'v}{c^2}$$

Isolando *dm* e considerando que o termo do numerador é o produto da força longitudinal pela velocidade:

$$dm = -\frac{Fvdt}{c^2}$$

Como o produto da velocidade pelo diferencial de tempo é o diferencial de espaço na direção *x,*

$$dm = -\frac{Fdx}{c^2}$$

Usando a definição de trabalho mecânico na direção longitudinal:

$$dm = -\frac{dW}{c^2}$$

Integrando a equação do repouso à uma velocidade arbitrária *v* e levando em consideração o teorema trabalho-energia:

$$m - m_o = -\frac{\Delta E}{c^2}$$

Isolando a energia, obtemos a relação massa-energia na variedade euclidiana:

$$\Delta E = -(m - m_o)c^2$$

$$\Delta E = -\Delta mc^2$$

Esse resultado confirma nossa hipótese que a inércia associada a energia é negativa e vice-versa.

Assim como ocorre na variedade minkowskiana, existe uma energia de repouso, porém essa energia de repouso é negativa:

$$E = -m_0 c^2$$

Como mostrou Minkowski (1909), a variedade euclidiana é difeomórfica a variedade galileana no limite de *c* tendendo ao infinito, isso significa que para uma vizinhança pequena, a variedade euclidiana se comporta como uma variedade galileana. Vamos verificar se a nossa equação satisfaz essa correspondência. Explicitando o fator gama, teremos:

$$\Delta E = m_o(1 - \Gamma)c^2$$

Vamos expandir o fator gama em uma série de Taylor:

$$\left(1+\frac{v^2}{c^2}\right)^{-1/2}=1-\frac{v^2}{2c^2}+\sum_{n=2}^{\infty}O\left(\frac{k_n}{c^{2n}}\right)$$

onde $k_n$ são funções constantes da velocidade.

Portanto, a variação da energia assume a seguinte forma:

$$\Delta E=m_o\left(1-1+\frac{v^2}{2c^2}-\sum_{n=2}^{\infty}O\left(\frac{k_n}{c^{2n}}\right)\right)c^2$$

$$\Delta E=m_o\left[\frac{v^2}{2}-\sum_{n=2}^{\infty}O\left(\frac{k_n}{c^{2(n-1)}}\right)\right]$$

Tomando $c$ tendendo ao infinito, o somatório tende a zero e recuperamos a expressão da energia cinética clássica:

$$\Delta E=\frac{m_o v^2}{2}$$

Após essa caracterização da energia na variedade euclidiana, Agora vamos mostrar que essas propriedades correspondem justamente mar de Dirac, previsto pela equação de Dirac.

**Equação de Dirac e a Energia Negativa**

Em 1928, Paul Dirac deduziu uma equação relativística para descrever o comportamento do elétron. A solução dessa equação incluí naturalmente a função de *spin,* e levou a previsão do pósitron (antipartícula do elétron) e da energia negativa. Uma apresentação mais detalhada está no livro do Eletrodinâmica Quântica (BASSALO, 2006), o qual o leitor deverá consultar caso sinta que falta algum detalhe. Para tornar o texto menos carregado, adotaremos o sistema de unidades naturais:

$$c=1 \qquad \hbar=1$$

Tomemos a equação de Dirac, em coordenadas naturais:

$$\left(\delta_i\hat{p}^i-m\hat{I}\right)|\psi\rangle=0$$

As componente do spinor de Dirac são:

$$|\psi\rangle=\begin{pmatrix}\psi_0\\ \psi_1\\ \psi_2\\ \psi_3\end{pmatrix}$$

A solução da equação de Schroedinger para um elétron livre é uma onda plana, dada por:

$$\psi(\vec{r}) = e^{-i\vec{r}\cdot\vec{p}}$$

Portanto, vamos procurar uma solução para equação de Dirac que corresponda a onda plana para velocidades pequenas.

$$\psi(r_i) = e^{-ir_i p^i} u(p^i)$$

Substituindo na equação de Dirac:

$$\left(i\delta_i \nabla^i - mI\right) e^{-ir_i p^i} u = 0$$
$$\left(i\delta_i \nabla^i e^{-ir_i p^i} - mIe^{-ir_i p^i}\right) u = 0$$
$$\left(-iip^i \delta_i e^{-ir_i p^i} - mIe^{-ir_i p^i}\right) u = 0$$
$$\left(p^i \delta_i e^{-ir_i p^i} - mIe^{-ir_i p^i}\right) u = 0$$

Evidenciando o exponencial:

$$\left(p^i \delta_i - mI\right) ue^{-ir_i p^i} = 0$$

Isso implica que as soluções que buscamos são da forma:

$$\left(p^i \delta_i - mI\right) u = 0$$

Expandindo a soma dentro do parêntesis:

$$\left(p^0 \delta_0 + p^1 \delta_1 + p^2 \delta_2 + p^3 \delta_3 - mI\right) u = 0$$

Substituindo as matrizes e as componentes do 4-vetor de momento:

$$\left[ E\begin{pmatrix} 1 & 0 & 0 & 0 \\ 0 & 1 & 0 & 0 \\ 0 & 0 & -1 & 0 \\ 0 & 0 & 0 & -1 \end{pmatrix} - p^x \begin{pmatrix} 0 & 0 & 0 & 1 \\ 0 & 0 & 1 & 0 \\ 0 & -1 & 0 & 0 \\ -1 & 0 & 0 & 0 \end{pmatrix} - p^y \begin{pmatrix} 0 & 0 & 0 & -i \\ 0 & 0 & i & 0 \\ 0 & i & 0 & 0 \\ -i & 0 & 0 & 0 \end{pmatrix} \right.$$
$$\left. - p^z \begin{pmatrix} 0 & 0 & 1 & 0 \\ 0 & 0 & 0 & -1 \\ -1 & 0 & 0 & 0 \\ 0 & 1 & 0 & 0 \end{pmatrix} - m \begin{pmatrix} 1 & 0 & 0 & 0 \\ 0 & 1 & 0 & 0 \\ 0 & 0 & 1 & 0 \\ 0 & 0 & 0 & 1 \end{pmatrix} \right] \cdot \begin{pmatrix} u_0 \\ u_1 \\ u_2 \\ u_3 \end{pmatrix} = \begin{pmatrix} 0 \\ 0 \\ 0 \\ 0 \end{pmatrix}$$

Efetuando essa soma, obtemos a seguinte matriz:

$$\begin{pmatrix} E-m & 0 & -p^z & -(p^x-ip^y) \\ 0 & E-m & -(p^x+ip^y) & p^z \\ p^z & (p^x-ip^y) & -(E-m) & 0 \\ (p^x+ip^y) & -p^z & 0 & -(E-m) \end{pmatrix} \cdot \begin{pmatrix} u_0 \\ u_1 \\ u_2 \\ u_3 \end{pmatrix} = \begin{pmatrix} 0 \\ 0 \\ 0 \\ 0 \end{pmatrix}$$

Realizando o produto das matrizes, obtemos as quatro equações diferencias:

$$\begin{aligned} &(E-m)u_0 + 0u_1 - p^z u_2 - (p^x - ip^y)u_3 = 0 \\ &0u_0 + (E-m)u_1 - (p^x + ip^y)u_2 + p^z u_3 = 0 \\ &p^z u_0 + (p^x - ip^y)u_1 - (E-m)u_2 - 0u_3 = 0 \\ &(p^x + ip^y)u_0 - p^z u_1 + 0u_2 - (E-m)u_3 = 0 \end{aligned}$$

Esse sistema de equações é homogêneo e pela regra de Crammer ele só terá solução se o determinante da matriz dos coeficientes que acompanham o *spinor u* for nula.

$$\det\begin{pmatrix} E-m & 0 & -p^z & -(p^x-ip^y) \\ 0 & E-m & -(p^x+ip^y) & p^z \\ p^z & (p^x-ip^y) & -(E-m) & 0 \\ (p^x+ip^y) & -p^z & 0 & -(E-m) \end{pmatrix} = 0$$

O cálculo desse determinante é bastante trabalhoso. O método mais simples é a aplicação da regra de Laplace, seguido da aplicação da regra de Sarrus. Outra forma é o uso de um software de matemática simbólica. Bassalo (2006, p. 123-124) apresenta o cálculo detalhado. Seja qual for o método adoto, esse determinante é igual à:

$$\begin{aligned} &(E^2 - m^2)^2 - 2(E^2 - m^2)\left(p^{x^2} + p^{y^2} + p^{z^2}\right) \\ &+ p^{x^4} + p^{y^4} + p^{z^4} + 2p^{x^2}p^{y^2} + 2p^{x^2}p^{z^2} + 2p^{y^2}p^{z^2} = 0 \end{aligned}$$

Levando em consideração que o quadrado e a quarta potência da norma do vetor momento são dadas por:

$$p^2 = p^{x2} + p^{y2} + p^{z2}$$

$$p^4 = p^{x^4} + p^{y^4} + p^{z^4} + 2p^{x^2}p^{y^2} + 2p^{x^2}p^{z^2} + 2p^{y^2}p^{z^2}$$

Substituindo na equação:

$$(E^2 - m^2)^2 - 2(E^2 - m^2)p^2 + p^4 = 0$$

Essa expressão pode ser fatorada e escrita como:

$$\left[\left(E^2 - m^2\right) - p^2\right]^2 = 0$$

Realizando a análise dimensional dessa expressão, podemos recuperar a velocidade da luz e escrever a equação na como:

$$E^2 - \left(m^2c^4 + p^2c^2\right) = 0$$

Isolando a energia e extraindo a raiz quadrada:

$$E = \left|\sqrt{\left(m^2c^4 + p^2c^2\right)}\right|$$

Portanto há dois estados de energia: um positivo e um negativo.

$$E_+ = +\sqrt{\left(m^2c^4 + p^2c^2\right)} \qquad E_- = -\sqrt{\left(m^2c^4 + p^2c^2\right)}$$

No referencial próprio, teremos além da relação massa-energia convencional, uma relação massa-energia negativa:

$$E_+ = +m_o c^2 \qquad E_- = -m_o c^2$$

A primeira solução corresponde a relação massa-energia de uma variedade lorentziana, a segunda equação corresponde a relação massa-energia em uma variedade euclidiana. Como o espaço que corresponde as energias negativas corresponde, na antiga teoria quântica de campos, ao mar de Dirac, então somos forçados a sugerir a seguinte conclusão:

*"O mar de Dirac é uma variedade euclidiana"*

Há outro resultado igualmente interessante que podemos extrair da análise hipercomplexa (CATONI *et al,* 2008, JANCEWICZ, 1988, ÖZDEMIR, 2018): o dual da unidade imaginária, é a unidade perplexa. Tomemos o elemento de linha de uma variedade espaço-temporal plana qualquer, onde *R* pode ser uma unidade imaginária ou uma unidade perplexa:

$$\begin{aligned} ds_R &= dr + Rcdt \\ ds_R^2 &= \left\langle dr + Rcdt, dr + \bar{R}cdt \right\rangle \\ ds_R^2 &= \left\langle dr + Rcdt, dr - Rcdt \right\rangle \\ ds_R^2 &= dr^2 - R^2c^2dt^2 \end{aligned}$$

Tomemos o elemento de linha do dual de *R*:

$$\begin{aligned} ds_{*R} &= dr + \left(*R\right)cdt \\ ds_{*R}^2 &= \left\langle dr + \left(*R\right)cdt, dr + \left(*\bar{R}\right)cdt \right\rangle \end{aligned}$$

$$ds_{*R}^2 = \left\langle dr + (*R)cdt, dr - (*R)cdt \right\rangle$$
$$ds_{*R}^2 = dr^2 - (*R)^2 c^2 dt^2$$

Sem perda de generalidade, tomemos que $R = p$, i. e, a variedade é minkowskiana,

$$ds_p^2 = dr^2 - p^2 c^2 dt^2$$
$$ds_p^2 = dr^2 - c^2 dt^2$$

Portanto $*R = *p$, i.e., $*R = i$, e o dual do elemento de linha será:

$$ds_i^2 = dr^2 - (i)^2 c^2 dt^2$$
$$ds_i^2 = dr^2 + c^2 dt^2$$

que corresponde a variedade euclidiana.

Assim podemos concluir que o dual da variedade minkowskiana é a variedade euclidiana e, por conseguinte, a energia negativa é o dual da energia positiva e as antipartículas são as duais das partículas ordinárias, o mar de Dirac é o dual do nosso espaço-tempo.

**A Força de Lorentz em Variedades Euclidianas: Uma Análise Heurística**

A força de Lorentz é uma expressão válida apenas para as variedades lorentzianas, porém, usando os mecanismos de análise de que desenvolvemos nas seções anteriores, podemos generalizar essa expressão para todas as variedades espaço-temporais planas, obtendo uma nova expressão da força que denominamos de *força de Poincaré*. Devido ao caráter nilpotente do número dual, para tornar a demonstração mais simples, mas sem perder o rigor, realizaremos o procedimento em duas partes. Inicialmente obtendo a expressão válida apenas para a variedade galileana e depois uma expressão geral que é válida para a variedade lorentziana e euclidiana, para então, por inspeção, buscar uma expressão geral que seja válida para as três variedades.

Tendo em mãos esse resultado, tomemos a lei de Faraday-Lenz para uma variedade arbitrária:

$$\oint_{R^2 \partial S(t)} \vec{E}(\vec{r},t) d\hat{l} = -R^2 \iint_{S(t)} \frac{d\vec{B}(\vec{r}.t)}{dt} d\hat{S}$$

Escrevendo o campo elétrico em função da força e da carga elétrica, e comutando os operadores diferenciais:

$$\oint_{R^2 \partial S(t)} \frac{\vec{F}}{q}(\vec{r},t) d\hat{l} = -R^2 \frac{d}{dt} \iint_{S(t)} \vec{B}(\vec{r},t) d\hat{S}$$

Para uma variedade galileana, $R^2 = 0$, portanto a expressão assume a seguinte forma:

$$\oint_{R^2\partial S(t)} \frac{\vec{F}}{q}(\vec{r},t)\,d\hat{l} = 0$$

que implica que o integrando deve se anular sobre todo o contorno:

$$\frac{\vec{F}}{q}(\vec{r},t) = 0$$

Portanto a força de Poincaré deve ser zero:

$$\vec{F}(\vec{r},t) = 0$$

Usaremos esse valor como referência para obter a expressão generalizada da força de Lorentz. Agora nos focaremos apenas na análise das variedades lorentzianas e euclidianas, o que implica que $R^2$ é igual a +1 ou a -1, respectivamente. Para isso tomaremos uma superfície imaginária *S(t)* cuja fronteira é orientada conforme o fator $R^2$. Para esta superfície podemos aplicar o Teorema de Stokes Generalizado:

$$\int_{R^2\partial S(t)} \omega = R^2 \int_{S(t)} d\omega$$

Multiplicando os dois lados da equação por $R^2$,

$$R^2 \int_{R^2\partial S(t)} \omega = R^4 \int_{S(t)} d\omega$$

Como $R^4$ é igual a unidade nas duas variedades, então o Teorema Generalizado de Stokes assume a seguinte forma:

$$\int_{S(t)} d\omega = R^2 \int_{R^2\partial S(t)} \omega$$

Agora, retomemos a Lei de Faraday-Lenz na forma de força:

$$\oint_{R^2\partial S(t)} \frac{\vec{F}}{q}(\vec{r},t)\,d\hat{l} = -R^2 \frac{d}{dt} \iint_{S(t)} \vec{B}(\vec{r},t)\,d\hat{S}$$

Usando o teorema generalizado de Helmholtz, a integral do lado direito pode ser escrita como:

$$\oint_{R^2\partial S(t)} \frac{\vec{F}}{q}(\vec{r},t)\,d\hat{l} = -R^2 \iint_{S(t)} \left( \frac{\partial \vec{B}(\vec{r},t)}{\partial t} + \left[\nabla \cdot \vec{B}(\vec{r},t)\right]\vec{v} \right) d\hat{S}$$
$$+ R^2 \oint_{R^2\partial S(t)} \left[\vec{v} \times \vec{B}(\vec{r},t)\right] \cdot d\hat{l}$$

Mas a divergência do campo magnético é nulo, então o segundo termo na integral dupla é zero:

$$\oint_{R^2\partial S(t)} \frac{\vec{F}}{q}(\vec{r}.t)\,d\hat{l} = -R^2 \iint_{S(t)} \frac{\partial \vec{B}(\vec{r}.t)}{\partial t} d\hat{S} + R^2 \oint_{\partial S(t)} \left[\vec{v} \times \vec{B}(\vec{r},t)\right] \cdot d\hat{l}$$

Utilizando a lei de Faraday-Maxwell,

$$\oint_{R^2\partial S(t)} \frac{\vec{F}}{q}(\vec{r}.t)\,d\hat{l} = \iint_{S(t)} \left(\nabla\times\vec{E}(\vec{r}.t)\right)d\hat{S} + R^2\oint_{\partial S(t)}\left[\vec{v}\times\vec{B}(\vec{r}.t)\right]\cdot d\hat{l}$$

Usando o Teorema de Stokes Generalizado, a integral dupla pode ser escrita como uma integral de linha sobre a fronteira de *S(t).*

$$\oint_{R^2\partial S(t)} \frac{\vec{F}}{q}(\vec{r},t)\,d\hat{l} = R^2\oint_{R^2\partial S(t)} \vec{E}(\vec{r},t)\,d\hat{l} + R^2\oint_{\partial S(t)}\left[\vec{v}\times\vec{B}(\vec{r},t)\right]\cdot d\hat{l}$$

que pode ser posta em uma única integral:

$$\oint_{R^2\partial S(t)} \frac{\vec{F}}{q}(\vec{r},t)\,d\hat{l} = R^2\oint_{R^2\partial S(t)} \left(\vec{E}(\vec{r}.t)+\left[\vec{v}\times\vec{B}(\vec{r}.t)\right]\right)d\hat{l}$$

Essa igualdade será satisfeita se o integrando do lado esquerdo for igual ao integrando do lado direito:

$$\frac{\vec{F}}{q}(\vec{r}.t) = R^2\left[\vec{E}(\vec{r}.t)+\vec{v}\times\vec{B}(\vec{r}.t)\right]$$

Multiplicando pela carga elétrica, obtemos a expressão da força de Poincaré (força de Lorentz generalizada):

$$\vec{F}(\vec{r}.t) = q\left(\vec{E}(\vec{r}.t)+R^2\left[\vec{v}\times\vec{B}(\vec{r}.t)\right]\right)$$

Ou em uma notação mais compacta:

$$\vec{F} = qR^2\left(\vec{E}+\left[\vec{v}\times\vec{B}\right]\right)$$

Se tomarmos $R^2 = 0$, obtemos que a força sobre a carga é nula:

$$\vec{F} = 0$$

que é o valor que calculamos anteriormente para as variedades galileanas. Portanto, a expressão da força de Poincaré que deduzimos é válida para todas as três variedades.

A partir dessa lei, podemos retirar as leis da força elétrica e da força magnética:

$$\vec{F}_e = R^2\left(q\vec{E}\right),$$
$$\vec{F}_m = R^2\left(q\vec{v}\times\vec{B}\right)$$

Como estamos interessados nas propriedades da variedade euclidiana, tomemos $R^2 = -1$. Nestas condições, a força de Poincaré adquire a seguinte configuração:

$$\vec{F} = -q\left(\vec{E} + \left[\vec{v} \times \vec{B}\right]\right)$$

Esse resultado nos informa que uma partícula negativa, como o elétron, irá se comportar na presença de um campo eletromagnético como uma partícula positiva. Da mesma forma que uma partícula positiva, como um próton, irá se comportar como uma partícula negativa. Se considerarmos o movimento quase estacionário, em que a variação da inércia é muito pequena, seriamos incapazes de dizer se uma partícula é um elétron em uma variedade euclidiana ou é um pósitron em uma variedade lorentziana.

Igualmente interessante é analisar o comportamento do nêutron. Essas partículas não interagem eletricamente, porém possuem um momento de dipolo magnético permanente e por isso são como pequenos imãs. Nas variedades euclidianas verifica-se a anti-lei de Lenz. Na prática, um observador lorentziano desavisado concluiria que o orientação do momento de dipolo magnético do nêutron está invertido e por isso, está ocorrendo uma inversão da lei de Lenz. Ele também poderia concluir que o nêutron está “voltando no tempo” e com isso salvar a lei de Lenz. Mas estas duas interpretações são justamente o que definem um anti-neutron.

Há outro resultado igualmente interessante que podemos extrair da análise hipercomplexa (CATONI *et al,* 2008, JANCEWICZ, 1988, ÖZDEMIR, 2018): o dual da unidade imaginária, é a unidade perplexa. Tomemos o elemento de linha de uma onde *R* pode ser uma unidade imaginária ou uma unidade perplexa:

$$\begin{aligned}
ds_R &= dr + Rcdt \\
ds_R^2 &= \left\langle dr + Rcdt, dr + \bar{R}cdt \right\rangle \\
ds_R^2 &= \left\langle dr + Rcdt, dr - Rcdt \right\rangle \\
ds_R^2 &= dr^2 - R^2c^2dt^2
\end{aligned}$$

Tomemos o elemento de linha do dual de *R*:

$$\begin{aligned}
ds_{*R} &= dr + (*R)cdt \\
ds_{*R}^2 &= \left\langle dr + (*R)cdt, dr + (*\bar{R})cdt \right\rangle \\
ds_{*R}^2 &= \left\langle dr + (*R)cdt, dr - (*R)cdt \right\rangle \\
ds_{*R}^2 &= dr^2 - (*R)^2 c^2dt^2
\end{aligned}$$

Sem perda de generalidade, tomemos que $R = p$, i. e, a variedade é lorentziana,

$$\begin{aligned}
ds_p^2 &= dr^2 - p^2c^2dt^2 \\
ds_p^2 &= dr^2 - c^2dt^2
\end{aligned}$$

Portanto $*R = *p$, i.e., $*R = i$, e o dual do elemento de linha será:

$$ds_i^2 = dr^2 - (i)^2 c^2 dt^2$$
$$ds_i^2 = dr^2 + c^2 dt^2$$

que corresponde a variedade euclidiana.

Portanto a nossa conclusão é que nas variedades euclidianas, as partículas lorentzianas se comportam como suas antipartículas, exatamente como ocorre no Mar de Dirac. E que o dual da variedade lorentziana é a variedade euclidiana e, por conseguinte, a energia negativa é o dual da energia positiva e as antipartículas são as duais das partículas, o mar de Dirac é o dual do nosso espaço-tempo

**A Força de Lorentz em Variedades Euclidianas: Uma Análise Heurística**

Outra evidência de que a anti-matéria pode ser compreendida como a matéria ordinária em uma variedade euclidiana, decorre da análise da força de Lorentz. Para obtermos a sua expressão mais geral, vamos utilizar um método heurístico, mas sem comprometer o rigor. A força de Lorentz pode ser escrita como uma combinação linear da força elétrica e da força magnética que atua sobre a partícula (ROSSER, 1968):

$$f_{Lorentz} = f_{elétrica} + f_{magnética}$$

Nosso objetivo é obter a expressão de cada uma dessas forças. Inicialmente, tomemos a equação de Faraday no vácuo:

$$\nabla \times \vec{E} = -R^2 \frac{\partial \vec{B}}{\partial t}$$

Por essa equação podemos inferir que a força magnética que atua sobre a partícula é expressa por (ROSSER, 1968)[16]:

$$f_{magnética} = qR^2 \left( \vec{v} \times \vec{B} \right)$$

O princípio da relatividade nos assegura que existe um referencial inercial onde a partícula está em repouso, portanto a força magnética deve ser de origem elétrica. Como $R^2$ é constante, então, podemos assumir que $qR^2$ é o equivalente de carga, portanto a equação da força elétrica pode ser escrita como (ROSSER, 1968):

$$f_{elétrica} = qR^2 \vec{E}$$

[16] A dedução de Rosser é feita para uma variedade minkowskiana, porém como $R^2$ apenas altera o sinal, o processo é o exatamente o mesmo.

Portanto, para uma partícula carregada em um referencial arbitrário, a força de Lorentz deve ser expressa pela relação:

$$f_{Lorentz} = qR^2\left(\vec{E} + \vec{v} \times \vec{B}\right)$$

Tomemos um elétron na presença de um campo eletromagnético, a expressão de suas força de Lorentz para uma variedade minkowskiana ($R^2$ = 1) será:

$$f_{Lorentz} = -e\left(\vec{E} + \vec{v} \times \vec{B}\right)$$

Entretanto, se tomarmos a expressão dessa força para um elétron em uma variedade euclidiana ($R^2$ = -1), teremos:

$$f_{Lorentz} = -e(-1)\left(\vec{E} + \vec{v} \times \vec{B}\right)$$

$$f_{Lorentz} = e\left(\vec{E} + \vec{v} \times \vec{B}\right)$$

Mas essa é a justamente a força de Lorentz que atua sobre um pósitron (anti-elétron).

Esse resultado é uma evidência favorável à nossa hipótese de que a anti-matéria pode ser compreendida como a matéria ordinária imersa em um espaço-tempo euclidiano.

**Propulsão de Alcubierre e as Variedades Euclidianas**

Em 1994, o físico mexicano Miguel Alcubierre estudou sobre quais condições físicas seriam necessárias para que observadores pudessem dobrar o espaço-tempo ou manter uma ponte de Einstein-Rosen estável. A conclusão de Alcubierre é que a componente energética ($T_{00}$) do tensor momento-energia deve ser negativa e expressa pela seguinte relação (ALCUBIERRE, 1994, L77):

$$-\frac{c^4}{8\pi G}\frac{v_s^2\left(x^2+y^2\right)}{4\left(\det g_{ij}\right)^2 r_s^2}\left(\frac{df}{dr_s}\right)^2$$

Sobre esse resultado, Alucbierre faz os seguintes comentários:

> O fato de essa expressão ser negativa em todos os lugares implica que as condições de energia fracas e dominantes são violadas. De maneira semelhante, pode-se mostrar que a forte condição de energia também é violada. Vemos então que, assim como acontece com os buracos de minhoca, é preciso matéria exótica para viajar mais rápido que a velocidade da luz. No entanto, mesmo que se acredite que a matéria exótica seja proibida classicamente, é sabido que a teoria quântica de campos permite a existência de regiões com densidades de energia negativas em algumas circunstâncias especiais (como, por exemplo, no efeito Casimir [4]). Portanto, a necessidade de matéria exótica não elimina necessariamente a possibilidade de usar uma distorção no espaço-tempo, como a descrita acima, para viagens interestelares hiper-rápidas. Como comentário final, mencionarei apenas o fato de que, embora o espaço-tempo descrito pela métrica (8) seja globalmente hiperbólico e, portanto, não contenha curvas causais fechadas, provavelmente não é muito difícil construir um espaço-tempo que contém essas curvas usando uma idéia semelhante à apresentada aqui. (ALCUBIERRE, 1994, L77).

O fato da propulsão de Alcubierre exigir energia negativa é compatível com a variedade euclidiana. De fato, uma das características do tempo euclidiano é que, ao contrário do tempo minkowskiano que é um eixo retilíneo ortogonal aos eixos espaciais, o seu eixo é uma circunferência fechada ortogonal as curvas espaciais em todos os pontos. De nossa análise anterior, concluímos que o mar de Dirac é uma variedade euclidiana, portanto os efeitos a ele relacionados também se relacionam com as propriedades da variedade. Portanto, podemos concluir que as condições de viagem do tempo e viagens hiperlumuninais discutidas por Alcubierre exigem que o espaço-tempo se comporte naquela região como uma variedade euclidiana.

## Considerações Finais

A questão de pesquisa que norteou esse ensaio era avaliar a possibilidade de construção de um programa de Erlangen para o espaço-tempo. A esta questão apresentamos uma resposta positiva e a justificativa provou-se também um método de se estudar as propriedades gerais para todo espaço-tempo, dos quais o espaço de Galileu e Minkowski são apenas casos particulares. Embora o objetivo modesto (se comparado com a pretensão inicial de F. Klein), a proposta de unificação não visa esgotar as questões sobre esse assunto, mas abrir outras perspectivas de pesquisa. Mencionamos algumas:

1) Nesse trabalho impusemos a condição de que $R$ é constante, porém, como $R$ é construído a partir das características vetoriais no quartenion híbrido $Z$, se os escalares reais $a, b, c, d$ forem parametrizados por uma variável $u$, então o valor de $R$ varia com $u$, e o espaço-tempo evolui com $u$. Em particular se $u$ for o tempo cosmológico, obtemos um modelo geodinâmico. Se $u$ é a distribuição de matéria e energia do espaço-tempo em diferentes estágios cosmológicos, o universo poderia passar de Minkowskiano para Euclidiano e Galileano. Por outro lado, se $u$ é algum parâmetro espacial, o espaço-tempo poderia apresentar diferentes métricas, em diferentes regiões. De forma mais geral, um parâmetro $u$ permitiria reduzir o número de cartas espaço-temporais de atlas topológico.
2) Trabalhos de Herranz, Ortega, Santander (1999), Kisil (2007, 2008), Catoni, Boccaletti, Cannata (2008) e, mais recentes, de Kisil (2012, 2012b), Zaripov (2016), Li, Yang, Qiao (2018) e Gerard et al (2018) indicam a possibilidade de cumprir as exigências de um programa completo de Erlangen, induzindo junto a dimensão temporal, a dimensão espacial. Alguns resultados preliminares de estudos que já realizamos indicam que essa possibilidade poderá ser cumprida se o fator indutor espacial for uma função de algum parâmetro $u$. É preciso de uma investigação mais detalhada para confirmar essa conjectura.
3) Em um trabalho já concluído, que se encontra em processo de revisão para a submissão e publicação, realizamos a construção de estruturas algébricas a partir das funções de Poincaré permite unificar os grupos SO (3), SO(4) e SO(1,3) e suas respectivas álgebras de Lie em uma superestrutura induzida por $R$. Desta estrutura também deduzimos propriedades importantes do espaço-tempo geral: os coeficientes de estrutura, representação infinitesimal e spinorial.

4) Em outro trabalho também já concluído, que se encontra em processo de revisão para a submissão e publicação, realizamos a reformulação do cálculo *K* de Bondi por meio das funções de Poincaré, que passam a incluir as variedades Galileana e Euclidiana e permite compreender melhor os processos de simultaneidade e sincronização de relógios em diferentes variedades.
5) Em um trabalho ainda em desenvolvimento, contrariando a tese de Poincaré que a escolha geometria do espaço é apenas convencional, apresentamos uma proposta empírica de se determinar a natureza do espaço. Nossa proposta é construída a partir da álgebra geométrica derivada do estudo das linhas coordenadas de Plücker e Cayley, em um sistema de coordenadas homogêneas. Nossa conclusão preliminar é que os fenômenos eletromagnéticos são equivalentes a estrutura geométrica do espaço-tempo, portanto a forma da luz e das equações do eletromagnetismo, principalmente a equação da indução de Faraday-Lenz.
6) Assim como o Cálculo *K*, essa abordagem também pode ser útil do ponto de vista do ensino de física e teoria de relatividade, à nível superior, pois sugere uma abordagem alternativa para a construção da Relatividade Especial.

Também Fomos capazes de construir uma Teoria da Relatividade Especial sem precisar postular a constância da velocidade da luz. Para decidirmos qual é a assinatura da métrica da variedade tangente plana a uma vizinhança infinitamente pequena do espaço-tempo é mais inteligível apenas recorremos a dados amplamente testados e aceitos como a lei de Lenz e a forma das ondas eletromagnéticas;

Tanto Poincaré (1902) como Einstein (1984) defendiam que o conteúdo empírico de uma teoria era uma medida de sua excelência. Ao rejeitarmos o postulado da constância da velocidade da luz e desenvolvermos um programa baseado apenas nas implicações do princípio da relatividade (isotropia) e da inércia (homogeneidade), somos levados a três variedades planas induzidas pela unidade hipercomplexa *R*. A determinação da variedade mais inteligível se torna um problema empírico associado a teoria eletromagnética, visto que as linhas coordenadas da variedade coincidem com as componentes do tensor eletromagnético, mais precisamente, os efeitos relacionados ao campo elétrico, como sua rotacionalidade, pois este depende explicitamente do fator *R*. Nestas condições, a constância da velocidade da luz se torna uma previsão teórica da teoria, que foi confirmada em 1919 por Quirino Majorana (MARTINS, 2015). Do ponto de vista epistemológico, essa nova abordagem é útil, pois aumenta o conteúdo empírico da teoria.

Ao escolhermos um sistema de unidades em que a velocidade da luz é a unidade todas as nossas variedades gozam do citério de inteligibilidade exigido por Minkowski. O mais curioso que somos capazes de preservar a variedade galileana sem que seja necessário exigir que a velocidade da luz no vácuo tenda a infinito. Isso exige que a escolha da variedade seja feito por critérios empíricos. Esses critérios são exatamente os mesmos que permitem transformar o postulado da constância da velocidade da luz, em uma consequência da teoria. Em nossa análise sugerimos dois fatos empíricos qualitativos: a lei de Lenz e a forma da onda luminosa.

Um outro ponto favorável a essa abordagem é que ela relaciona as propriedades geométricas do espaço-tempo a uma unidade hipercomplexa *R.* A relação entre números hipercomplexos e as propriedades geométricas é um objeto de estudo matemático que ainda está sendo explorado pelos pesquisadores:

> Tais geometrias multidimensionais não foram completamente investigadas e isso nos permite afirmar a seguinte consideração: o tipo de números bidimensionais deriva das soluções de uma equação de grau 2. Encontramos a mesma classificação em outros campos matemáticos. Temos:
>
> • Soluções imaginárias → números complexos → geometria euclidiana → geometria diferencial de Gauss (formas diferenciais quadráticas definidas) → equações diferenciais parciais elípticas;
>
> • Soluções reais → números hiperbólicos → geometria de Minkowski (espaço-tempo) → geometria diferencial nas superfícies de Lorentz (formas diferenciais quadráticas não definidas) → equações diferenciais parciais hiperbólicas.
>
> Além disso, em mais de duas dimensões, sugerimos os seguintes elos gerais:
>
> • O tipo de soluções de uma equação algébrica de grau N → sistemas de números hipercomplexos → grupo multiplicativo → geometrias → geometrias diferenciais.
>
> Dessa maneira, a geometria diferencial em um espaço N-dimensional derivaria de uma forma diferencial de grau N, em vez das formas diferenciais quadráticas euclidianas ou pseudo-euclidianas. Essas propriedades peculiares podem abrir novos caminhos para aplicações em teorias de campo. (CATONI et al, 2008, p. 24-25).

Nesse sentido, essa proposta é a primeira abordagem relativística que associa explicitamente as propriedades físicas do espaço-tempo aos números hipercomplexos, cuja escolha é determinada empiricamente por meio da análise de fenômenos que são induzidos pela unidade hipercomplexa *R.*

Por fim vale a pena parafrasear Silva e Bagdonas (2012, p. 211) "se não há uma postura única para ser defendida como 'o que todos deveriam fazer'" podemos problematizar os métodos existentes, pois "mesmo quando não há consenso, pode-se apresentar uma pluralidade de visões, uma vez que o objetivo do ensino não é doutrinar, mas indicar razões para que se aceite uma visão particular (*Ibid,* p. 211), até porque a história da ciência já nos provou que a confluência de diferentes estilos de pensamento é essencial para a evolução e construção do conhecimento científico.

## REFERÊNCIAS